\documentclass{article}

\usepackage{graphicx}
\usepackage{amssymb}
\usepackage{amsthm}
\usepackage{amsmath}
\usepackage{lineno}
\usepackage{url}

\newcommand\bx{{\bf x}}
\newcommand\by{{\bf y}}
\newcommand\bu{{\bf u}}
\newcommand\bv{{\bf v}}
\newcommand\bp{{\bf p}}
\newcommand\bq{{\bf q}}
\newcommand\bg{{\bf g}}

\newcommand\bX{{\bf X}}
\newcommand\bpsi{\psi}

\newtheorem{definition}{Definition}
\newtheorem{theorem}{Theorem}
  
\begin{document}

%% Title, authors and addresses

\title{
      Partial Optimal Transport \\
  for a Constant-Volume Lagrangian Mesh \\
       with Free Boundaries
}

\author{Bruno L\'evy \\ Université de Lorraine, CNRS, Inria, LORIA, F-54000 Nancy, France}

\maketitle

\begin{abstract}
This article introduces a representation of dynamic meshes, adapted to some numerical simulations that require controlling the volume of objects with free boundaries, such as incompressible fluid simulation, some astrophysical simulations at cosmological scale, and shape/topology optimization. The algorithm decomposes the simulated object into a set of convex cells called a Laguerre diagram, parameterized by the position of $N$ points in 3D and $N$ additional parameters that control the volumes of the cells.  These parameters are found as the (unique) solution of a convex optimization problem  -- semi-discrete Monge-Amp\`ere equation -- stemming from optimal transport theory.  In this article, this setting is extended to objects with free boundaries and arbitrary topology, evolving in a domain of arbitrary shape, by solving a partial optimal transport problem. The resulting Lagrangian scheme makes it possible to accurately control the volume of the object, while precisely tracking interfaces, interactions, collisions, and topology changes.
\end{abstract}

%\begin{keyword}
%Fluids \sep Optimal Transport \sep Free Surface
%% MSC codes here, in the form: \MSC code \sep code
%% or \MSC[2008] code \sep code (2000 is the default)
%\end{keyword}

\section{Introduction}

Some numerical simulations require to control the volume of an object while allowing it to change shape and topology. For instance, in incompressible fluid simulation (see e.g. \cite{Frey2018}), the volume of fluid is conserved, while the shape of the fluid can considerably vary throughout the simulation, and can change topology (split and merge). In some astrophysic simulations \cite{EUR,EURNature,levy:hal-03081581}, the Universe considered at a cosmological scale can be modeled as a "fluid"\footnote{The "atoms" of this "fluid" correspond to galaxy clusters !}. With the simplifying assumption that this fluid obeys the incompressible Euler equation, one can reconstruct the full dynamics of the Universe from the knowledge of the density fluid at current time  \cite{EUR,EURNature,levy:hal-03081581}. To name another example, in shape and topology optimization (see e.g. \cite{geoffroydonders:hal-01939201}), one wants to find the shape of a given volume with maximum resistance (minimum compliance). Again, in this example, the considered shape can have an arbitrary topology, that can change during computation. There are several difficulties in the three applications mentioned above: (1) choosing a representation that can account for changes of topology, (2) controlling the volume, (3) tracking the interfaces and the changes of topology, and (4) tracking the interactions and the collisions between the simulated object and the boundary of the domain. 

Clearly, to account for changes of topology, it is possible to represent the object as a density supported by a Eulerian grid (see for instance the \emph{homogeneization method} for shape optimization \cite{alma991038414809703276}) and enforce the constraint of volume conservation using Lagrange multipliers or similar techniques, then track isosurfaces in this grid. However, we think it is interesting to experiment with an alternative Lagrangian representation that directly represents the simulated object and its boundary. The proposed alternative representation has the following properties:
\begin{itemize}
 \item The new representation is a Lagrangian mesh that continuously depends on a set of $N$ points in 3D space;
 \item each cell of the mesh has a prescribed volume for any position of the $N$ points;
 \item changes of topology, interfaces, interactions and collisions are accurately tracked.  
\end{itemize}

The approach builds on recent advances in numerical optimal transport, that resulted in Lagrangian schemes for fluid simulation \cite{DBLP:journals/corr/MerigotMT17} or early Universe reconstruction \cite{levy:hal-03081581}. In the works cited above, the object fills the entire simulation domain. In the rest of this article, after summarizing the underlying semi-discrete optimal transport method (Section \ref{sect:semi_discrete_OT}), I show how the mathematical setting can be extended it to objects with free boundaries, by solving a \emph{partial} optimal transport problem (Section \ref{sect:SDPOT}). Then I detail the numerical algorithm that can solve this partial optimal transport problem (Section \ref{sect:algo}). Finally, I demonstrate some applications of the method to free-surface fluid simulation (Section \ref{sect:results}).

\section{Volume control through optimal transport}
\label{sect:OT}

In this section, I summarize the existing semi-discrete optimal transport framework with the objective of giving an intuition on the aspects that are important to control the volume of the cells in a Lagrangian mesh \footnote{The reader is referred to \cite{merigot2020optimal, DBLP:journals/cg/LevyS18, ComputationalOT, Santambrogio, OTON, opac-b1122739} for an extensive and general introduction on optimal transport.}. Then, in the next section, I explain how to extend this framework to simulated objects with free boundaries.

We suppose for now that the simulated object entirely fills a simulation domain $\Omega$, that can be the 2D unit square or the 3D unit cube\footnote{We will consider later in this article the case where $\Omega$ is an arbitrary simplicial set.}. The goal is now to find a partition of $\Omega$ into a set of $N$ cells $V_i$ with the following properties:
\begin{itemize}
    \item the set of cells $(V_i)_{i=1}^N$ depends on a set of parameters $(\bx_i)_{i=1}^N$. These parameters can be interpreted as 2D (resp. 3D) points (hence a \emph{Lagrangian} representation);
    \item the area (resp. volume) of each $V_i$'s is controlled: $|V_i| = \nu_i$ for some prescribed $\nu_i's$ such that $\sum_{i=1}^N \nu_i = | \Omega | = 1$;
    \item the cells $V_i$ \emph{continuously} depend on the $\bx_i$'s;
\end{itemize}

\subsection*{The Monge-Ampère equation}

Such a Lagrangian mesh parameterization can be obtained through a specific (semi-discrete) version of the Monge-Ampère (MA) equation. In the general (continuous) setting, the MA equation may be thought of as seeking for an application $T: \Omega \rightarrow \Omega$ with controlled Jacobian, by solving for a potential $\psi: \Omega \rightarrow \mathbb{R}$:

\begin{equation}
  \left\{
  \begin{array}{l}
  \mbox{det}(\mbox{Hess}(\psi)(\bx)) = \nu(\bx)\quad \forall \bx \in \Omega \\[2mm]
  \mbox{s.t. } \psi^{cc} = \psi \quad \mbox{where\ } 
  \psi^c(\bx) = \inf_{\by \in \Omega} [ \| \bx - \by \|^2 - \psi(\by)]
  \end{array}
  \right.
  \label{eqn:MA}
\end{equation}
where $\nu: \Omega \rightarrow \mathbb{R}^+$ is a square-integrable density\footnote{In its general form, the Monge-Ampère equation also has a density $\mu$ on the left-hand side, but in the context of this article, we consider a uniform density $\mu=1$.}.
In the constraint (second line), $\psi^c$ corresponds to the Legendre-Fenchel transform of $\psi$, and the condition that applying it twice to $\psi$ does not change $\psi$ means that $\psi$ is convex (because the graph of $\psi^{cc}$ corresponds to the convex hull of the graph of $\psi$). From the solution of Eq. \ref{eqn:MA}, one can deduce an \emph{optimal transport map} $T: \bx \mapsto \bx - {\small \frac{1}{2}} \nabla \psi^c(\bx)$ that minimizes a "transport cost" while satisfying mass conservation \cite{BrenierPFMR91}:
\begin{equation}
\left\{
\begin{array}{l}
    \inf_T \left[ \int_{\Omega} \| T(\bx) - \bx \|^2 d\bx \right] \quad \mbox{s. t.} \\[2mm] 
    \int_{T^{-1}(B)} d\bx = \int_B \nu(\bx) dx \quad \forall\  \mbox{measurable set } B \subset \Omega.
\end{array}
\right.
\label{eqn:Monge}
\end{equation}
With the viewpoint of optimal transport, $\psi$ can be considered as the Lagrange multiplier of the volume conservation constraint (second line). The optimal transport map $T = \mbox{Id} - {\small \frac{1}{2}} \nabla \psi^c$ is deduced from the gradient of the Legendre transform of the solution $\psi$ of the Monge-Ampère equation (\ref{eqn:MA}). The  left-hand side of the Monge-Ampère equation corresponds to the Jacobian of $T$. Hence, the Monge-Ampère equation is a mean of controlling the Jacobian of an application. The Kantorovich dual of (a relaxed version of) the optimization problem in Eq. \ref{eqn:Monge} writes:
\begin{equation}
    \left\{
    \begin{array}{l}
    \sup_{\psi} \left[ K(\psi) = 
       \int_{\Omega} \psi^c(\bx)d\bx +
       \int_{\Omega} \psi(\bx) \nu(\bx)d\bx
     \right]   
       \\[2mm]
       \mbox{s. t. } \psi^{cc} = \psi
    \end{array}
    \right.
    \label{eqn:kanto}
\end{equation}

The Kantorovich dual is smooth and convex. These properties are interesting because they can be used to establish the existence and uniqueness of the optimal transport map and the associated Lagrange multiplier $\psi$. They can be also exploited to design a numerical solution mechanism. In the next subsection, I present the so-called \emph{semi-discrete} setting, where the target density $\nu$ is replaced with a discrete probability measure, supported by a pointset $(\bx_i)_{i=1}^N$.

\begin{figure}[t]
  \centerline{
    \includegraphics[height=45mm]{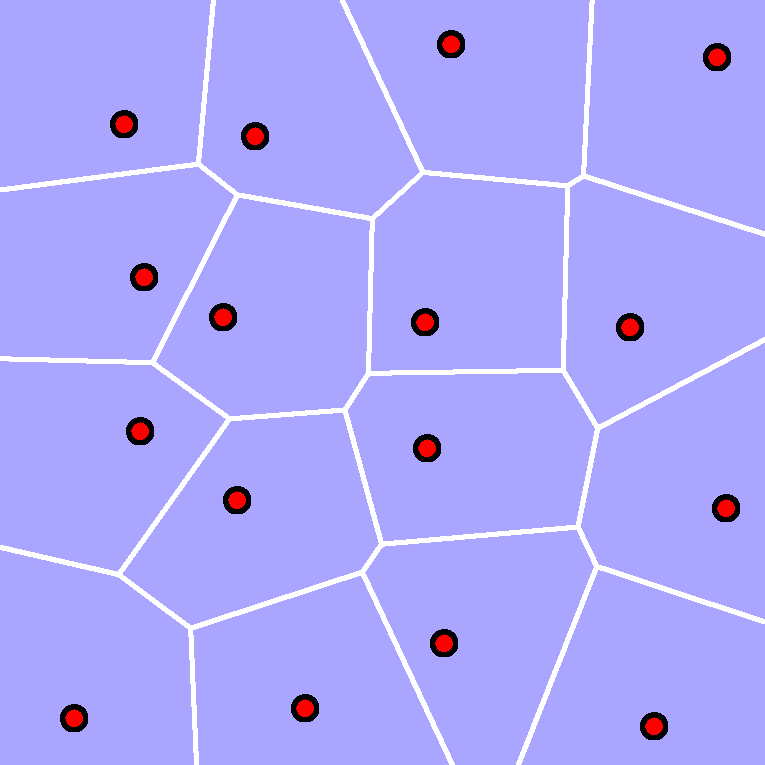}
    \hspace{2mm}
    \includegraphics[height=45mm]{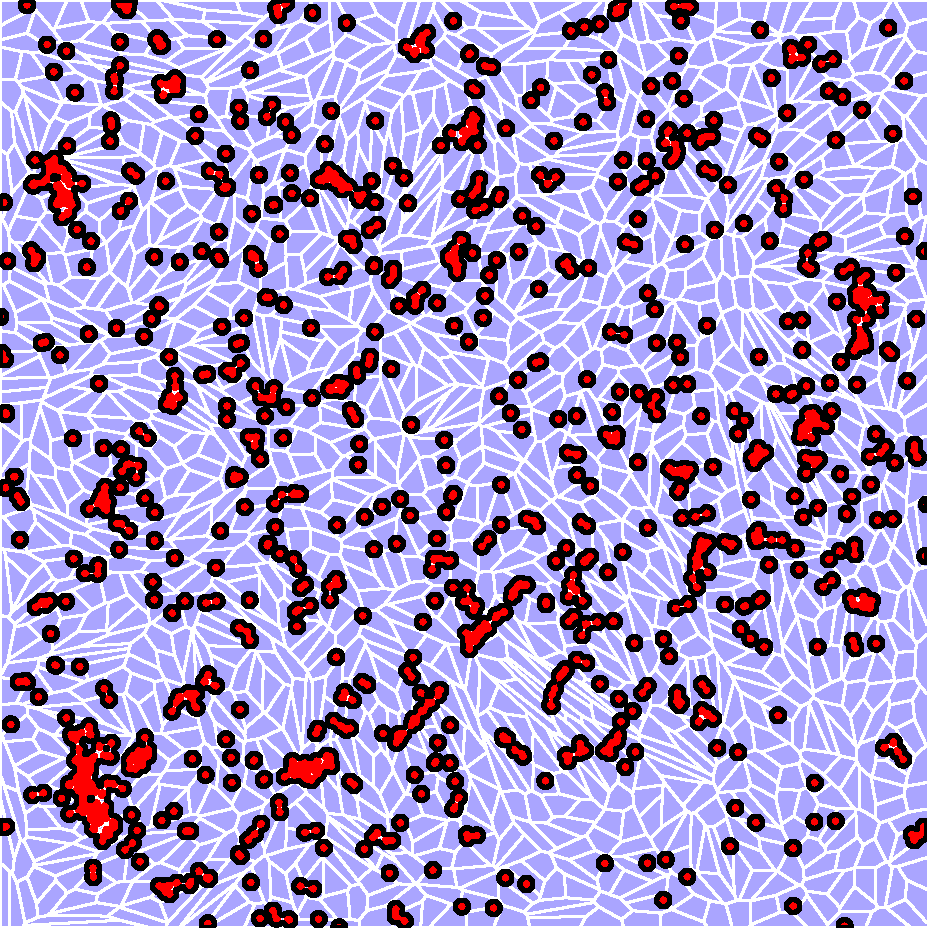}    
    \hspace{2mm}
    \includegraphics[height=45mm]{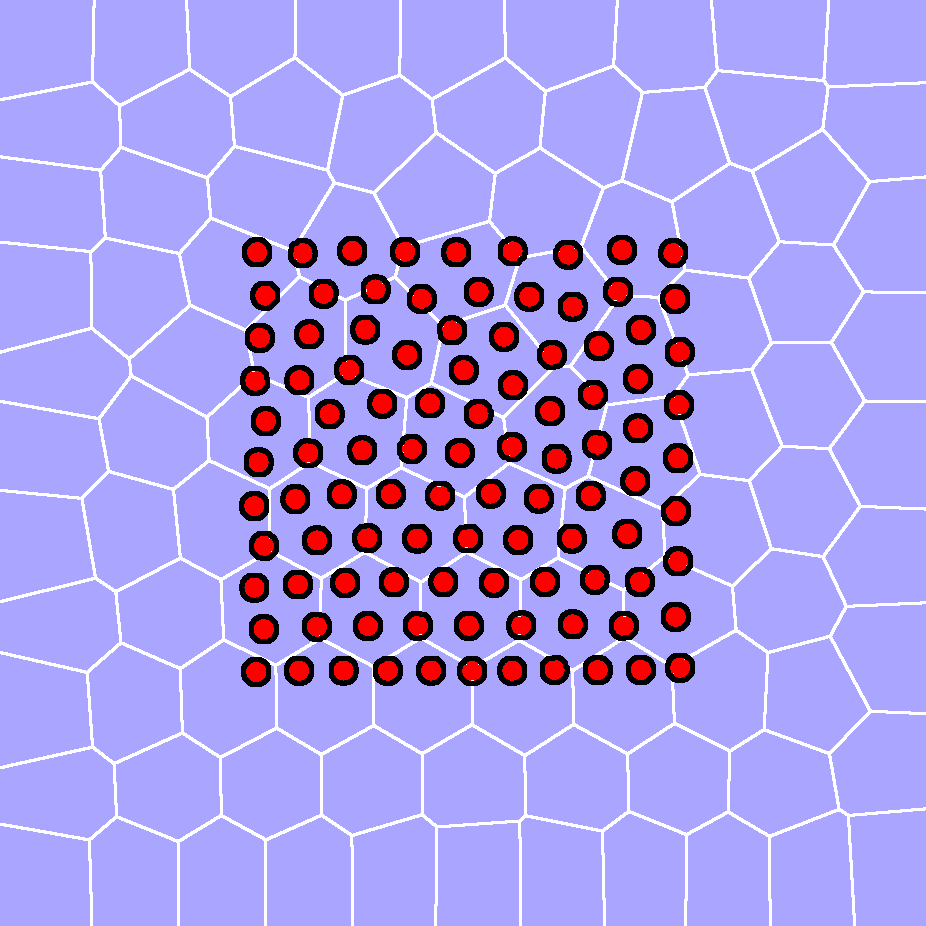}    
  }
  \caption{Left: Example of Laguerre diagram. Center and Right: unlike Voronoi cells, Laguerre cells do not necessarily contain the points they are associated with, depending on both the distribution of the points (center) and the values of the $\psi_i$'s (right). 
  }
  \label{fig:laguerre}
\end{figure}

\subsection*{Semi-discrete Optimal Transport}
\label{sect:semi_discrete_OT}

The equivalence between the Kantorovich dual problem (Eq. \ref{eqn:kanto}) and the optimal transport problem (Eq. \ref{eqn:Monge}), established in Brenier's polar factorization theorem \cite{BrenierPFMR91} characterizes mathematical objects (probability measures) that can be less regular than the densities involved in the Monge-Ampère equation (Eq. \ref{eqn:MA}). For instance, the density $\nu$ can be replaced with a weighted sum of Dirac masses 
$\nu = \sum_{i=1}^N \nu_i \delta_{\bx_i}$. A subset $B$ of $\Omega$ is measured as follows by a so-defined $\nu$:
$$
   \int_B \nu(\bx) d\bx = \sum_{j | \bx_j \in B} \nu_j.
$$

In this setting, the potential function $\psi$ is parameterized by a vector $(\psi_i)_{i=1}^N$. The funtional $K$ of the dual problem (Eq. \ref{eqn:kanto}) becomes a function $K: \mathbb{R}^N \rightarrow \mathbb{R}$ that depends on the vector $(\psi_i)_{i=1}^N$:

\begin{equation}
    \left\{
    \begin{array}{lclll}
    K(\psi) &=& \int_{\Omega}  \psi^c(\bx)d\bx & + & \int_{\Omega} \psi(\bx) \nu(\bx)d\bx \\[2mm]
            &=& \int_{\Omega}  \psi^c(\bx)d\bx & + & \sum_i \psi_i \nu_i \\[2mm]
            &=& \int_{\Omega} \inf_{i} \left[ \| \bx - \bx_i \|^2 - \psi_i \right] d\bx & + & \sum_i \psi_i \nu_i.
    \end{array}
    \right.
    \label{eqn:SDkanto1}
\end{equation}
The second line is obtained by taking into account the discrete definition of $\nu$, and the third one by replacing $\psi^c$ by its
definition (in Equation \ref{eqn:MA}). The integral in the first term of $K(.)$ can be rearranged:
\begin{equation}
    K(\psi) = \sum_i \int_{Lag_i\psi} (\| \bx - \bx_i \|^2 - \psi_i) d\bx +
    \sum_i \psi_i \nu_i
\end{equation}
where the partition of $\Omega$ into the sets $Lag_i^\psi$, called a \emph{Laguerre diagram}, is defined as follows:

\begin{definition}
Given the unit square $\Omega$ (resp. unit cube), a set $\bX = (\bx_1, \ldots \bx_N)$ of points in $\Omega$, a vector
${\bpsi} = (\psi_1, \ldots \psi_n) \in \mathbb{R}^N$, the Laguerre diagram is the partition
of $\Omega$ into the $N$ regions $Lag^\bpsi_i$ defined by:
$$
   Lag^\bpsi_i = \left\{ \bx \in \Omega \quad | \quad
     \begin{array}{lcl}
       \|\bx - \bx_i\|^2 - \psi_i & \le & \|\bx - \bx_j\|^2  -\psi_j \quad \forall j \neq i 
     \end{array} \right\}
$$
\end{definition}   

Each region $Lag^\psi_i$ is referred to as a \emph{Laguerre cell}. Laguerre diagram can be considered as a generalization of Voronoi diagram (they are equivalent in the specific case where all the $\psi_i$'s have the same value). Laguerre diagrams have been extensively studied and characterized in computational geometry \cite{DBLP:journals/siamcomp/Aurenhammer87} (in this context, if the cost corresponds to the squared distance, like in our case, Laguerre diagrams are called \emph{power diagrams}). Two examples of Laguerre diagrams are shown in Figure \ref{fig:laguerre}. It is worth mentioning that depending on the $\psi_i$'s, a Laguerre cell \emph{does not necessarily contain the point} it is associated with (unlike Voronoi cells). For instance, it is easy to check that one can translate the entire diagram by an arbitrary vector $\bu$, without changing the $\bx_i$'s, by setting $\psi_i = \sqrt{\max_j(\bu \cdot \bx_j) - \bu \cdot \bx_i}$,
hence the cells can be artitrarily far away from the $\bx_i$'s.

\begin{figure}[t]
  \centerline{
    \includegraphics[height=47mm]{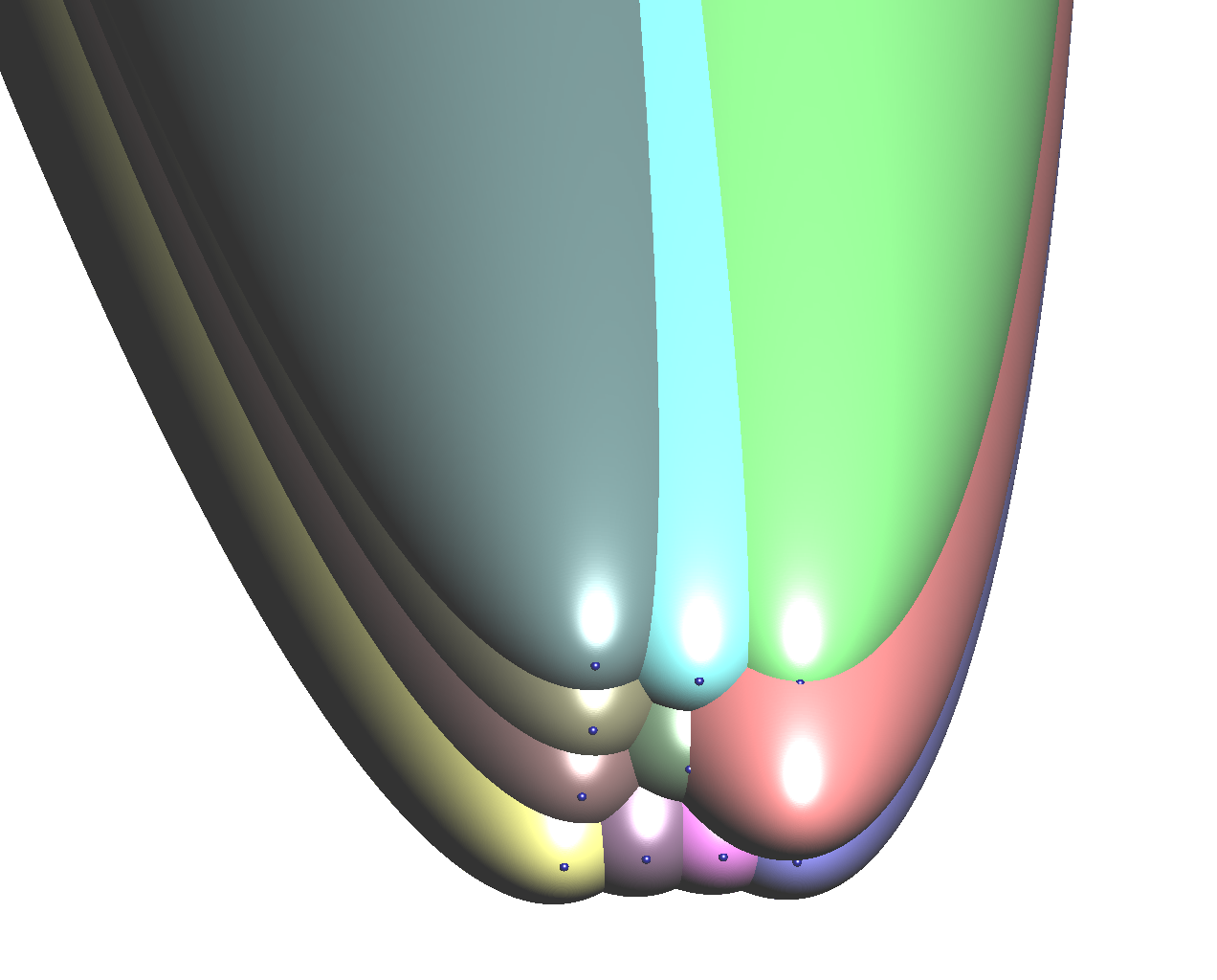}
    \includegraphics[height=47mm]{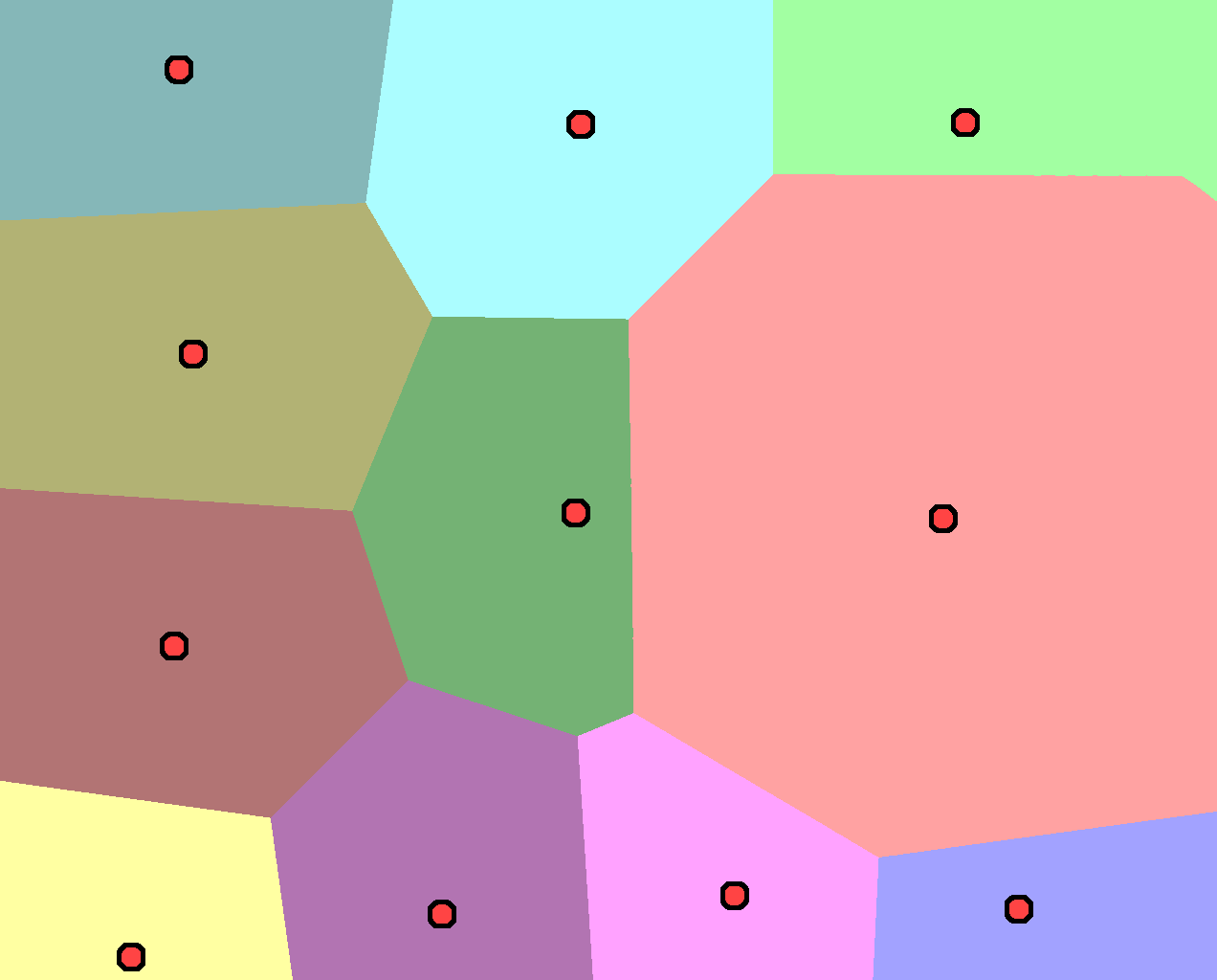}    
  }
  \caption{Laguerre diagrams considered from an alternative viewpoint. Left: the Laguerre diagram is the minimization diagram of the familly of functions
    $f_i(\bx) = \|\bx, \bx_i\|^2 - \psi_i$. Right: looking at these graphs
    from above, one sees a set of convex polygonal cells. 
  }
  \label{fig:laguerre_min_diagram}
\end{figure}
 
The Laguerre diagram can also be considered as the minimization
diagram of the familly of functions $f_i(\bx) = \| \bx - \bx_i\|^2 -
\psi_i$. The graphs of these functions (paraboloids) for a 2D diagram is pictured in Figure
\ref{fig:laguerre_min_diagram}-Left. The coefficient $\psi_i$ 'shifts' the associated
paraboloid along the $Z$ axis. Seen from above (Figure
\ref{fig:laguerre_min_diagram}-Right), the function graphs appear as convex polygonal
cells. The common boundary between two cells corresponds to the
projection of a parabola included in a plane orthogonal to the
picture, hence a straight segment.

Still considering Figure \ref{fig:laguerre_min_diagram}-Left, each coefficient $\psi_i$ corresponds to the amount of translation of the associated paraboloid along the Z axis. As one can imagine, lowering a paraboloid (that is, \emph{increasing} $\psi_i$, because of the "-" sign) increases the size of the associated Laguerre cell. Conversely, raising it decreases the size of the cell\footnote{A cell can even completely disappear if the corresponding paraboloid is shifted above all the other graphs.}.
One can also see that the Laguerre diagram does not change when adding a constant to all $\psi_i$'s (the entire diagram is shifted along the Z axis and the image does not change when viewed from above).

Now the question is: given a vector of prescribed volumes $\nu_i \ge 0$, such that $\sum_i \nu_i = | \Omega | = 1$, is it possible to find the values of $\psi_i$ such that $| Lag_i^\psi | = \nu_i$ for all $i$ ? In other words, is it possible to tune the heights of the paraboloids in such a way that the areas of the cells seen from above match the prescribed areas ? The (positive) answer is given by the following theorem \cite{DBLP:conf/compgeom/AurenhammerHA92} (it is also a direct consequence of Brenier's more general polar factorization theorem \cite{BrenierPFMR91} considered in the specific semi-discrete setting). We summarize below the main argument of the proof in \cite{DBLP:conf/compgeom/AurenhammerHA92}:
\begin{theorem}
  Given a set of $n$ points $\bX = (\bx_1, \bx_2, \bx_n)$ in $\Omega$, a set of positive prescribed volumes
  $(\nu_1, \nu_2, \ldots \nu_n)$ such that $\sum_i \nu_i = 1$,
  there exists a unique (up to a translation) set of scalars $(\psi_1, \psi_2, \ldots \psi_n)$ such that each
  Laguerre cell $Lag_i^\bpsi$ has the prescribed volume $V_i$.
  \label{thm:AHA}
\end{theorem}
\paragraph*{\bf proof}(summarized, see \cite{DBLP:conf/compgeom/AurenhammerHA92,DBLP:journals/cg/LevyS18} for a complete proof)
  Consider the Kantorovich dual function $K(\bpsi) : \mathbb{R}^N \rightarrow \mathbb{R}$ defined by:
$$  
\begin{array}{lcl}  
   K(\bf \bpsi) & = & \sum\limits_i \int\limits_{Lag^\psi_i}  (\|\bx - \bx_i\|^2 - \psi_i) d\bx + \sum\limits_i \psi_i \nu_i \\[5mm]
                      & = & \int\limits_{\Omega} \inf\limits_i \left[ \|\bx - \bx_i\|^2 - \psi_i \right] d\bx + \sum\limits_i \psi_i \nu_i
\end{array}.   
$$
The function $K$ has the following properties:
\begin{enumerate}
  \item The function $K$ is concave;
  \item the function $K$ is differentiable up to the second order;
  \item the components of the gradients of $K$ are given by: \\ $\quad \quad \partial K / \partial \psi_i = \nu_i - | Lag^\bpsi_i |$.
\end{enumerate}
We explain below how to obtain the concavity of $K$ (the two other
properties can be obtained by direct calculation. The second-order differentiability of $K$ with respect to $\psi$ is studied in \cite{journals/M2AN/LevyNAL15,DBLP:journals/corr/KitagawaMT16}, and its differentiability with respect to $\bx$ in \cite{degournay:hal-01721681}, as well as the expression of the second-order derivatives.
 
\begin{figure}
    \centering
    \includegraphics[height=40mm]{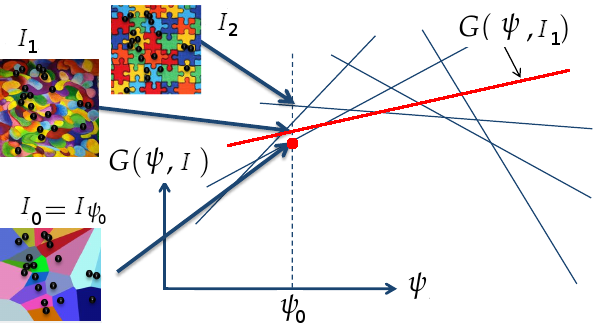}
    \includegraphics[height=40mm]{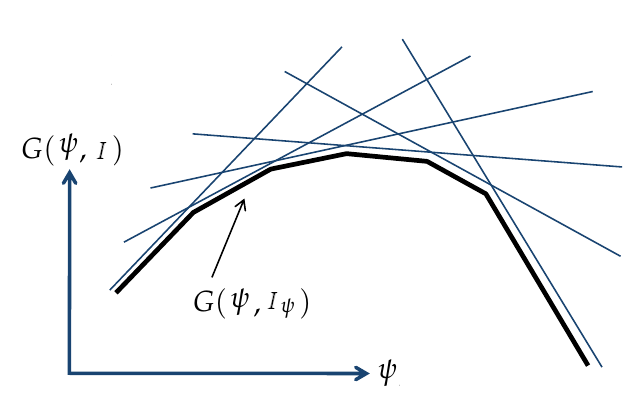}
    \caption{Convexity of $K$:  the graph of $K$ is the lower envelope of a family of affine functions $G(\psi,I)$, parameterized by the Lagrange multipliers $\psi$ and an index map $I$, that maps each point of $\Omega$ to an index in $[1\ldots N]$ (symbolized by colors on the left).}
    \label{fig:convexity}
\end{figure}

To show the concavity of $K$, we introduce the following
functional $G$, parameterized by $\bpsi$ and by an \emph{index map} $I : \Omega \rightarrow [1,\ldots N]$ that associates one of the points $\bx_{I(\bx)}$ to each point $\bx \in \Omega$.
$$
   G(\bpsi,I) = \int\limits_{\mathbb{R}^3} (\|\bx - \bx_{I(\bx)}\|^2 - \psi_{I(\bx)}) d\bx.
$$
Clearly, it is easy to verify that for a fixed index map $I$, $G$ \emph{affinely} depends on $\bpsi$. Among all the possible index maps $I$, let us consider now $I_\psi$ that maps
each point $\bx$ to the index of the Laguerre cell it belongs to:

$$
I_\psi(\bx) = i \quad \forall \bx \in Lag^\psi_i
$$
We also have (from the definition of $Lag^\psi_i$):
$$
   I_\psi(\bx) = \mbox{arg min}_i [ \|\bx - \bx_i\|^2 - \psi_i ]
$$
which implies that for a given $\psi_0$, among all possible index maps $I$, $I_{\bpsi_0}$ minimizes $G(\psi_0, I)$. Thus, the graph of $G(\bpsi,I_\bpsi)$ is the lower envelope of a
family of affine functions, hence $\bpsi \mapsto G(\bpsi, I_\bpsi)$ is
a concave function (see Figure \ref{fig:convexity}). Since $G(\bpsi,I_\bpsi)$ corresponds to the
first term of $K$, and since the second term of $K$ is linear in $\bpsi$,
$K$ is also concave. $K$ has a unique maximizer $\bpsi^*$, where its gradient
vanishes, and the expression of the derivatives indicates that the
volumes of the Laguerre cells match the prescribed volumes. 
$\blacksquare$ \\

It is easy to verify that the optimal transport map $T_{\psi^*} = \mbox{Id} - {\small \frac{1}{2}} \nabla \psi^c$ maps a point $\bx$ of $\Omega$ to the point $\bx_{I^{\psi^*}}(\bx)$:
$$
\begin{array}{lcll}
   T_{\psi^*}(\bx) & = & \bx - \frac{1}{2} \nabla \left( \inf_i [ \| \bx - \bx_i \|^2 - \psi_i ] \right) \\[2mm]
                   & = & \bx - \frac{1}{2} \nabla \left(  \| \bx - \bx_i \|^2 - \psi_i   \right) 
                        & \mbox { where } i=I^{\psi^*}(\bx) \\[2mm]
                   & = & \bx - \frac{1}{2} ( 2 \bx - 2 \bx_i )   \\[2mm]
                   & = & \bx_i  
\end{array}
$$
In other words, for all $\bx \in Lag^{\psi^*}_i$, $T(\bx) = \bx_i$. Thus, the Laguerre cell $Lag^\psi_i$ corresponds of the set of points $\bx$ of $\Omega$ mapped to $\bx_i$ through the optimal transport map. The theorem implies that the following three statements are equivalent:
\begin{enumerate}
    \item The application $T_{\psi^*}$ is the solution to the optimal transport problem (Eq \ref{eqn:Monge});
    \item the $\psi^*_i$'s are such that $|Lag_i^{\psi^*}| = \nu_i$ for all $1 \le i \le n$;
    \item the vector $(\psi^*)_{i=1}^N$ is the unique (up to a translation) maximizer of the Kantorovich dual $K$. 
\end{enumerate}

In our context, our initial motivation (controlling the volumes of a Lagrangian mesh) is satisfied by the second statement. The optimal transport map $T_{\psi^*}$ in the first one can be considered as a "by product" that we are not going to use directly. In the (2D) context depicted in Figure \ref{fig:laguerre}, the theorem means that by shifting correctly the paraboloids along the $Z$ axis, one can make the areas of the Voronoi cells match prescribed areas. In addition, the third statement and the second-order differentiability of $K$   \cite{journals/M2AN/LevyNAL15,degournay:hal-01721681} can be exploited to design a Newton algorithm \cite{DBLP:journals/corr/KitagawaMT16}. It means that from any set of points $(\bx_i)_{i=1}^N$, it is possible to compute a partition of $\Omega$ into a set of cells $V_i = Lag^\psi_i$ of prescribed volumes, that is $|V_i| = \nu_i$ for a set of positive $\nu_i$'s such that $\sum_i \nu_i = |\Omega_i|$. About the convexity constraint $\psi^{cc}=\psi$ in Eq. \ref{eqn:kanto}, in the semi-discrete setting, it is equivalent to the absence of empty Laguerre cells in the diagram ($Lag_i^\psi \neq \emptyset \quad \forall 1 \le i \le N$), as can be shown by direct computation (see \cite{DBLP:journals/cg/LevyS18} for details).

Before detailing the numerical solution mechanism that computes the $\psi_i$'s, I shall explain how the geometric setting can be extended to objects with free boundaries. 

\section{Free boundaries through partial optimal transport}
%----------------------------------------------------------
\label{sect:SDPOT}

Let us now consider a different setting, where the simulated object does not fill the volume $\Omega$ entirely. That is, the sum of 
the prescribed volumes $\nu_i$ is smaller than the volume of $\Omega$. This problem is referred to as a \emph{partial optimal transport} problem. I show how
partial optimal transport can be considered as a particular instance of optimal transport. To do so, we still consider the optimal transport problem (Eq. \ref{eqn:Monge}) 
and its Kantorivich dual (Eq. \ref{eqn:kanto}), with the difference that this time transport is done towards a set of "objects" (pointsets) $\mathcal{O}_i$:

% TODO: some revision needed here !!!
\begin{equation}
    \begin{array}{l}
    \inf_I \left[ \int_{\Omega} d^2(\bx,\mathcal{O}_{I(\bx)})d\bx\right] \\[2mm]
    \mbox{subject to } | I^{-1}(\mathcal{O}_i) | = \nu_i \quad \forall i
    \end{array}
\end{equation}
where:
\begin{itemize}
\item the index map $I: \Omega \rightarrow [1..N]$ assigns an object $\mathcal{O}_i$ to each point $\bx$ of $\Omega$;
\item each $\mathcal{O}_i$ is a set of points of $\Omega$ (that can contain either a single point, or an integer number of points, or be a continuous subset of $\Omega$);
\item each $\mathcal{O}_i$ is supposed to receive a prescribed quantity of matter $\nu_i$ (constraint);
\item the distance between a point $\bx$ and an object $\mathcal{O}$ is defined by $d(\bx, \mathcal{O}) = \inf_{\by \in \mathcal{O}} \left[ \| \bx - \by \| \right]$;
\item $I^{-1}(\mathcal{O}_i) = \{ \bx \ | \ I(\bx) = i \}$;
\item Given an index map $I$, the corresponding map $T_I: \Omega \rightarrow \Omega$ is defined by $T_I(\bx) = \mbox{arg inf}_{\by \in \mathcal{O}_{I(\bx)}} \| \bx - \by \|^2$.
\end{itemize}

Consider now the dual problem:
$$
\begin{array}{l}  
   \sup_\psi \left[
   K(\psi) = \int\limits_{\Omega} \inf\limits_i \left[ d^2(\bx, \mathcal{O}_i) - \psi_i \right] d\bx + \sum\limits_i \psi_i \nu_i \right] \\[2mm]
   \mbox{subject to } \psi^{cc} = \psi
\end{array}.   
$$

\begin{figure}[t]
  \centerline{
    \includegraphics[height=47mm]{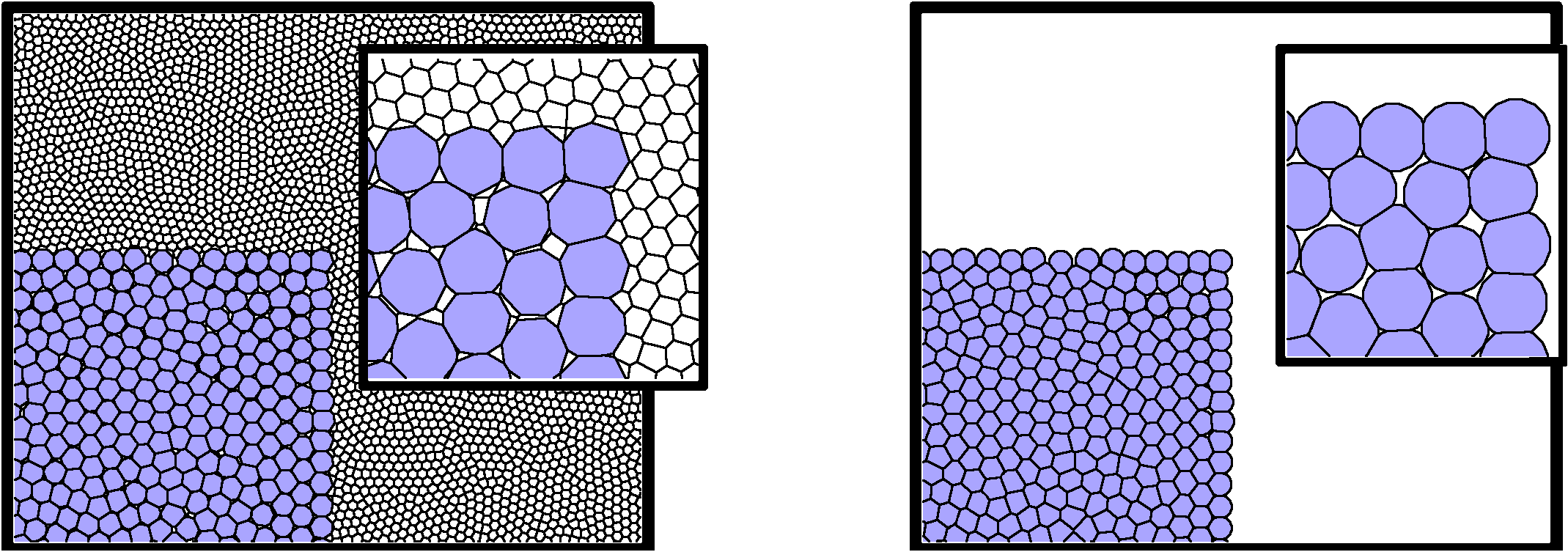}
  }
  \caption{Left: approximation of partial optimal transport using a     
     discretization of the background ("ghost cells"). 
     Right: exact solution to partial optimal transport, where the background is considered as a continuum. Each cell $V_i^\psi$ is the intersection between the laguerre cell $Lag_i^\Psi$ and the ball centered on $\bx_i$ of radius $\sqrt{\psi_i}$.
  }
  \label{fig:POT}
\end{figure}

It keeps the structure of our initial Kantorovich dual (Eq. \ref{eqn:kanto},\ref{eqn:SDkanto1}), and the argument for the convexity, smoothness, existence and uniqueness of the solution still hold, but now we have a more general setting that we can use to account for free boundaries: we consider now a set of $N+1$ objects $\mathcal{O}_{i=0}^N$:
\begin{itemize}
\item The first object $\mathcal{O}_0$ is a set of $M$ points $\{\by_1, \ldots \by_M\}$ uniformly distributed in $\Omega$;
\item each $\mathcal{O}_i$ for $1 \le i \le N$ is the singleton $\{\bx_i\}$; 
\item the object $\mathcal{O}_0$ is associated with $\nu_0 = |\Omega| - \sum_{i=1}^N \nu_i$, that is all the volume of $\Omega$ not affected to the points $\{\bx_i\}$.
\end{itemize}

This configuration is depicted in Figure \ref{fig:POT}-Left: the cells of the fluid are displayed in blue. Each of them is a convex polygon. The background points $\by_{i=1}^M$ fill the entire domain $\Omega$, including the part occupied by the fluid. All the $\by_i$ points  share the same Lagrange multiplier $\psi_0$. This may be interpreted as the fact that it takes no cost for background cells to exchange matter (air). For a part occupied by the fluid, each cell $Lag^\psi_i$ that correspond to the fluid has a Lagrange multiplier $\psi_i$ larger than $\psi_0$, then the fluid cell "shadows" the background cells. On the boundary of the fluid, the presence of the background cells limit the size of the fluid cells. The external boundary of the fluid is formed by straight segment shared by a fluid cell and a background cell (closeup in Figure \ref{fig:POT}-Left).  

A similar technique was used in \cite{DBLP:journals/tog/GoesWHPD15}, where it is proposed to insert points $\by_i$ in the vicinity of the boundary of the fluid (in the cited article, the corresponding cells are called "ghost cells"). The main advantage it that it allows directly reusing a discrete optimal transport implementation to solve transport problems with free boundaries. However, the "ghost cells" technique has two drawbacks: 
\begin{itemize}
    \item first, it is difficult to know \emph{in advance} where to insert the ghost cells to make sure the boundary of the fluid is accurately represented, that is, there is no easy way of predetermining which ghost cells influence the fluid;
    \item second, the ghost cells have a significant computational cost: computing a Laguerre diagram costs $O(N\sqrt[d]{N})$ (where $d \in \{2,3\}$ is the dimension), and the Newton algorithm converges in $O(N \log(N))$ iterations (empirical results in \cite{levy:hal-03081581}). The "ghost cells" techniques computes a diagram with $N+M$ vertices (instead of $N$), where $M$ needs to be sufficiently large to accurately capture the free boundary.  
\end{itemize}

\begin{figure}[t]
  \centerline{
    \includegraphics[height=47mm]{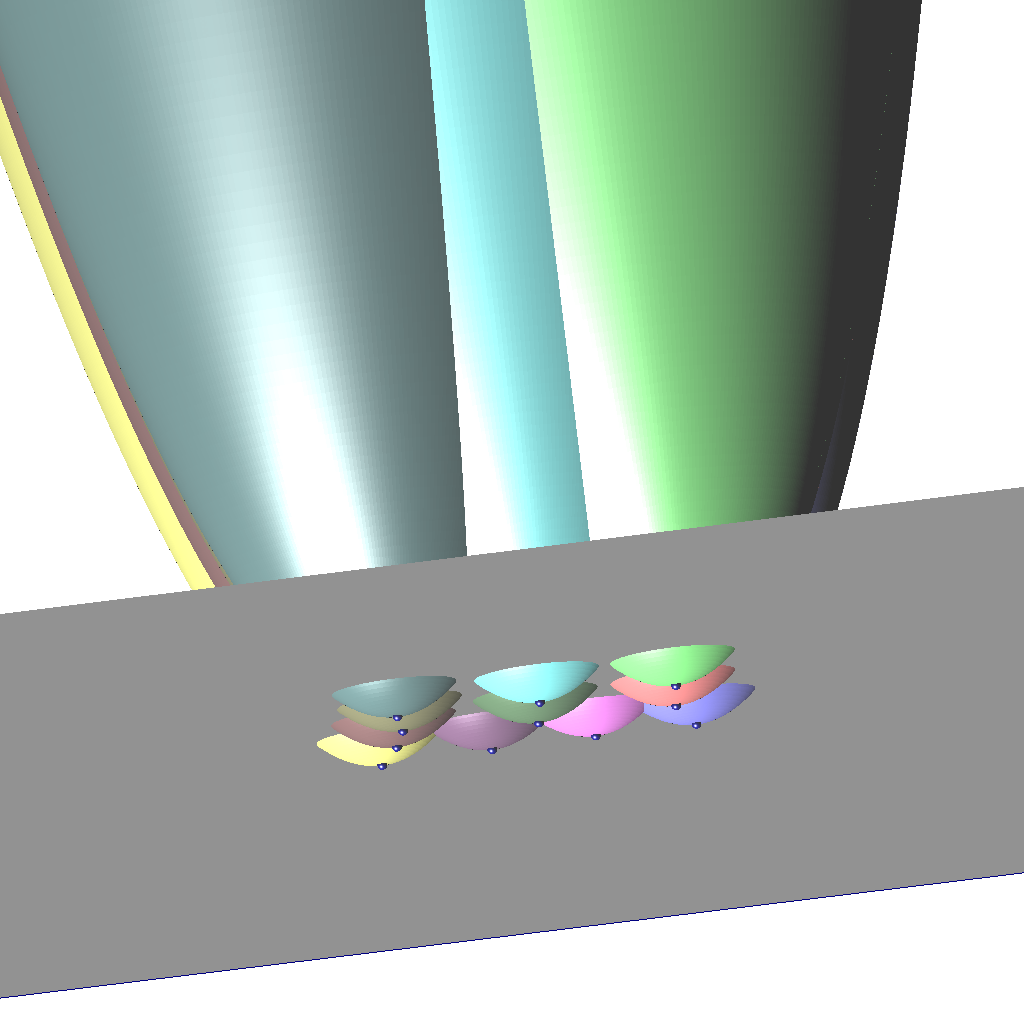}
    \includegraphics[height=47mm]{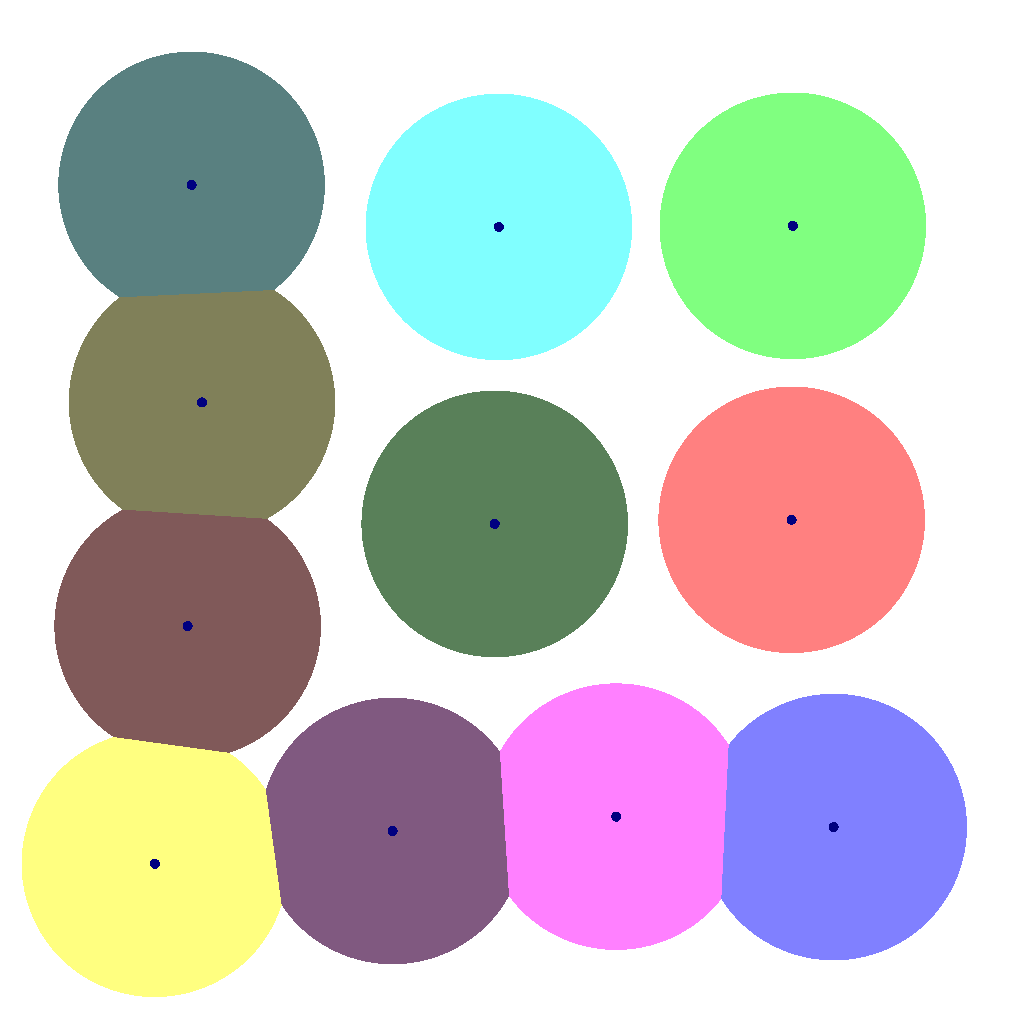}    
  }
  \caption{Left: The minimization diagram of the familly of functions
    $f_i(\bx) = \| \bx - \bx_i \|^2 - \psi_i$ augmented with the graph of the function
    $f(\bx) = 0$ (horizontal plane). Right: seen from below, the obtained cells
    $V_i^\psi$ are  the intersections between the
    Laguerre cells $Lag^\bpsi_i$ and the disks of radii $\sqrt{\psi_i}$ centered
    on the $\bx_i$'s (or the empty set if $\psi_i$ is negative).
  }
  \label{fig:laguerre2}
\end{figure}

To overcome these limitations, I propose an alternative technique: let us now imagine that the number of background points tends to infinity. More simply put, the background object $\mathcal{O}_0$ becomes the entire $\Omega$ domain (and it is still associated with $\nu_0 = |\Omega| - \sum_{i=1}^N \nu_i$). Let us also remember that the vector $\psi_{i=0}^N$ is independent on a translation, thus, w.l.o.g. we can choose $\psi_0 = 0$. Let us now consider a cell $V_i^\psi$ associated with a fluid particle:
$$
\begin{array}{lcllcll}
   V_i^\psi & = & \{\bx \quad | \quad & d^2(\bx, \mathcal{O}_i) - \psi_i & < & d^2(\bx, \mathcal{O}_j) - \psi_j) \quad \forall j \neq i & \} \\[3mm]
       & = &  \{\bx \quad | \quad & \| \bx - \bx_i \|^2 - \psi_i & < & \| \bx - \bx_j \|^2 - \psi_j) \quad \forall 1 \le j \neq i \le N \\
       &   &     \mbox{  and} & \| \bx - \bx_i \|^2 - \psi_i & < &  d^2(\bx, \mathcal{O}_0) - \psi_0 & \} \\ [3mm]
       & = &  \{\bx \quad | \quad & \| \bx - \bx_i \|^2 - \psi_i & < & \| \bx - \bx_j \|^2 - \psi_j) \quad \forall 1 \le j \neq i \le N \\
       &   &     \mbox{  and} & \| \bx - \bx_i \|^2  & < &  \psi_i & \}. 
\end{array}
$$

The last line is obtained by remembering that $\psi_0 = 0$, and by noticing 
that $d^2(\bx, \mathcal{O}_0) = 0$ for all $\bx$ in $\Omega$ since $\mathcal{O}_0 = \Omega$. In other words, $V_i^\psi$ is the intersection between the Laguerre cell 
$Lag^\psi_i$ and a disk of radius $\sqrt{\psi_i}$ centered on $\bx_i$, as shown in Figure \ref{fig:POT}-Right\footnote{or the empty set if $\psi_i$ is negative. Note that a negative $\psi_i$ cannot happen in our context, since it would contradict the convexity of $\psi$ ($\psi^{cc} = \psi \Leftrightarrow V_i^\psi \neq \emptyset \ \forall 1 \le i \le N$), guaranteed by all the iterations of the KMT Newton algorithm \cite{DBLP:journals/corr/KitagawaMT16} (more on this below).}. 
From this observation, not only "ghost cells"
are no longer needed, but also we can much more accurately compute the boundary of the fluid, by computing the intersection between the Laguerre cells and a set of disks. Put differently, what we compute is the limit case where the number of ghost cells $M$ tends to infinity. Not only the result will be more precise, but the overall computational cost will be significantly reduced, since the number of vertices in the Laguerre diagram is not increased.

As shown in Figure \ref{fig:laguerre2}, to gain more intuition about this setting, it is also possible to take the "minimization diagram" point of view: the distance to $\mathcal{O}_0 = \Omega$
is equal to zero on $\Omega$, and $\psi_0=0$, thus the graph of the associated function $f_0$ is the horizontal plane $Z=0$. The other functions $f_i$
are shifted paraboloids just like before. Now, looking at the diagram from above, one will see the tip of the paraboloids intersected by the horizontal planes
(disks). The paraboloids can also intersect each other, forming polygonal cells with straight edges (that correspond to projected parabolas), just like in the previous configuration.

\section{Numerical Solution Mechanism}
\label{sect:algo}

I shall now explain how to design a numerical solution mechanism. The associated algorithm takes the following inputs and produces the following outputs:
$$
\begin{array}{ll}
   \mbox{\bf Input:}    &- \mbox{ The domain } \Omega\ (\mbox{ can be } [0,1]^d \mbox{ or a simplicial mesh})\\
                        &- \mbox{ a set of } N \mbox{ points } \bx_i \in \Omega \\
                        &- \mbox{ prescribed volumes } (\nu_i)_{i=1}^N \mbox{ such that } \sum_i \nu_i \le |\Omega| \\[2mm]
   \mbox{\bf Output:} &- \mbox{the (unique) vector } \psi^* \in \mathbb{R}^N \mbox{ that maximizes } K(.) \\
                      &- \mbox{the cells } (V_i^{\psi^*})_{i=1}^N
                         \mbox{ such that } |V_i^{\psi^*}| = \nu_i \\
\end{array}   
$$

The core of the algorithm solves the following optimization problem:
\begin{equation}
\begin{array}{l}
    \sup_\psi \left[ K(\psi) = 
       \sum\limits_{i=1}^N\ \int\limits_{V_i^\psi} (\| \bx - \bx_i \|^2 - \psi_i) d\bx + \sum\limits_{i=1}^N \nu_i \psi_i
    \right] \\[4mm]
    \mbox{where } V_i^\psi = Lag_i^\psi \cap \{ \bx \ | \ \| \bx - \bx_i \|^2 \le \psi_i \} \\[2mm]
    \mbox{subject to } V_i^\psi \neq \emptyset \quad \forall 1 \le i \le N
\end{array}
\label{eqn:KPOT}
\end{equation}

The KMT algorithm due to Kitagawa, Merigot and Thibert \cite{DBLP:journals/corr/KitagawaMT16} is a Newton algorithm that produces a series of iterates $\psi^{(k)}$ that provably converges to the solution of Eq. \ref{eqn:KPOT}. Each iterate satisfies the convexity constraint $\psi^{cc} = \psi$ (that is, $V_i^\psi \neq \emptyset \quad \forall k, \forall 1 \le i \le N$, see also \cite{DBLP:journals/cg/LevyS18}). Adapting the KMT algorithm to our specific context has numerical and geometrical aspects detailed below.

\subsection{Numerical aspects}

The KMT algorithm follows the classical structure of a Newton algorithm:
$$
\begin{array}{ll}
   \ (1): & \psi \leftarrow [ 0 \ldots 0 ] \\
   \ (2): & \mbox{Loop} \\
   \ (3): & \quad \mbox{Compute the cells } (V_i)_{i=1}^N \\
   \ (4): & \quad \mbox{Compute the gradient } \nabla K(\psi) \\
   \ (5): & \quad \mbox{If } \| \nabla K(\psi) \|_\infty < \epsilon_K \mbox{ then Exit loop} \\
   \ (6): & \quad \mbox{Compute the Hessian matrix } \nabla^2 K (\psi) \\
   \ (7): & \quad \mbox{Solve for } \bp \in {\mathbb R}^n \mbox{ in  } \nabla^2 K(\psi) \bp = -\nabla K(\psi) \\
   \ (8): & \quad \mbox{Find the descent parameter } \alpha \\
   \ (9): & \quad \psi \leftarrow \psi + \alpha \bp \\
   \ (10): & \mbox{End loop}
\end{array}
$$

The algorithm above needs to evaluate the gradient and Hessian matrix of $K(.)$ at each main loop iteration. The coefficients of the gradient and Hessian matrix can be deduced from the cells $(V_i^\psi)_{i=1}^N$ that are computed at step (3). The algorithm that computes the cells is detailed later in the next subsection on the geometric aspects. 
The components ${\partial K}/{\partial \psi_i}$ of the gradient $\nabla K(.)$ are given by the following expression \cite{DBLP:journals/corr/KitagawaMT16,DBLP:journals/cg/LevyS18}:
\begin{equation}
  \frac{\partial K}{\partial \psi_i}  =  \nu_i - |{V_i^\psi}|. 
\end{equation}
In other words, each component of the gradient corresponds to the prescribed volume $\nu_i$ associated with a point $\bx_i$ minus the volume of the cell $V_i^\psi$. For the vector $\psi^*$ that maximizes $K(.)$, all components of the gradient vanish, which means that each cell $V^{\psi^*}_i$ has exactly the prescribed volume $\nu_i$.  

This expression of the gradient leads also to a natural stopping criterion (line 5), the largest component of the gradient corresponds to the maximum volume error. We stop the algorithm as soon as it is smaller than a prescribed $\epsilon_K$ (typically one percent of $\nu_i$). \\

We now consider the Hessian matrix computed at step (6). Adapting the formulas in \cite{DBLP:journals/corr/KitagawaMT16,DBLP:journals/cg/LevyS18} to our context, the coefficients of the Hessian are given by:
\begin{eqnarray}
  \nonumber
  \frac{\partial^2 K}{\partial \psi_i \partial \psi_j} & = & \frac{1}{2} \frac{|V^\psi_{ij}|}{\| \bx_j - \bx_i \|} \quad \mbox{if } i \neq j \\
  \frac{\partial^2 K}{\partial \psi_i^2} & = & - \left(\sum_{j \neq i} \frac{\partial^2 K}{\partial \psi_i \partial \psi_j}\right)
                                               - \frac{1}{2}\frac{|V^\psi_{i0}|}{\sqrt{\psi_i}}
\end{eqnarray}
where $|V_{ij}^\psi| = |\partial V_i^\psi \cap \partial V_j^\psi|$ denotes the area of the intersection between the border of the cell
$V_i^\psi$ and the border of the cell $V_j^\psi$, that is, a polygonal facet in the Laguerre diagram clipped by the two balls $B(\bx_i,\sqrt{\psi_i})$ and $B(\bx_j,\sqrt{\psi_j})$. The term $|V^\psi_{i0}|$ corresponds to the free surface area in $\partial V_i^\psi$, that is the portion of the border of $V_i^\psi$ included in the sphere $S(\bx_i, \sqrt{\psi_i})$, that touches the unoccupied portion of $\Omega$. 

Step (7) of the algorithm computes the Newton step vector $\bp$, by solving a linear system. Except the boundary term in $\partial^2{K}/\partial \psi_i^2$, this linear system is identical as the one solved in \cite{levy:hal-03081581}: it corresponds to a Poisson equation discretized with finite elements. The matrix is sparse, with a non-zero coefficient at coefficient $(i,j)$ if and only if the cells $V^\psi_i$ and $V^\psi_j$ touch each other. The additional term $(1/2) \left( | V_{i0}^\psi | / \sqrt{\psi_i} \right)$  in $\partial^2 K / \partial \psi_i^2$ does not change the sparsity pattern of the Hessian: since the Lagrange multiplier associated with $\psi_0$ is fixed and equal to 0, there is no partial derivative with respect to it except the diagonal term. The same linear solver as in \cite{levy:hal-03081581} can be used (Jacobi-preconditioned conjugate gradient \cite{Hestenes1952} with sparse matrix stored in CRS format, with optional GPU acceleration), as well as the same stopping criterion $\| H \bp - \bg \| / \| \bg \| \le 10^{-3}$. \\

Once the step vector $\bp$ is computed, we need to find a good descent parameter $\alpha$ (step (8)). In the KMT algorithm \cite{DBLP:journals/corr/KitagawaMT16}, provably convergent, the descent parameter $\alpha$ is determined as follows:
$$
\begin{array}{ll}
   \ (1): & \alpha \leftarrow 1 \\
   \ (2): & \mbox{Loop} \\
   \ (3): & \quad \mbox{If  } \inf_i | V^{\psi + \alpha \bp}_i | >a_0 \\
   \ (4): & \quad \mbox{and } \| \nabla K(\psi + \alpha \bp) \| \le (1 - \alpha / 2) \| \nabla K(\psi) \| \\
   \ (5): & \quad \quad \mbox{then Exit loop} \\
   \ (6): & \quad \alpha \leftarrow \alpha / 2 \\
   \ (7): & \quad \mbox{Compute the cells } (V^{\psi+\alpha \bp}_i)_{i=1}^N \\
   \ (8): & \mbox{End loop}
\end{array}
$$
where $a_0 = \frac{1}{2}\min\left( \inf_i  |V^{\psi^(0)}_i|  , \inf_i(\nu_i) \right)$. \\

The KMT algorithm iteratively halves the descent parameter $\alpha$ until two criteria are met: the volume of the smallest cell needs to be larger than a threshold $a_0$ (line 3), and the norm of the gradient needs to decrease sufficiently (line 4). The threshold $a_0$ for the minimum cell volume corresponds to (half) the minimum cell volume for $\psi = 0$ (also called Voronoi diagram) and minimum prescribed area $\nu_i$. \\

Equipped with the KMT algorithm above, we can now compute the descent parameter $\alpha$, by plugging the algorithm above into line (8) of the Newton algorithm at the beginning of this section. \\

\subsection{Geometrical aspects}

\begin{figure}
    \centering
    \includegraphics[width=\textwidth]{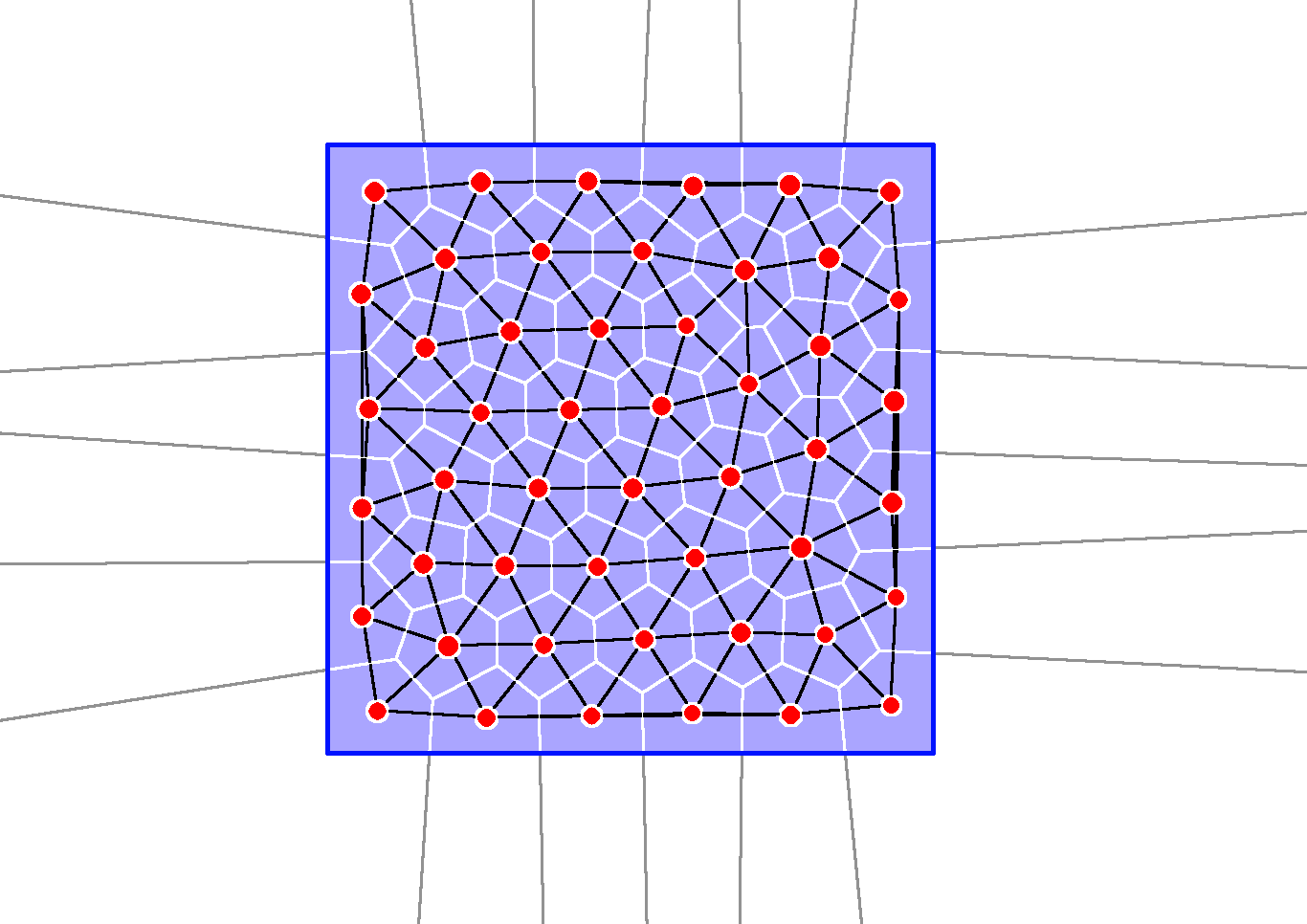}
    \caption{The Bowyer-Watson algorithm computes the Laguerre diagram (cell boundaries displayed in white and grey). It uses internally a dual representation (triangulation in black). One also needs to compute the intersection with $\Omega$ (blue square in this example).}
    \label{fig:triangulation}
\end{figure}

To evaluate the components of the gradient and coefficients of the Hessian of $K(.)$, the KMT algorithm needs to construct the cells $V_i^\psi = Lag_i^\psi \cap B(\bx_i,\sqrt{\psi_i})$, that is, the intersection between Laguerre cells and balls centered on the $\bx_i$'s for each $\psi^{(k)}$ iterate. To do so, we first compute the Laguerre diagram, using the classical Bowyer-Watson algorithm \cite{DBLP:journals/cj/Bowyer81,journals/cj/Watson81},
see Figure \ref{fig:triangulation}. A readily-available open-source implementation is available in our GEOGRAM library \cite{WEB:GEOGRAM} (and also in the CGAL library \cite{DBLP:journals/comgeo/BoissonnatDPTY02} for programmers who have a taste for C++ templates). The details of our highly optimized implementation in GEOGRAM are given in \cite{levy:hal-03081581}.  

Given the points $\bx_i$ and the vector $\psi$, the Bowyer-Watson algorithm computes the Laguerre diagram. Internally, it uses the dual triangulation, formed by the black triangles in Figure \ref{fig:triangulation} (in 3D, they become tetrahedra). Each vertex of the Laguerre diagram is deduced from one of the triangles (resp. tetrahedra), and the Laguerre cells are obtained by traversing the triangles (tetrahedra) incident to a given $\bx_i$. Note that the Boywer-Watson algorithm computes the Laguerre diagram in the entire $\mathbb{R}^d$ domain, with infinite Laguerre cells (grey straight lines in the Figure). To obtain the $V_i$ cells, one needs to compute intersections with the domain $\Omega$ (and also with the balls centered on the $\bx_i$'s):
$$
   V_i^\psi = Lag_i^\psi \cap \Omega \cap B(\bx_i,\sqrt{\psi_i})
$$
 
In 2D, constructing the cells $V_i^\psi$ means computing intersections between convex polygons and disks, which does not represent too much difficulty (see Figure \ref{fig:POT}). However, it becomes a challenging task in 3D:
as can be seen, we need to compute a large number of intersections, and 3D mesh intersection is known to be a delicate operation, subject to numerical precision problems: the intersection algorithm depend on a small set of functions, called \emph{geometric predicates}, that take as input some points, and that returns a discrete value. For instance, such a geometric predicate can tell whether a point $\bp_1$ is above or below another point $\bp_2$. These geometric predicates are often polynomials in the coordinates of the points. The limited precision of floating point numbers can have catastrophic consequences, for instance when the algorithm estimates that a certain point $\bp_1$ is above another point $\bp_2$, and later 
inconsistently estimates that $\bp_2$ is above $\bp_1$. Such inconsistent behavior can result in invalid combinatorics (see e.g. \cite{DBLP:conf/compgeom/Shewchuk96}). \\

It is important to stress that we are going to compute a \emph{huge} number of intersections (billions), hence these numerical issues that correspond to what may me initially thought of as very unlikely corner cases occur in fact \emph{thousands times} in a typical simulation. The goal now is to find a strategy to overcome this robustness issue while keeping reasonable performance. 
The strategy is to express the problem in terms of a small number of "atomic" operations, that can be used as solid foundations to build the rest of the algorithm. These "atomic" operations are simple and can be robustly implemented:
\begin{itemize}
    \item (1) the algorithm will solely manipulate \emph{convex polytopes} and their intersections. A convex polytope $C$ is defined 
    as the intersection of $N_v$ half-spaces $\Pi^+_i$:
$$
    C = \bigcap_{i=1}^{N_v} \Pi^+_i = \bigcap_{i=1}^{N_v} \{ \bx \ | \ a_i x + b_i y + c_i z + d_i \le 0 \}
$$
    \item (2) to avoid inconsistencies, all the used predicates will be computed 
      in arbitrary precision \cite{DBLP:conf/compgeom/Shewchuk96}, with arithmetic filters to speed-up
      computations in the easy cases \cite{meyer:inria-00344297}.  
\end{itemize}

A consequence of $(1)$ is that the balls $B(\bx_i,\sqrt{\psi_i})$ will be approximated by convex polytopes:
$$
  B(\bx,R) \simeq \hat{B}(\bx,R) = \bigcap_{i=1}^{N_u} \{ \by \ | \ \| \bx - \by \|^2 \le \| \bx + 2\bu_i - \by \|^2 \}  
$$
where $(\bu_i)_{i=1}^{N_u}$ is a predetermined set of unit vectors $\bu$ that uniformly sample the unit sphere. 
While this strategy shares with the "ghost cells" approach the fact that it approximates the balls as convex polytopes, it   
differs from it in two important aspects: 
\begin{itemize}
\item first, computations are \emph{local} to a given
$V_i^\psi$ cell, with a small number of vertices (typically 50 to 200, as compared to the total number of points $N$), whereas the "ghost cells" 
strategy needs inserting all additional vertices into the \emph{global} Laguerre diagram, with $N+M$ vertices;
\item second, each $\hat{B}(\bx,R)$ can be instanced from a \emph{precomputed} unit sphere template $\hat{B}\left({\bf 0},1\right)$ without needing to compute any intersection, by scaling and translating the unit sphere template (no combinatorics needs to be computed). 
\end{itemize}

This means a large number $N_u$ (typically 100 to 200) of unit directions $\bu_i$ can be used without significantly 
increasing the computation time: the "partial optimal transport" algorithm computes a solution that is equivalent to using $N_u \times N$ "ghost cells", without the associated computational cost. Not only it significantly reduces the complexity, but also it makes it possible to use a large value for $N_u$, that gives a precise approximation of the balls $B(\bx_i, \sqrt{\psi_i})$. Alternatively, one can also use a data structure for convex polytopes with spherical faces \cite{DBLP:journals/siamnum/LeclercMSS20}, but this comes with longer execution times, and robustness is more difficult to ensure.\\

The domain $\Omega$ can be either a convex polytope, or a simplicial mesh $\Omega = \bigcup_{i=1}^{N_T} T_i$. In both cases the intersections between 
$\Omega$, $Lag_i^\psi$ and $\hat{B}(\bx_i, \sqrt{\psi_i})$ involved in the definition of $V_i^\psi$ can be computed as intersections between convex polytopes.

\subsubsection{Convex polytopes: H and V representation}

The "atomic" operation computes the intersection between a convex polytope $C$ and a half-space $\Pi^+$:
$$
   C \leftarrow C \cap \Pi^+.
$$

\begin{figure}
    \centerline{
    \includegraphics[height=55mm]{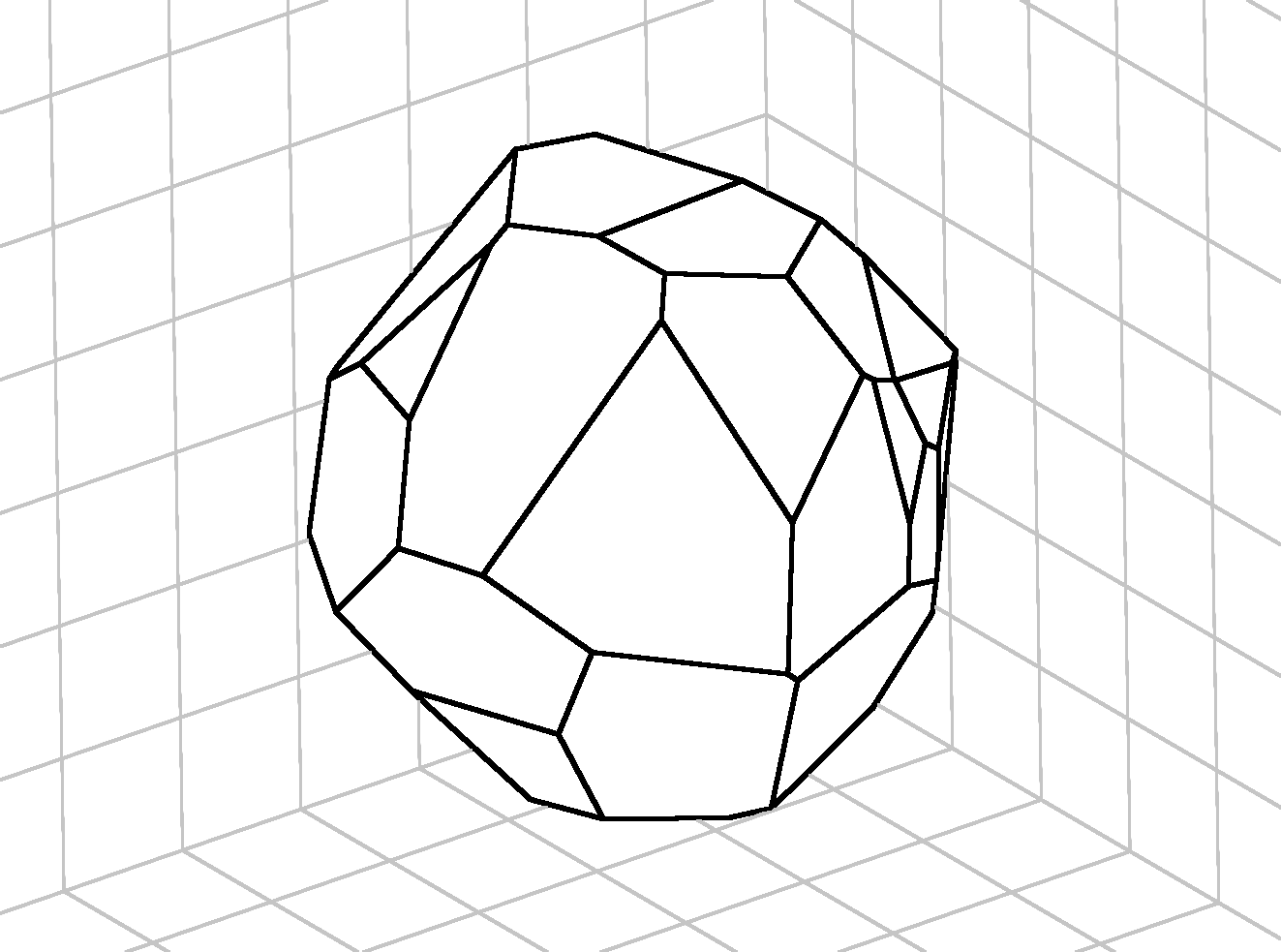}
    \includegraphics[height=55mm]{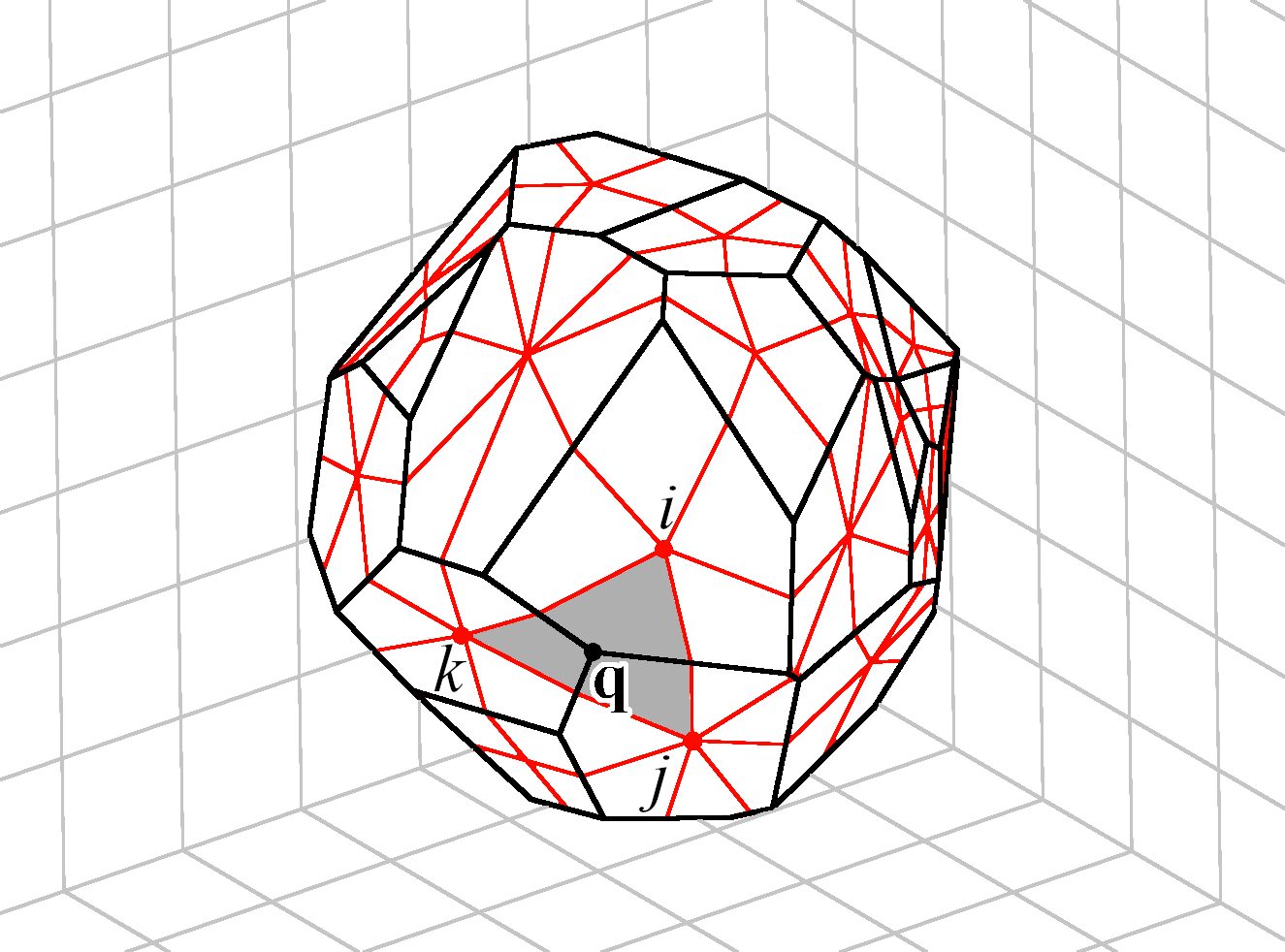}
    }
    \caption{Representation of convex polytopes.}
    \label{fig:cells}
\end{figure}

In the end, to estimate the gradient and Hessian of $K(.)$, we will need to compute the volume of the cells and the areas of the 
cell borders. To do so, we need to convert from the initial representation of the cell, that is, as an intersection of half-spaces" (H-representation), $(\Pi_i^+)_{i=1}^{Nv}$ into an explicit representation of all the vertices and facets of the cell (V-representation). As shown in Figure \ref{fig:cells}, if the planes are in generic position\footnote{we will discuss degenerate cases further on.}, each vertex (black) is shared by three planes $\Pi_i, \Pi_j, \Pi_k$. It is then natural to represent the polytope in dual form (see Figure \ref{fig:cells}), by a triangulation
$(T_l = \{ i_l, j_l, k_l \})_{l=1}^{N_t}$ shown in red in the Figure (the same representation is exploited by the Boywer-Watson algorithm that we use to compute the Laguerre diagram). The coordinates $q_x, q_y, q_z$ of the vertex $\bq_l$ associated with triangle $T_l$ are the solution of:
$$
\left[ \begin{array}{ccc}
   a_i & b_i & c_i \\
   a_j & b_j & c_j \\
   a_k & b_k & c_k
\end{array} \right]
\left[ \begin{array}{c}
  q_x \\ q_y \\ q_z
\end{array} \right]
= -
\left[ \begin{array}{c}
  d_i \\ d_j \\ d_k
\end{array} \right]
$$

Using Cramer's formula, the coordinates $q_x, q_y, q_z$ are given by:
\begin{equation}
\left[ \begin{array}{c}
  q_x \\ q_y \\ q_z
\end{array} \right] = 
{\tiny
-\frac{1}{\left| 
\begin{array}{ccc}
   a_i & b_i & c_i \\
   a_j & b_j & c_j \\
   a_k & b_k & c_k
\end{array} \right|}
\left[
\begin{array}{c}
\\
\left| 
\begin{array}{ccc}
   d_i & b_i & c_i \\
   d_j & b_j & c_j \\
   d_k & b_k & c_k
\end{array} \right| \\ \\
\left| 
\begin{array}{ccc}
   a_i & d_i & c_i \\
   a_j & d_j & c_j \\
   a_k & d_k & c_k
\end{array} \right| \\ \\
\left| 
\begin{array}{ccc}
   a_i & b_i & d_i \\
   a_j & b_j & d_j \\
   a_k & b_k & d_k
\end{array} \right| \\ \\
\end{array}
\right]}
\label{eqn:q}
\end{equation}
(the i-th component of the solution corresponds to the determinant of the system with its i-th column replaced with the r.h.s divided by the determinant of the system, see \cite{DBLP:journals/siamcomp/Aurenhammer87,DBLP:journals/tog/EdelsbrunnerM90} for similar computations).

\begin{figure}
    \centerline{
    \includegraphics[height=55mm]{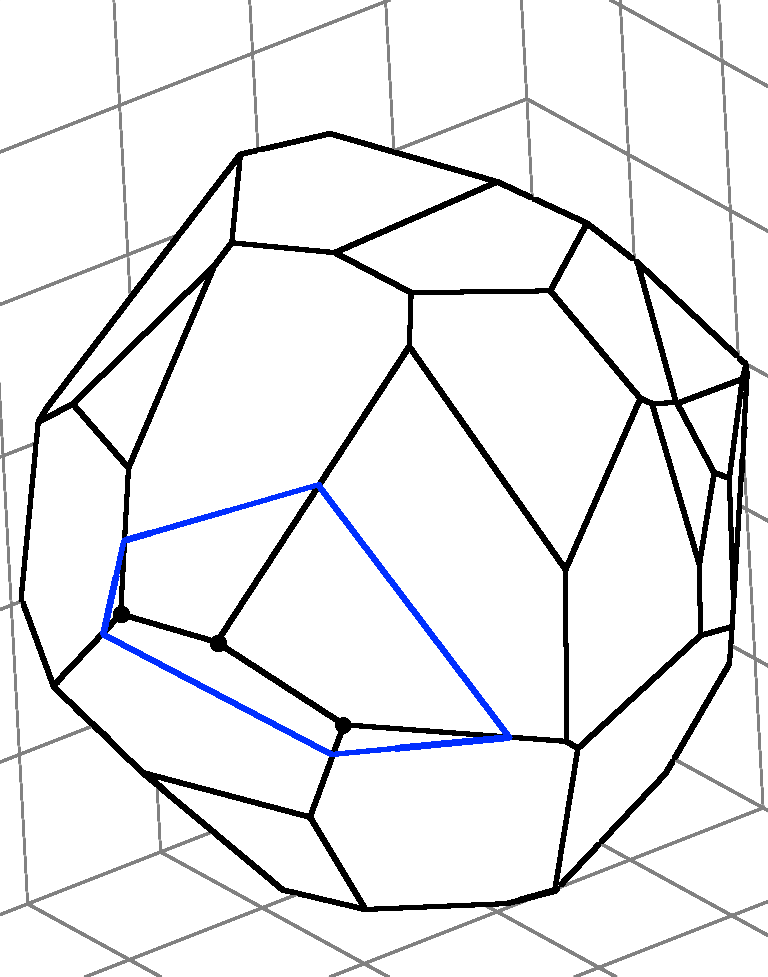}
    \includegraphics[height=55mm]{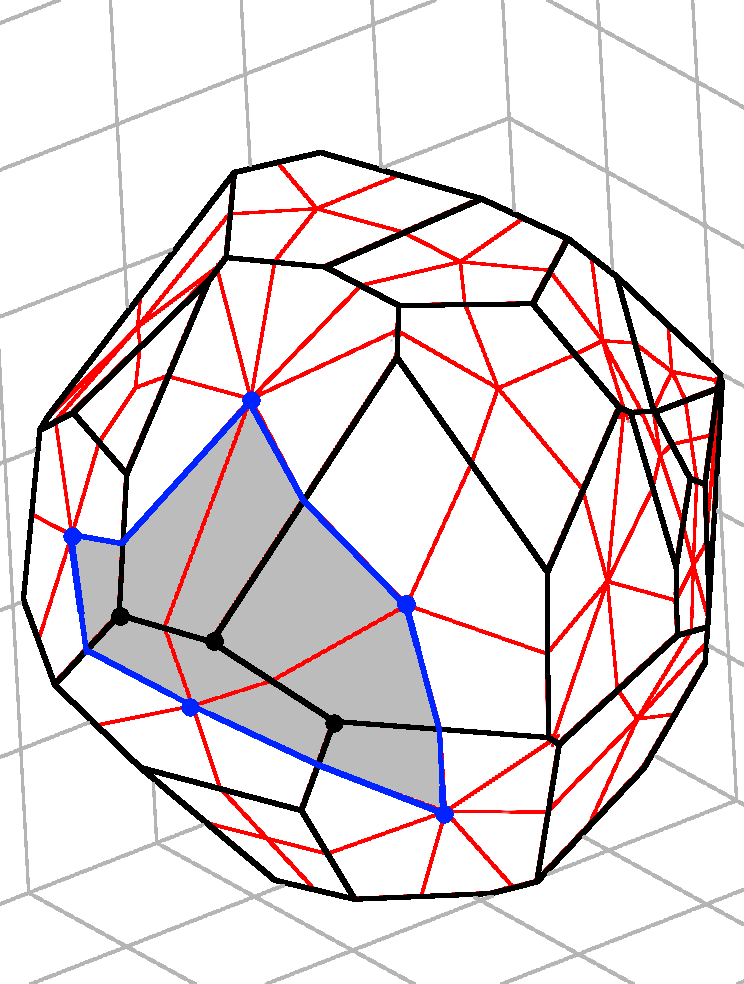}
    \includegraphics[height=55mm]{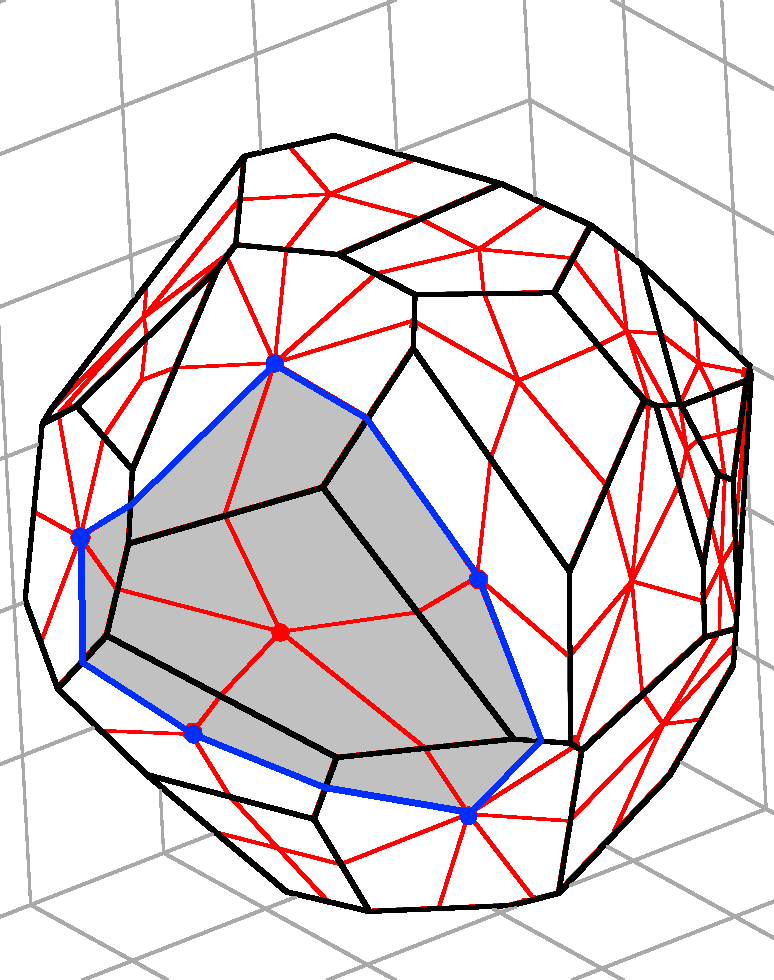}
    }
    \caption{Clipping a convex polytope by a half-space.}
    \label{fig:clipping}
\end{figure}

The algorithm to compute the intersection between a cell $C$ and a half-space $\Pi^+(a,b,c,d)$ works in three phases (see Figure \ref{fig:clipping}):
$$
\begin{array}{ll}
   \mbox{\bf Input:}    &- \mbox{ A convex polytope } C  \mbox{ and a half-space } \Pi^+  \\
   \mbox{\bf Result:}   &- \mbox{ replaces } C \mbox { with } C \cap \Pi^+   \\
\ (1): & \mbox{ classify the vertices } \\
\ (2): & \mbox{ discard the triangles } \\
\ (3): & \mbox{ triangulate the hole } \\
\end{array}
$$
\textbf{The first phase} determines the triangles that correspond to vertices $\bq_l$ that are on the negative side of $\Pi^+$ (shown in black in Figure \ref{fig:clipping}-Left), that is, $a q_x + b q_y + c q_z + d < 0$. This can be rewritten as:
$$
{\tiny
\left|
\begin{array}{ccc}
  a_i & b_i & c_i \\
  a_j & b_j & c_j \\
  a_k & b_k & c_k 
\end{array}
\right|
\times
\left|
\begin{array}{cccc}
  a_i & b_i & c_i & d_i \\
  a_j & b_j & c_j & d_j \\
  a_k & b_k & c_k & d_k \\
  a   & b   & c   & d
\end{array}
\right|} \le 0
$$
by replacing the coordinates of $\bq$ by their expression (one obtains an expression of the determinant above, developed w.r.t. to its last row). Note that the vertices of the triangles can be consistently oriented in such a way that the sign of the first term (the $3\times3$ determinant) is always positive. In the end, to classify a vertex, one only needs to compute the sign of the $4 \times 4$ determinant
$| \Pi_i, \Pi_j, \Pi_k, \Pi |$ formed by the coefficients of the planes. This is the only predicate\footnote{besides {\tt orient3d} and {\tt in\_weighted\_sphere} used by the Bowyer-Watson algorithm that constructs the Laguerre diagram.} used by the algorithm. Note also that during the computation of the intersections, only the equations of the $\Pi_i$ planes are used, the $\bq$ points do not need to be explicitly computed. To compute the exact sign of these determinants, I use expansion arithmetics \cite{DBLP:conf/compgeom/Shewchuk96}, 
and arithmetic filters \cite{meyer:inria-00344297} to quickly determine the sign in the easy cases. The implementation in \cite{DBLP:journals/cad/Levy16}
generates the code that computes the predicate from its formula. This strategy ensures that all combinatoric decisions taken by the algorithm will remain consistent. The points $\bq$ are computed once all clipping operations are computed, right before evaluating the volumes and areas involved in the gradient and Hessian of $K(.)$.\\
 
\textbf{The second phase} discards the triangles $(i,j,k)$ such that $| \Pi_i, \Pi_j, \Pi_k, \Pi |$ $<$ $0$ (shown in grey in 
Figure \ref{fig:clipping}-Center). This leaves a hole in the triangulation (the border of the hole is shown in blue). As often done in implementations of the Bowyer-Watson algorithm, the discarded triangles are kept in a linked list, so that they can be reused in the subsequent steps; \\

\textbf{The third phase} constructs a new triangle for each edge on the border of the hole created at the previous phase
(see Figure \ref{fig:clipping}-Right). \\ 

The algorithm has the same structure as the Bowyer-Watson algorithm, with the difference that the $\tt in\_sphere$ predicate, that determines which triangle to discard, is replaced with ${\tt sign}|\Pi_i, \Pi_j, \Pi_k, \Pi|$. Another difference is the way the Bowyer-Watson determines the list of triangles to be discarded. In our case we test all the triangles. We could instead adapt the Bowyer-Watson strategy, that starts from a random vertex, and walks along the edges of the dual graph until it finds a triangle to be discarded, but since the number of triangles per polytope remains small in our case (typically $ < 100$), we did not observe a speedup using this technique. 

\subsubsection{Computing the cells $V_i^\psi$}

I shall now explain how to use the algorithm of the previous paragraph to compute the cells. Each cell 
$V_i^\psi = Lag_i^\psi \cap \Omega \cap \hat{B}(\bx_i, \sqrt{\psi_i})$ is the intersection between three
objects: the Laguerre cell $Lag_i^\psi$, the domain $\Omega$, represented by a simplicial mesh, that is,
a set of tetrahedra, and the (approximated) ball $\hat{B}(\bx_i, \sqrt{\psi_i})$. 
Note that the domain $\Omega$ is not necessarily convex, but since it is decomposed into simplices, we only
need to compute intersections between convex objects. The algorithm that computes a cell $V_i^\psi$ is detailed below:
 
$$
\begin{array}{ll}
   \mbox{\bf Input:}    &- \mbox{ a simplicial mesh } \Omega \\
                        &- \mbox{ the set of points } (\bx_i)_{i=1}^N \mbox{ and the vector } (\psi_i)_{i=1}^N \\
                        &- \mbox{ the index } i \mbox{ of a cell} \\
   \mbox{\bf Output:}   &- \mbox{ A set of convex polytopes } \mathcal{\bf C} = \{ C_k \} \mbox{ such that } V_i^\psi = \bigcup C_k\\[2mm]
%   \hline \\[4mm}
\end{array}
$$

$$
\begin{array}{ll}
%\hline \\
   \ (1):  & \mathcal{T} \leftarrow \{ t \subset \Omega \ | \ B(\bx_i,\sqrt{\psi_i}) \cap t \neq \emptyset \} \\
   \ (2):  & \mbox{If } \exists t \in \mathcal{T} \ | \ \partial t \cap \partial \Omega \neq \emptyset \\
   \ (3):  & \quad \mbox{For each } t \in \mathcal{T} \\
   \ (4):  & \quad \quad C \leftarrow t \\
   \ (5):  & \quad \quad C \leftarrow C \cap Lag_i^\psi \\
   \ (6):  & \quad \quad C \leftarrow \hat{B}(\bx_i, \sqrt{\psi_i}) \cap C \\
   \ (7):  & \quad \quad {\bf C} \leftarrow {\bf C} \cup \{ C \} \\
   \ (8): & \quad \mbox{End for}  \\
   \ (9): & \mbox{Else}  \\
   \ (10): & \quad C \leftarrow Lag_i^\psi \\
   \ (11): & \quad C \leftarrow \hat{B}(\bx_i, \sqrt{\psi_i}) \cap C \\
   \ (12): & \quad \mathcal{\bf C} \leftarrow \mathcal{\bf C} \cup \{ C \} \\
   \ (13): & \mbox{End if}   \\
\end{array}
$$

In line (1), we first determine the set of simplices $\mathcal{T}$ that have a non-empty intersection with the ball 
$B(\bx_i, \sqrt{\psi_i)}$. It is done using an axis-aligned bounding box tree (AABB tree), see e.g \cite{WEB:OPCODE}. 

Then we distinguish two different cases: \textbf{first case:} if one of the intersected tetrahedra touches the boundary $\partial \Omega$ of the domain, then $V_i^\psi$ is not necessarily convex, but can be easily decomposed into convex objects: to do so, we compute the intersections $Lag_i^\psi \cap \hat{B}(\bx_i, \sqrt{\psi_i}) \cap t$ for each $t$  in $\mathcal{T}$. \textbf{Second case:} the ball $B(\bx_i,\sqrt{\psi_i})$ is included in $\Omega$ (no tetrahedron in $\mathcal{T}$ touches $\partial \Omega$), then $V_i^\psi = Lag_i^\psi \cap B(\bx_i, \sqrt{\psi_i})$ is convex and can be directly computed (lines (9) to (13)). 

In lines (6) and (11), we compute the intersection between a convex polytope and the approximated ball $\hat{B}(\bx_i, \sqrt{\psi_i})$. If $C$'s bounding sphere centered on $\bx_i$ has a radius smaller than $\sqrt{\psi_i}$ (that is, $\max_{\bq \in C} \{ \| \bx_i - \bq \|$ $<$ $\sqrt{\psi_i}$), then $C$ is entirely contained in the ball and one can skip the intersection computation. Else, the approximated ball $\hat{B}(\bx_i, \sqrt{\psi_i})$ is 
instanced (copied from the unit ball, translated and scaled), and then clipped by all the $\Pi_i$'s from the H-representation of $C$ (in this order, because $\hat{B}(\bx_i, \sqrt{\psi_i})$ has a larger number of facets than $C$ in general).
 
In line (4), the convex polytope $C$ is initialized from a tetrahedron of $\Omega$. Combinatorics are initialized from a fixed tetrahedron template, and the plane equation of each facet $\bp_1, \bp_2, \bp_3$ is
given by $ax + by + cz + d = 0$, where $(a,b,c) = (\bp_2-\bp_1) \times (\bp_3-\bp_1)$ and 
$d = -(a,b,c) \cdot \bp_1$.

In line (10), $C$ is initialized from a Laguerre cell. Combinatorics are copied from the Laguerre diagram, also stored in dual representation (tetrahedra). The triangles of $C$ correspond to the facets of the tetrahedra incident to vertex $i$ opposite to $i$. The plane equations $\Pi_{i,j}$ are given by: 
$
\bx \in \Pi_{ij} \quad \Leftrightarrow \quad
\| \bx - \bx_i \|^2 - \psi_i = \| \bx - \bx_j \|^2 - \psi_j
$
or: $ax + by + cz + d = 0$ where $(a,b,c) = 2(\bx_i - \bx_j)$ and $d = \bx^2_j -  \bx_i^2 - \psi_j + \psi_i$. \\

Up to know we have supposed that all planes where in generic position (that is, each vertex of the polytope is shared by exactly three planes). To handle the specific configurations where more than three planes meet at a single vertex, one would think of using the symbolic perturbations, as in the "simulation of simplicity" technique \cite{DBLP:journals/tog/EdelsbrunnerM90}. In our case, the symbolic perturbation is trivial: one can use the predicate $|\Pi_i, \Pi_j, \Pi_k, \Pi| \le 0$ as is (provided that it is implemented with exact arithmetics, see \cite{DBLP:journals/cad/Levy16} and references herein). It means that each time a vertex $\bq$ is exactly located on the boundary of a clipping half-space $\Pi^+$ the associated triangle will be discarded. To evaluate the efficiency of this technique, degenerate configurations are tested in the next section. \\
 
It is not necessary to store the set of convex polytopes $\mathcal{\bf C}$: one can process them in a "streaming" manner, when assembling the gradient and Hessian of $K(.)$. Then, the algorithm stores a single $C$ at any time, and has only a small memory overhead besides the storage taken by the Laguerre diagram (stored as a 3D triangulation). Moreover, the algorithm is very simple to parallelize (then it stores one convex polytope per thread). When assembling the gradient and Hessian, it just needs a synchronization primitive (spinlock) to ensure that two threads do not modify the same entry at the same time.

%$$
%\begin{array}{ll}
%   \  & \quad \mbox{If } \mbox{Radius}(\bx_i,C) \ge \sqrt{\psi_i} \\
%   \  & \quad \quad C \leftarrow \hat{B}(\bx_i, \sqrt{\psi_i}) \cap C \\
%   \  & \quad \mbox{End if} \\
%\end{array}
%$$
 
\section{Numerical experiments}
\label{sect:results}

I shall now test the different components of the algorithm, first the lowest-level one (convex polytopes clipping), then the partial optimal transport, and finally demonstrate the algorithm used to implement a basic free-surface fluid simulator in 3D. 
 
\subsection{Testing convex polytope clipping}

\begin{figure}
    \centerline{
    \includegraphics[height=85mm]{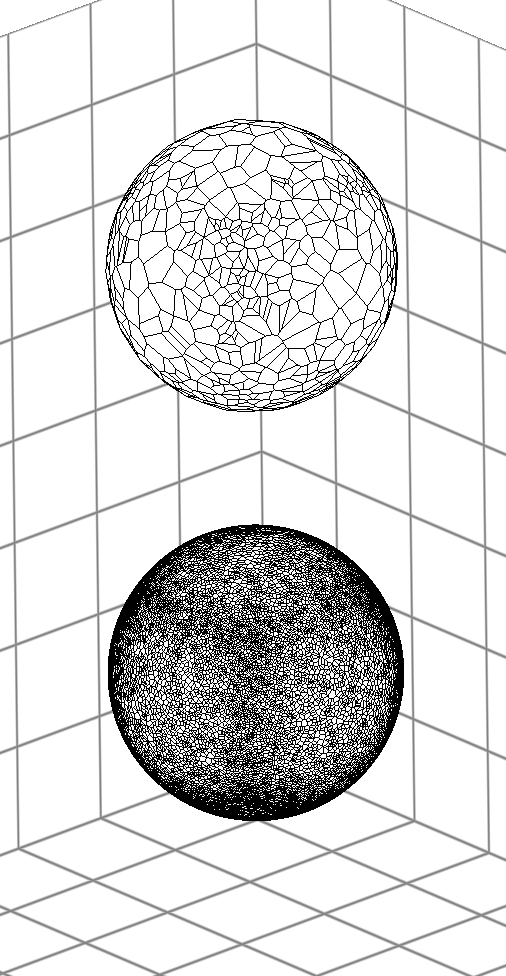}    
    \hspace{10mm}
    \includegraphics[height=85mm]{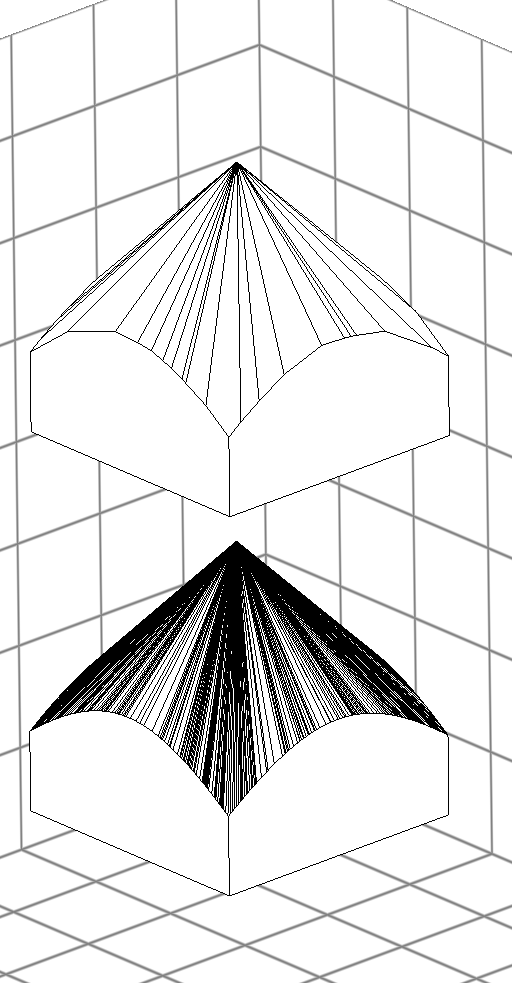}   
    }
    \caption{Testing the robustness of the convex polytope clipping algorithm.}
    \label{fig:clipping_test}
\end{figure}

The robustness of the convex polytope clipping algorithm is tested using some highly degenerate configurations that are created on purpose. In Figure \ref{fig:clipping_test}-Left, the algorithm computes the intersection between 10000 (top) and 30000 (bottom) half-spaces tangent to a sphere with random normal vector. The same sequence of half-spaces is fed to the algorithm 5 times, which creates degenerate configurations with half-spaces that exactly correspond to existing facets. On the right, a cube is clipped by 30 (top) and 500 (bottom) random half-spaces tangent to a cone. This creates a highly degenerate vertex, common to all facets. Again, the sequence of half-spaces is fed to the algorithm 5 times. In both configurations, the implemented exact predicates robustly handles all the degeneracies.
Execution time is smaller than 1 second for both tests.

\subsection{Testing partial optimal transport}

\begin{figure}
    \centerline{
    \includegraphics[height=50mm]{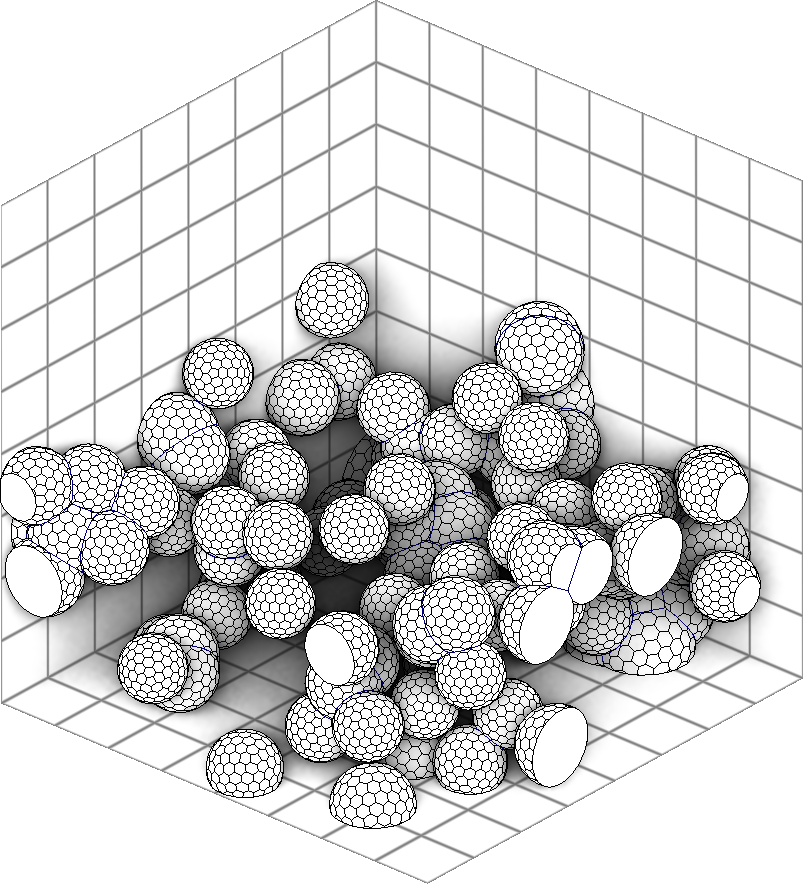}    
    \hspace{2mm}
    \includegraphics[height=50mm]{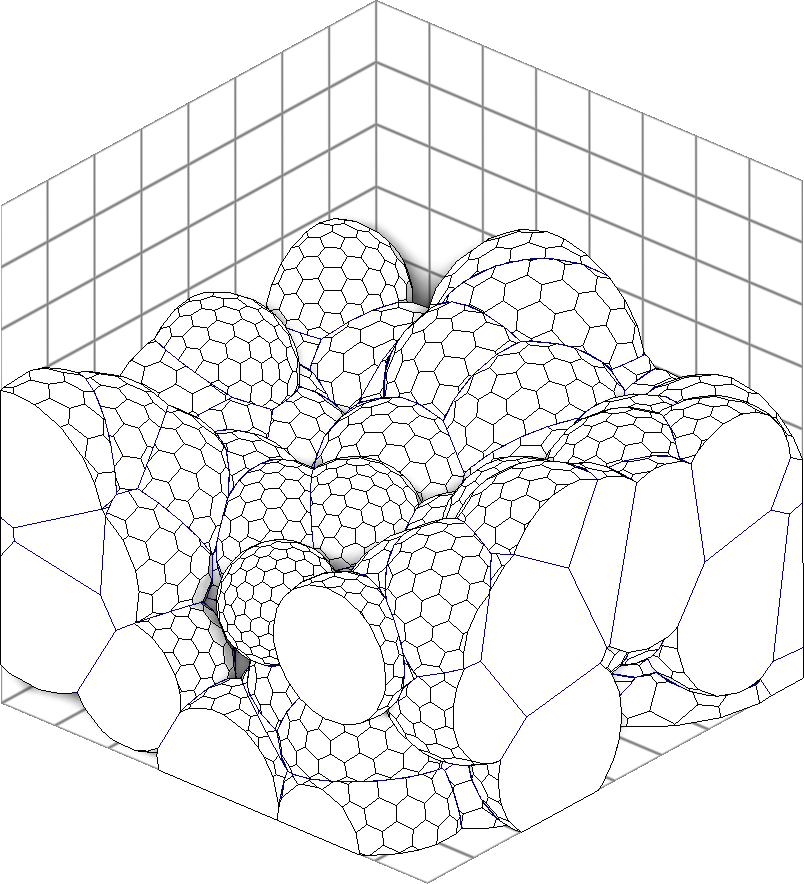}  
    \hspace{2mm}
    \includegraphics[height=50mm]{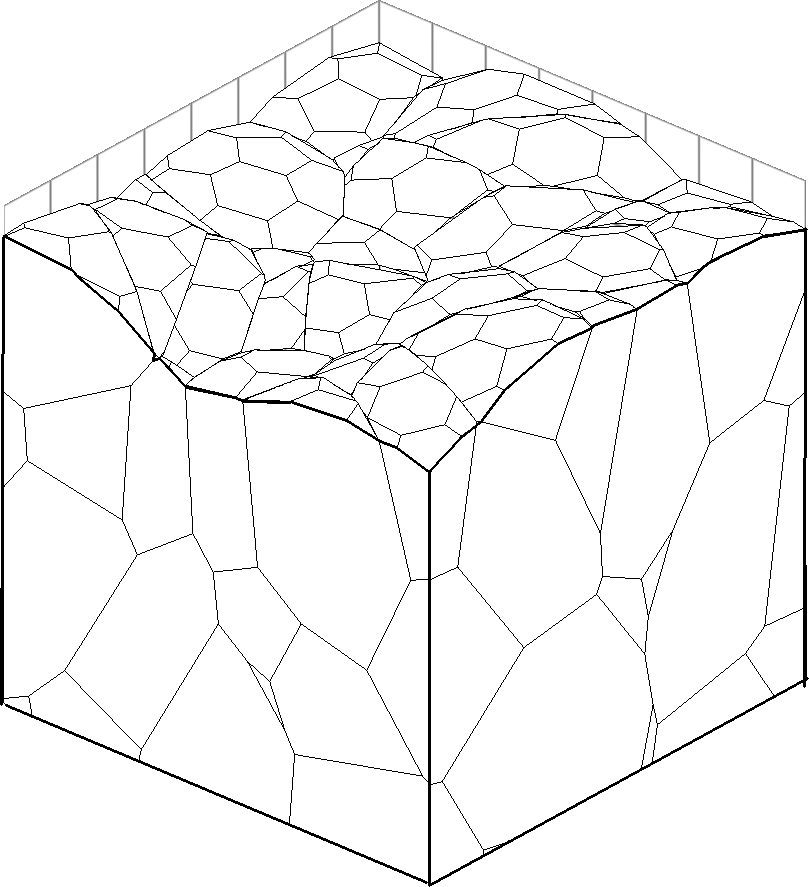}  
    }
    \caption{Testing partial optimal transport in a cube, with 100 random points in the lower half of the cube. From left to right: fluid volume = 10\%, 50\% and 90\%.}
    \label{fig:POT_cube}
\end{figure}

We now test partial optimal transport. In the first test, shown in Figure \ref{fig:POT_cube}, we compute partial optimal transport between a cube and a set of 100 points randomly distributed in the lower half of a cube. The cells are constrained to occupy 10\%, 50\% and 90\% of the cube respectively. The balls are approximated by $N_u=162$ half-spaces.

$$
  \begin{array}{c|c|c|c|c}
     \mbox{fraction} & \mbox{nb iter} & \mbox{max err} & \mbox{avg err} & \mbox{time} \\
    \hline  
     10\% & 4 & 0.19\% & 0.006\% & 0.28s \\
     50\% & 5 & 0.54\% & 0.1\% & 0.39s \\
     90\% & 7 & 0.008\% & 0.001\% & 0.58s\\
   \end{array}
$$
 
The table above reports the number of Newton iteration (second column) spent to reach convergence, that is, when the largest difference between requested and actual cell volume among all cells (third column) is smaller than $\epsilon_K = 1\%$. The average volume difference is also reported (fourth column), as well as total execution time (fourth column). As can be seen, when the fraction of the volume occupied by the cells is larger, the algorithm takes a larger number of iterations to converge. This is because the interactions between the cells introduce mode dependencies in $K(.)$.

\begin{figure}
    \centerline{
    \includegraphics[height=40mm]{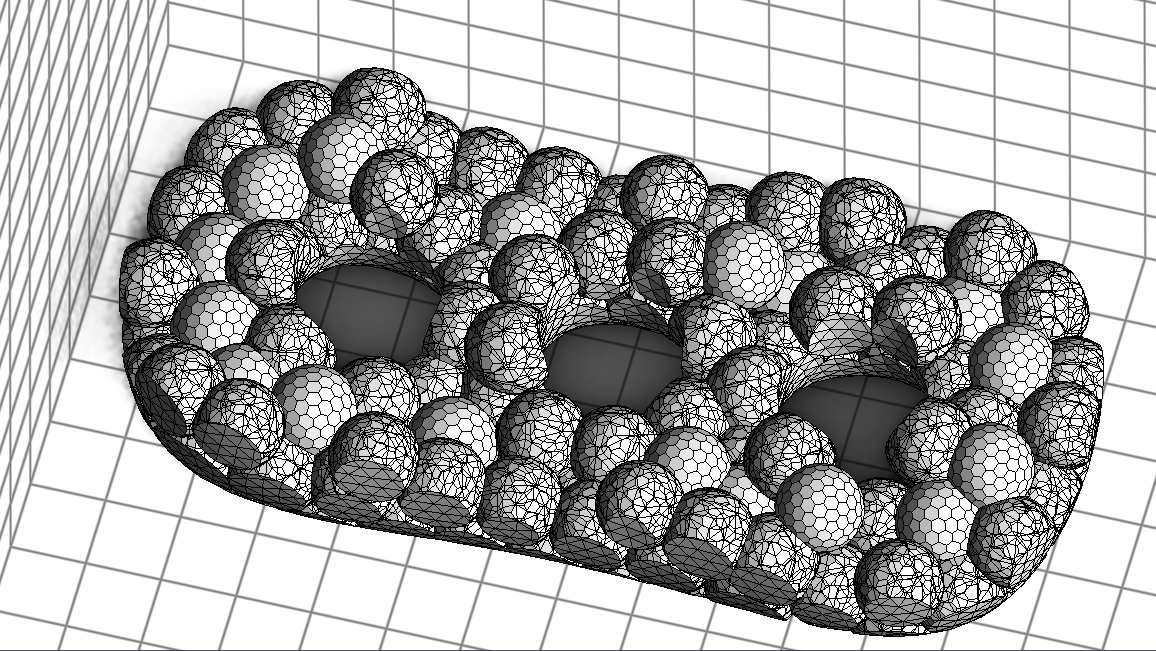}    
    \hspace{2mm}
    \includegraphics[height=40mm]{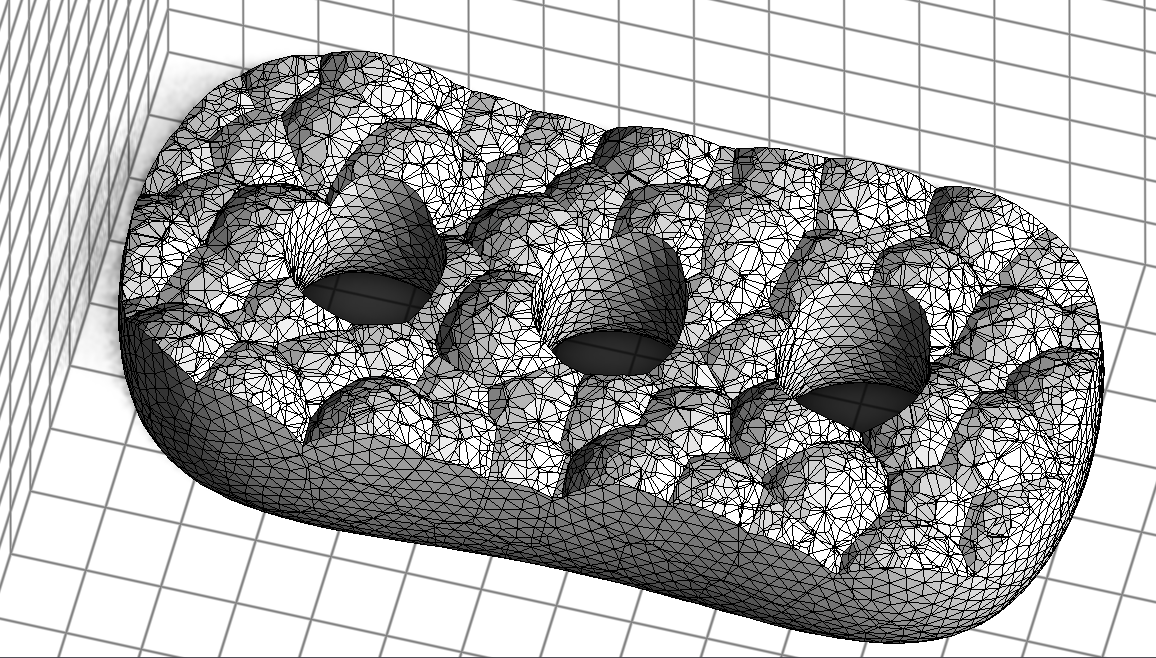}  
    }
    \caption{Testing partial optimal transport in a tet mesh, with 100 random points in the lower half of the domain. Left: fluid volume = 50\%, Right: 75\%.}
    \label{fig:POT_three_holes}
\end{figure}

We now test partial optimal transport in a domain $\Omega$ of arbitrary shape, here a genus 3 torus represented by a tessellated mesh (see Figure \ref{fig:POT_three_holes}. The tetrahedral mesh has 56K elements. The points are distributed in the bottom half of $\Omega$. Statistics and timings are reported in the table below.
$$
  \begin{array}{c|c|c|c|c}
     \mbox{fraction} & \mbox{nb iter} & \mbox{max err} & \mbox{avg err} & \mbox{time} \\
    \hline  
     10\% & 3 &  0.0032\% &  0.0031\% &  1.9s \\
     50\% & 4 &  0.9\% &  0.2\% &  4s \\
     75\% & 5 &  0.45\% & 0.12\% &  4s \\
     90\% & 6 &  0.02\% & 0.003\% &  5.88s \\
   \end{array}
$$

\subsection{Testing the Lagrangian mesh representation: a toy free-boundary fluid simulator}
 
I shall now demonstrate how the algorithm can be used to implement a Lagrangian mesh with controlled volume that can change topology, with a very simple incompressible Navier-Stokes simulator for fluids with free boundaries.
In the context of this article, this "toy" simulator is meant to demonstrate that our Lagrangian mesh can reproduce some of the typical behaviors of a fluid while accurately tracking collisions and changes of topology. A more extensive quantitative evaluation of the simulator and the calibration of its parameters will be the subject of another article.

The simulator is obtained by adding viscosity and surface tension terms to the Gallouet-Merigot scheme \cite{Gallouet2017}, that simulates an incompressible Euler fluid. Inspired by Brenier and Benamou's point of view on incompressible fluids \cite{BrenierPFMR91,DBLP:journals/nm/BenamouB00}, the Gallouet-Merigot scheme softly projects the motion onto the incompressibility constraint by adding a "spring" force, that moves each point $\bx_i$ towards the centroid of its cell $V_i$. This additional "pressure" force makes the motion of the fluid tangent to the manifold of incompressible motions, and the computed fluid motion provably converges to the solution of an incompressible Euler fluid, that is, a geodesic on the manifold of incompressible motions (see \cite{WEB:Tao,Marsden} on geodesic flows and the Euler-Arnold equation). The reader is referred to \cite{Gallouet2017} for all the details. \\

The fluid is modeled as a set of $V_i$ cells, parameterized by the $\bx_i$ points. The volume of each $V_i$ cell is controlled and remains constant throughout the whole simulation. All the cells have the same mass $m$. Each cell is subjected to the following forces:

$$
  \begin{array}{lclcl}
     F_p(\bx_i) & = & \frac{1}{\epsilon_p^2} (\bx_i - \bg_i) 
     & & \mbox {where } \bg_i = \frac{1}{|V_i|} \int_{V_i} \bx dx  
     \\[4mm]
     F_g(\bx_i) & = & -m g {\bf z} \\[4mm]
     F_v(\bx_i) & = & \mu \hat{\Delta} \bv (\bx_i) & = & \mu \sum_{j \in N_i} \frac{1}{2\| \bx_j - \bx_i \|} \ |V_{ij}|\ (\bv_j - \bv_i) \\[4mm]
     F_t(\bx_i) & = & \gamma \hat{\Delta} \bx (\bx_i) & = & \gamma \sum_{j \in N_i} \frac{1}{2\| \bx_j - \bx_i \|} \ |V_{ij}|\ (\bx_j - \bx_i).
  \end{array}
$$
Following \cite{Gallouet2017}, the "pressure" $F_p$ may be thought of as a "spring" that connects each point $\bx_i$ to the centroid $\bg_i$ of the associated cell $V_i$ with a stiffness $\epsilon_p$. The additional forces are viscosity and surface tension: viscosity $F_v$ is proportional to the Laplacian of the velocity. Surface tension $F_t$ is derived in a volumetric manner, from the mutual attraction in the fluid. Inside the fluid, the net result is zero. On the boundary, the net result is a force that pulls towards the interior of the fluid. Both viscosity and surface tension use the $\mathbb{P}_1$ Laplacian: 
 
$$
 \hat{\Delta} f(\bx) = \sum_{j \in N_i} \frac{1}{2 \| \bx_j - \bx_i \| } \ | V_{ij} |\ (f(\bx_j) - f(\bx_i))
$$
where $|V_ij| = |\partial V_i \cap \partial V_j|$ denotes the area of the facet common to the cells $V_i$  and $V_j$, and where $N_i$ denotes the set of cells touching $V_i$. In the expressions of $F_v$ and $F_t$, it is computed component-wise. \\
 
The simulation results shown below use a semi-implicit time-stepping, with implicit integration for the viscosity, as follows:

$$
  \begin{array}{lcl}
    \bx^{(k+1)} & = & \bx^{(k)} + \delta t \bv^{(k)} \\[2mm]
    \bv^{(k+1)} & = & \bv^{(k)} + \frac{\delta t}{m} \left( 
       F_p + F_g + F_t + \mu \hat{\Delta} \bv^{(k+1)} 
    \right) \\
    \mbox{or: }  
    \left( \mbox{Id} - \mu \frac{\delta t}{m} \hat{\Delta}\right) \bv^{(k+1)} & = &
       \bv^{(k)} + \frac{\delta t}{m} \left( F_p + F_g + F_t\right) \\[2mm]
    \mbox{or: }  
    \left( 
      \hat{\Delta} - \frac{m}{\mu \delta t} \mbox{Id}
    \right) \bv^{(k+1)} & = & - \frac{1}{\mu} \left( 
       \frac{m}{\delta t} \bv^{(k)} + F_p + F_g + F_t 
    \right).
  \end{array}
$$
The linear system is solved for each component $x,y,z$ separately. Note that the matrix of this linear system corresponds to $\hat{\Delta}$ plus a diagonal term. To assemble it, the same code that constructs the Hessian of $K()$ can be reused. \\
Sub-stepping is used whenever the CFL condition is violated, that is whenever a displacement $\delta t\ \bv$ is such that more than one cell is crossed, that is whenever there exists an $i$ such that $\delta_t \ |\bv_i| > \sqrt{\psi_i}$. \\
 
\begin{figure}
    \centerline{
    \includegraphics[height=45mm]{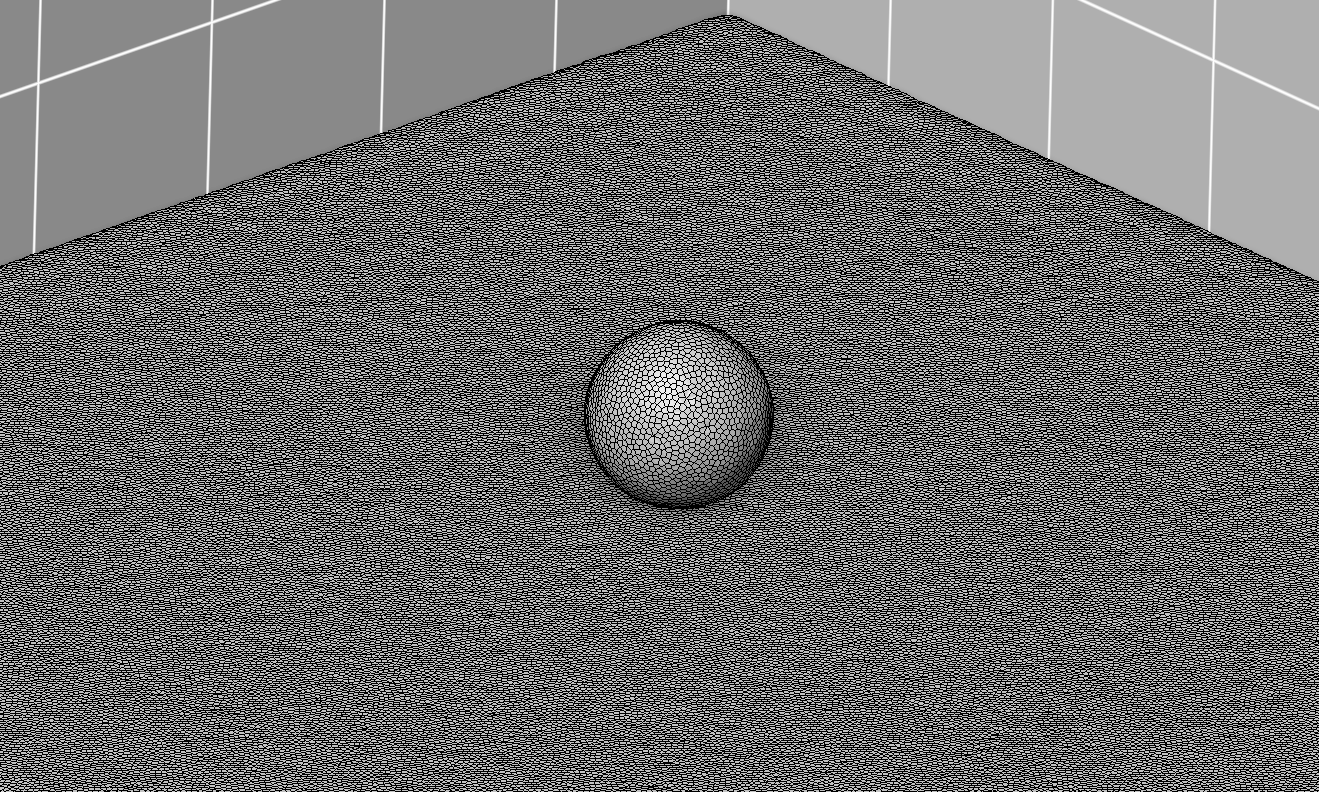}    
    \includegraphics[height=45mm]{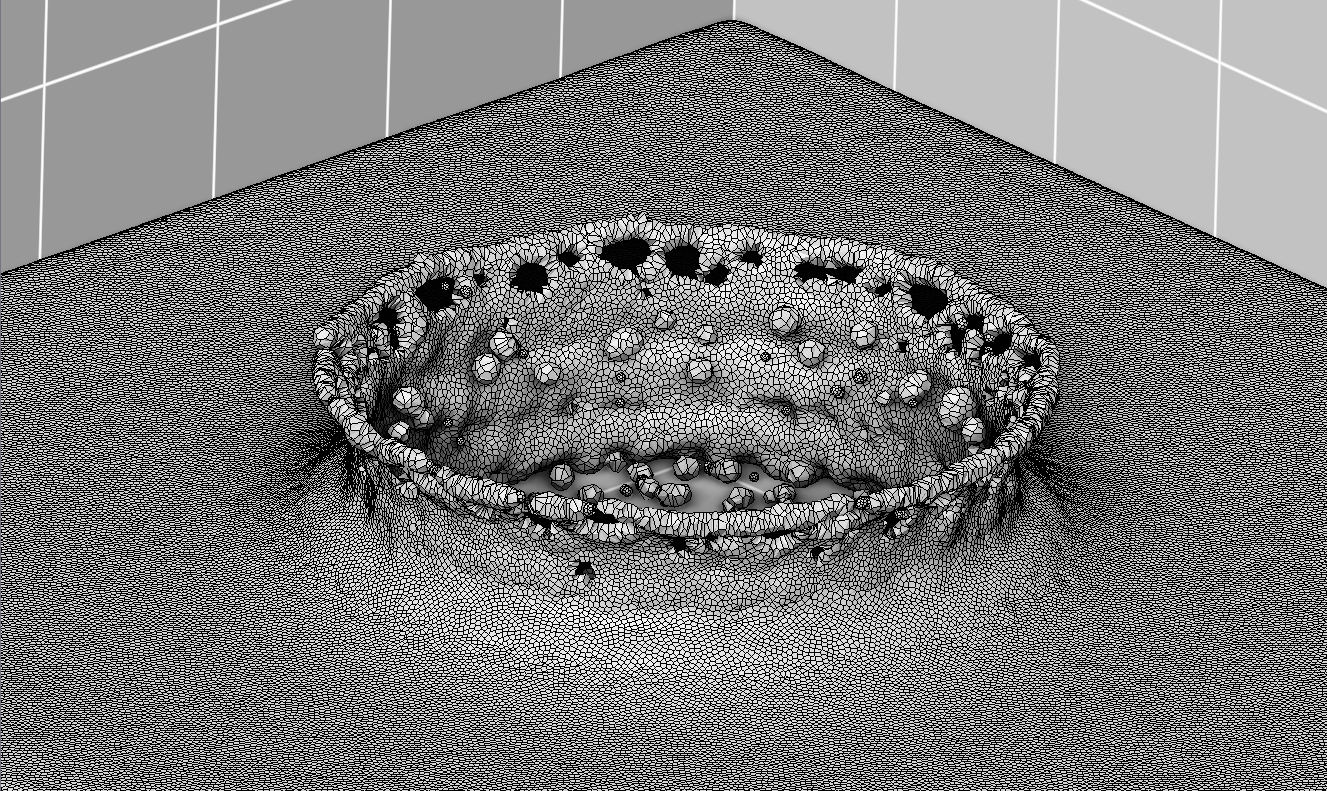}
    }
    \vspace{2mm}
    \centerline{
    \includegraphics[height=45mm]{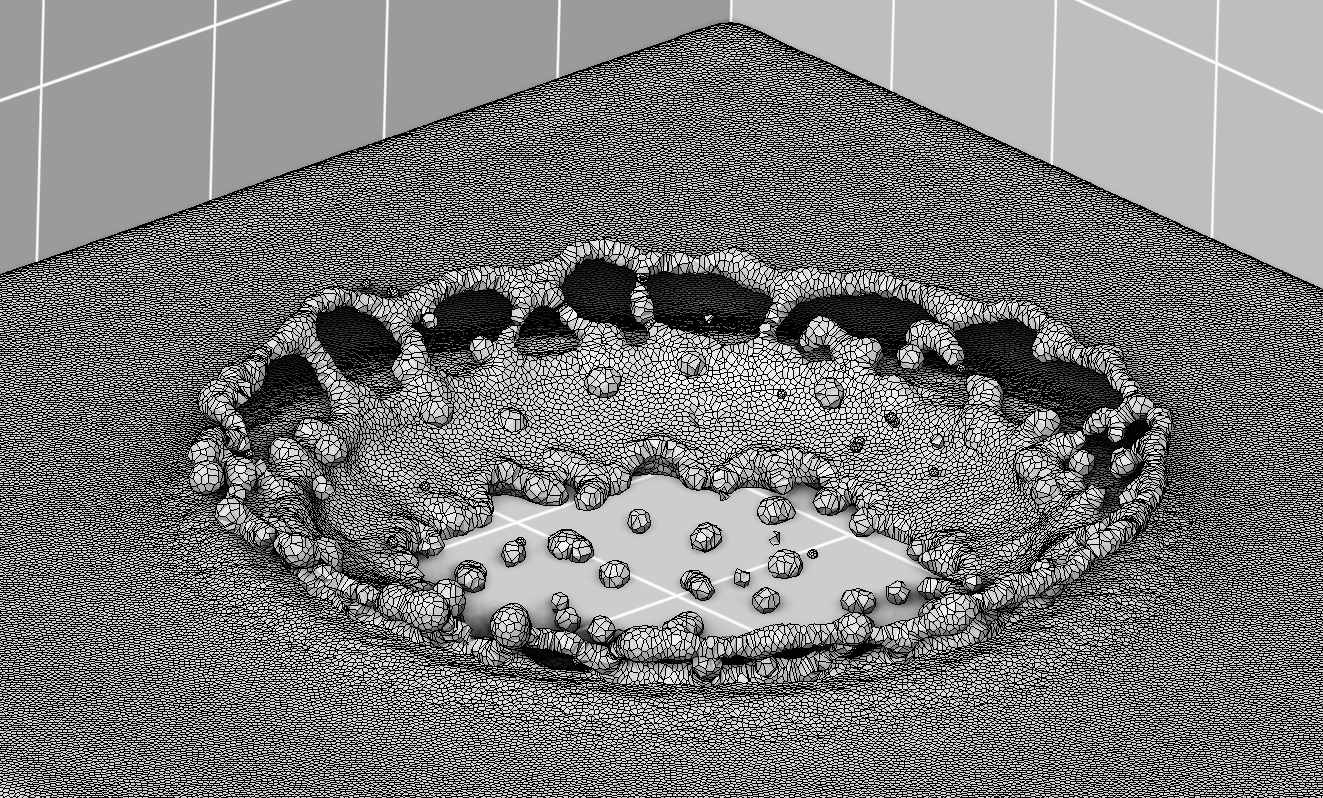}
    \includegraphics[height=45mm]{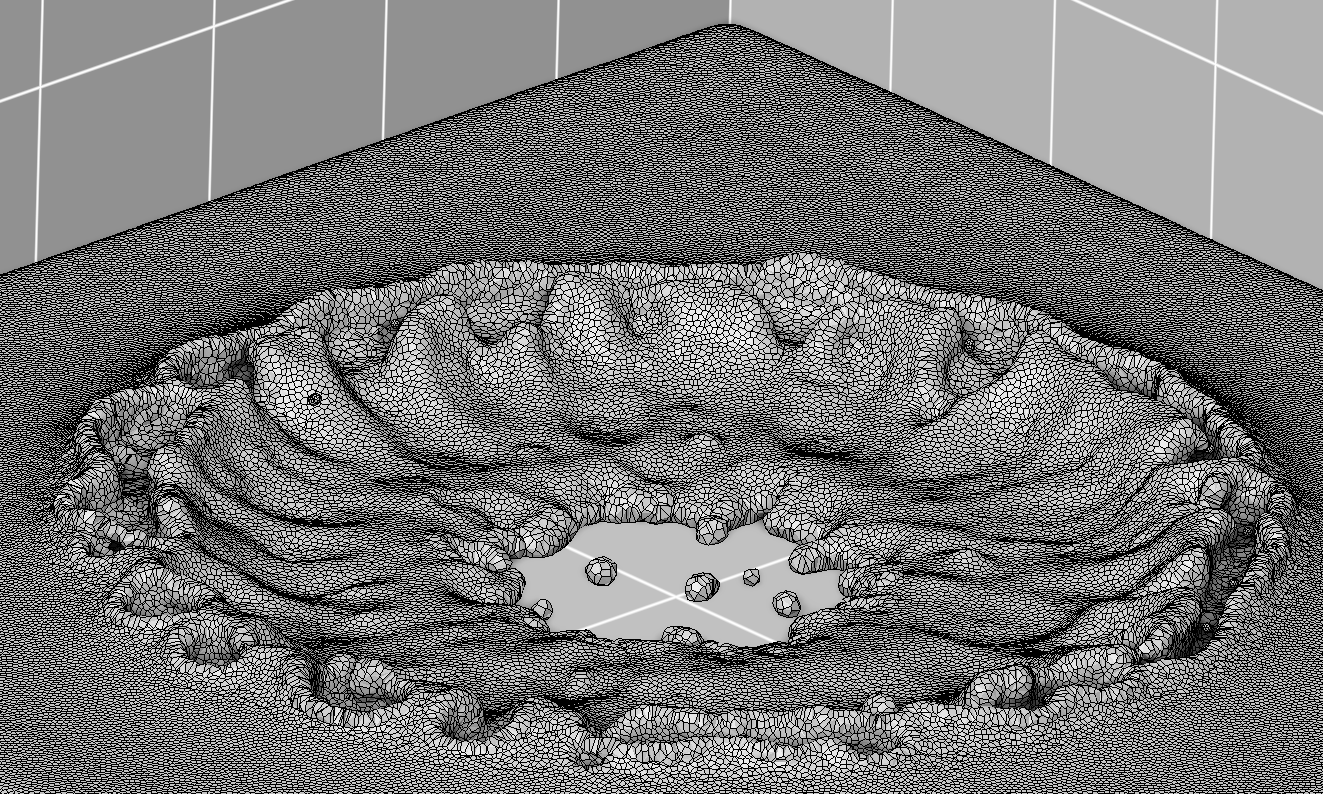}
    }
    \vspace{2mm}
    \centerline{
    \includegraphics[height=45mm]{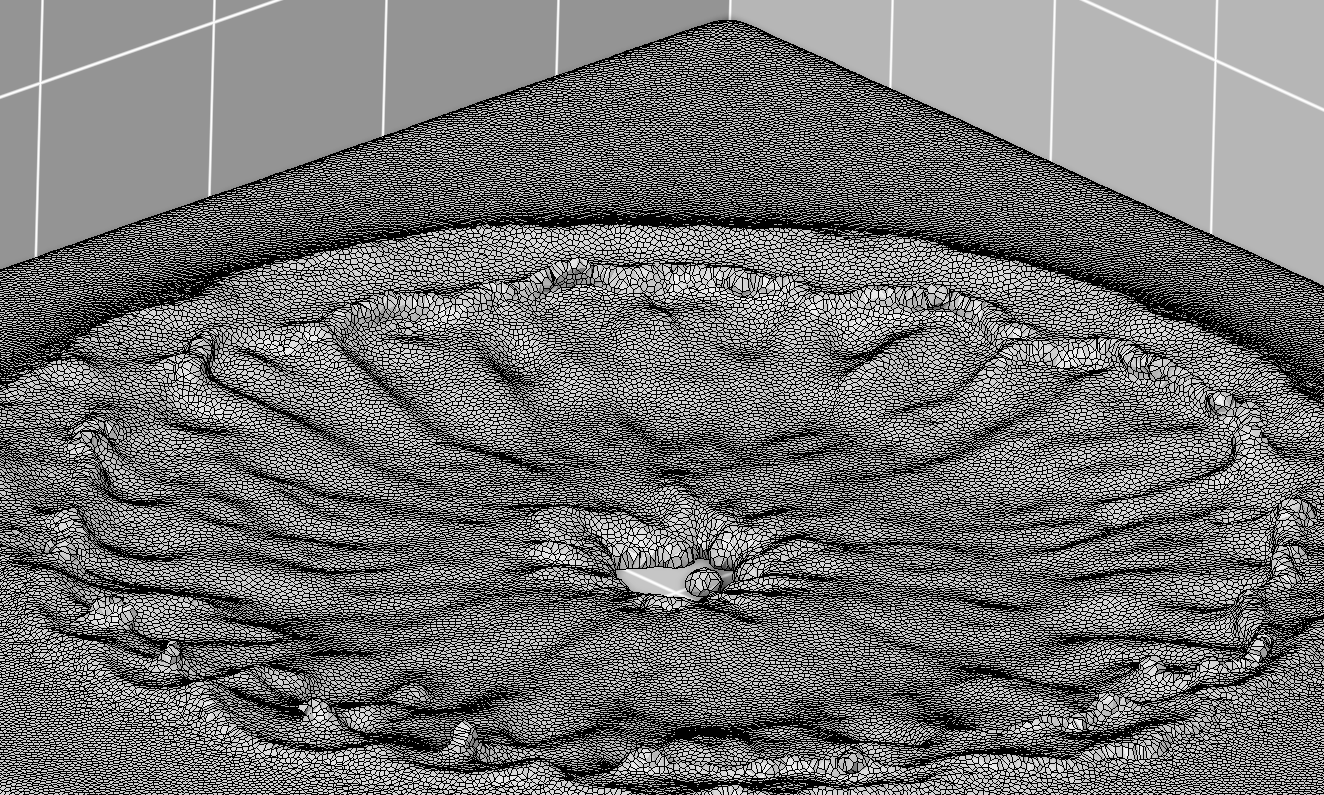}
    \includegraphics[height=45mm]{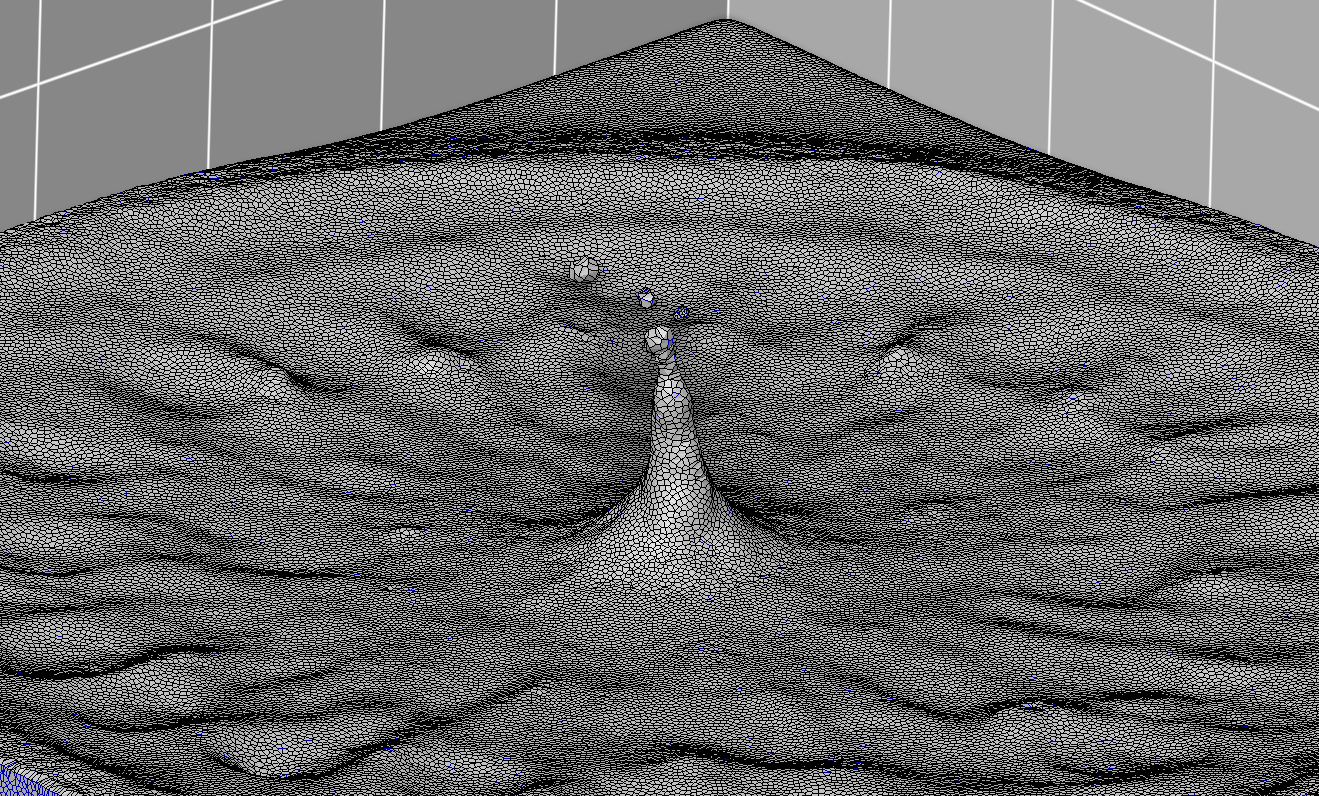}
    }
    \caption{Testing surface tension and topology changes: crown splash}
    \label{fig:crown}
\end{figure}

\begin{figure}
    \centerline{
    \includegraphics[width=60mm]{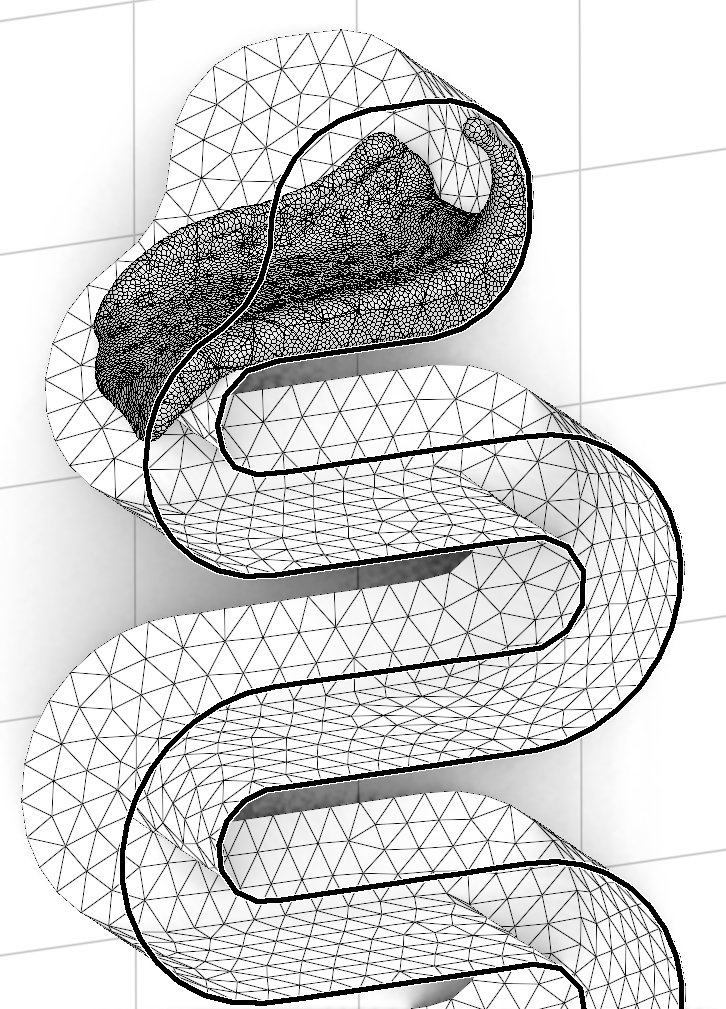}
    \hspace{2mm}
    \includegraphics[width=60mm]{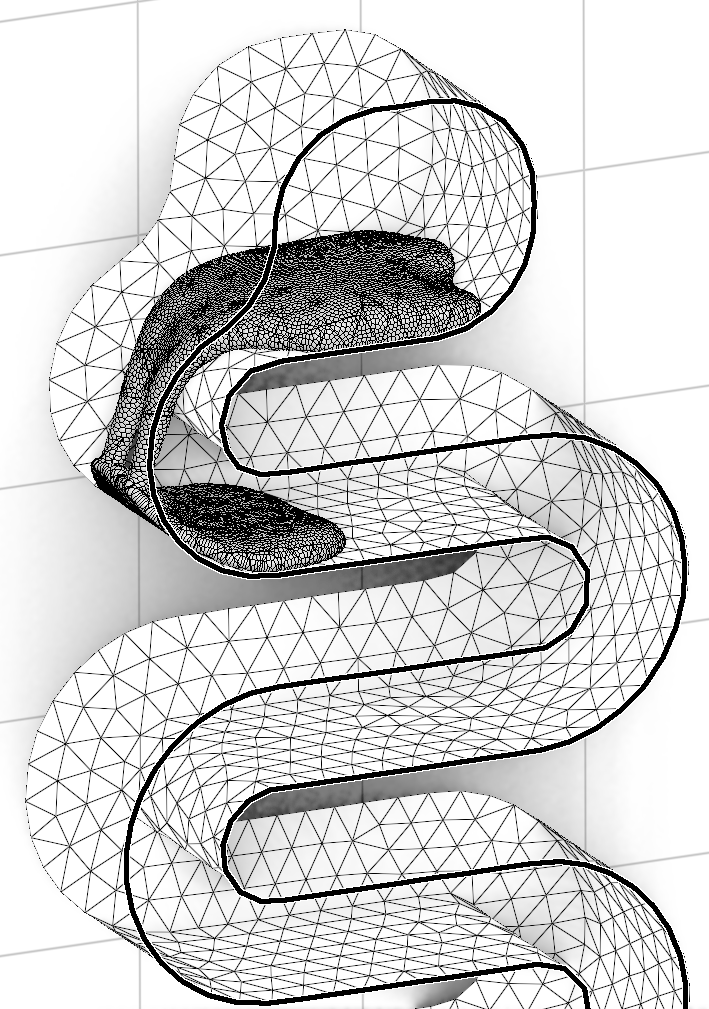}
    }
    \vspace{2mm}
    \centerline{
    \includegraphics[width=60mm]{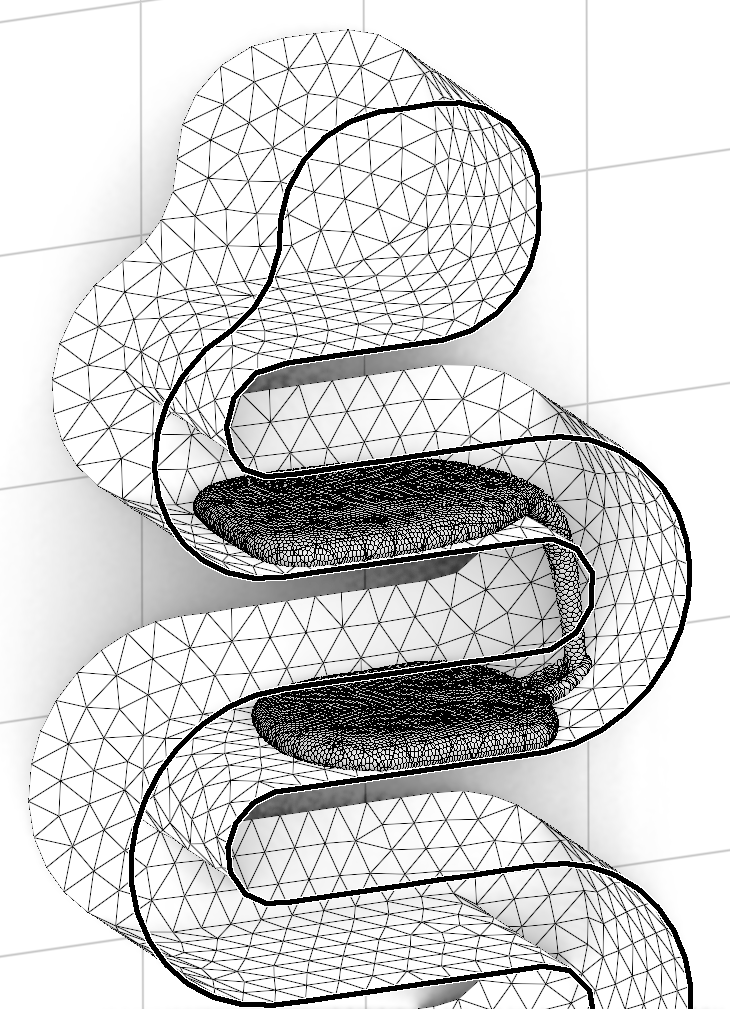}
    \hspace{2mm}
    \includegraphics[width=60mm]{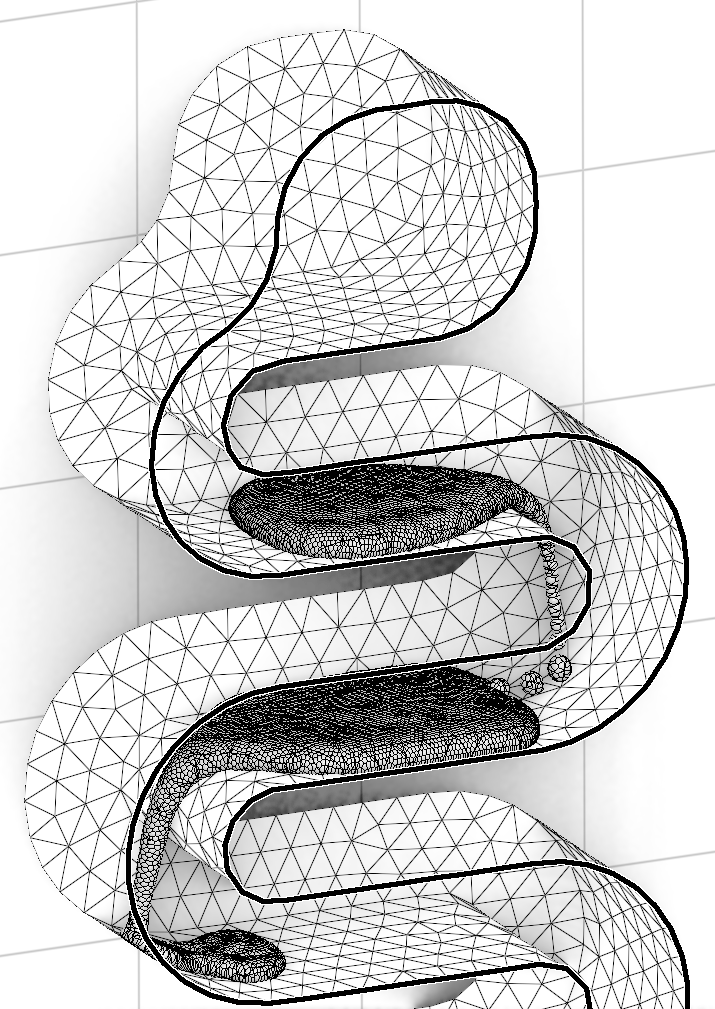}
    }
    \caption{Free-surface flow and interactions with boundaries: moderate surface tension.}
    \label{fig:honey}
\end{figure}

\begin{figure}
    \centerline{
    \includegraphics[width=60mm]{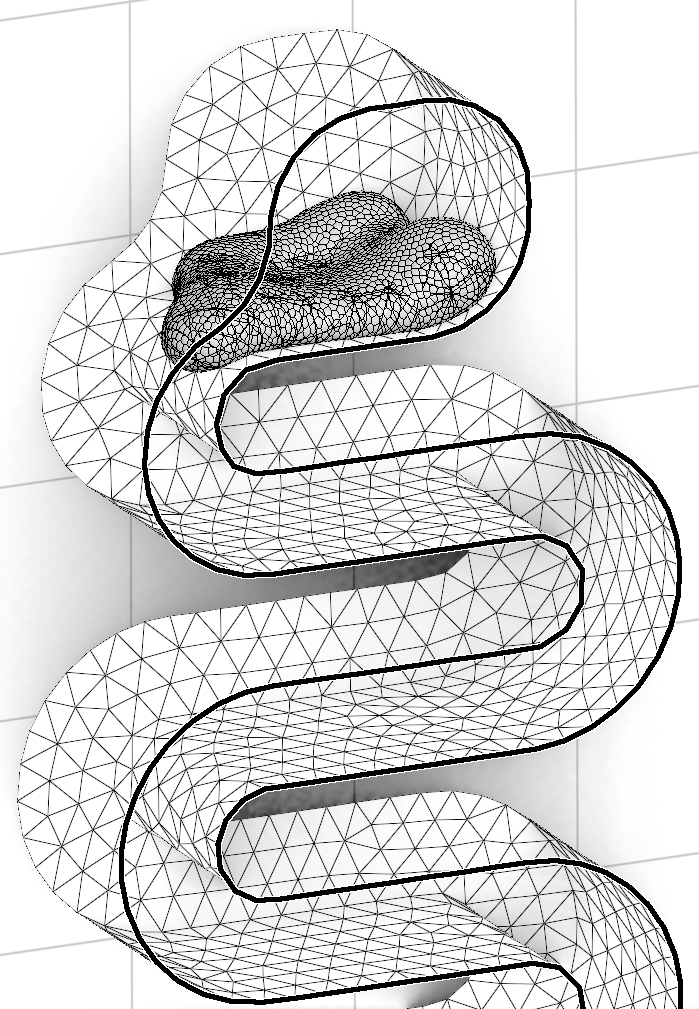}
    \hspace{2mm}
    \includegraphics[width=60mm]{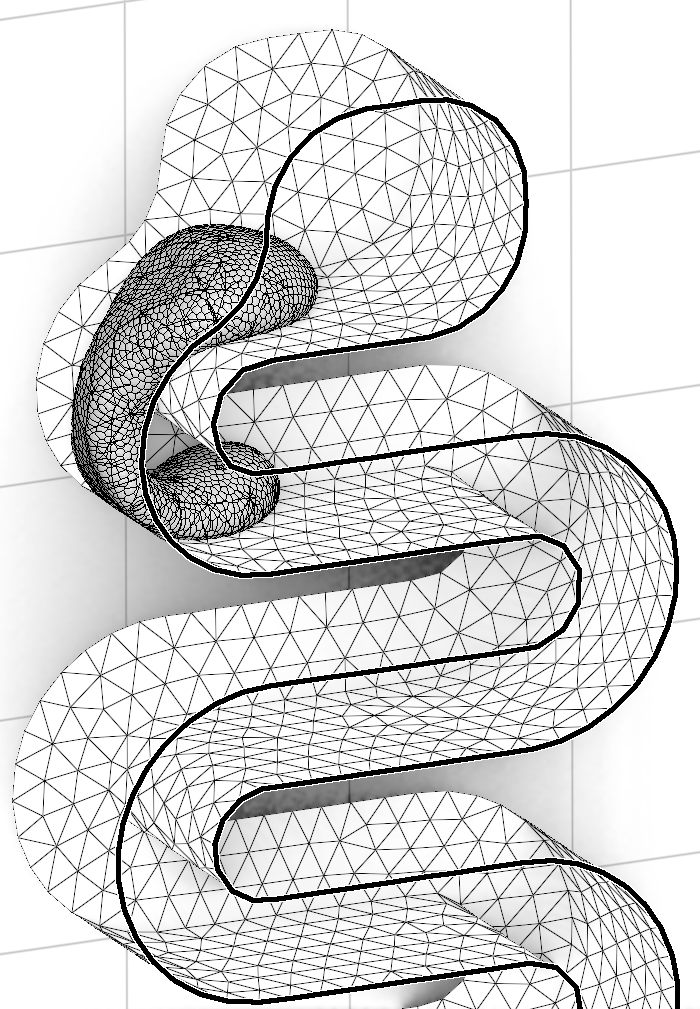}
    }
    \vspace{2mm}
    \centerline{
    \includegraphics[width=60mm]{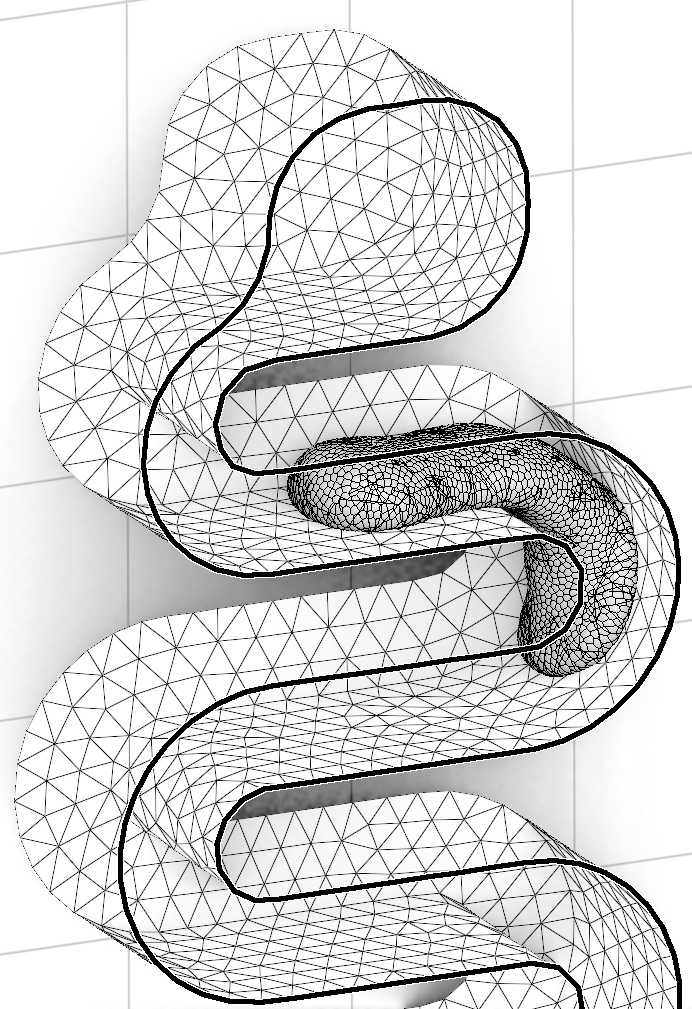}
    \hspace{2mm}
    \includegraphics[width=60mm]{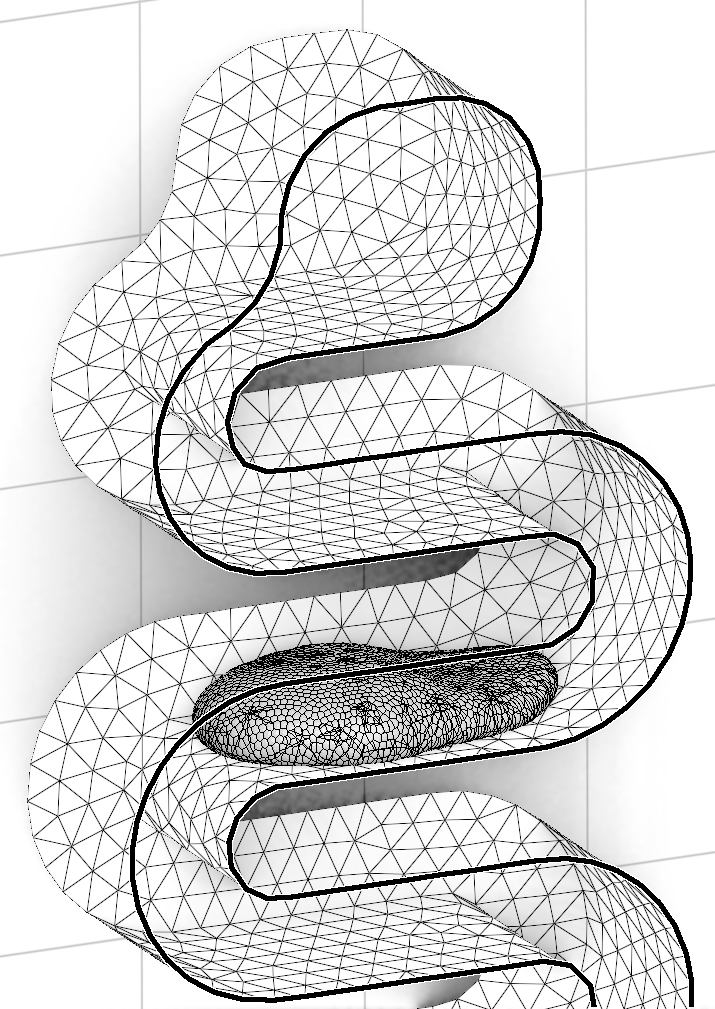}
    }
    \caption{Free-surface flow and interactions with boundaries: higher surface tension.}
    \label{fig:mercury}
\end{figure}

\begin{figure}
    \centerline{
    \includegraphics[width=146mm]{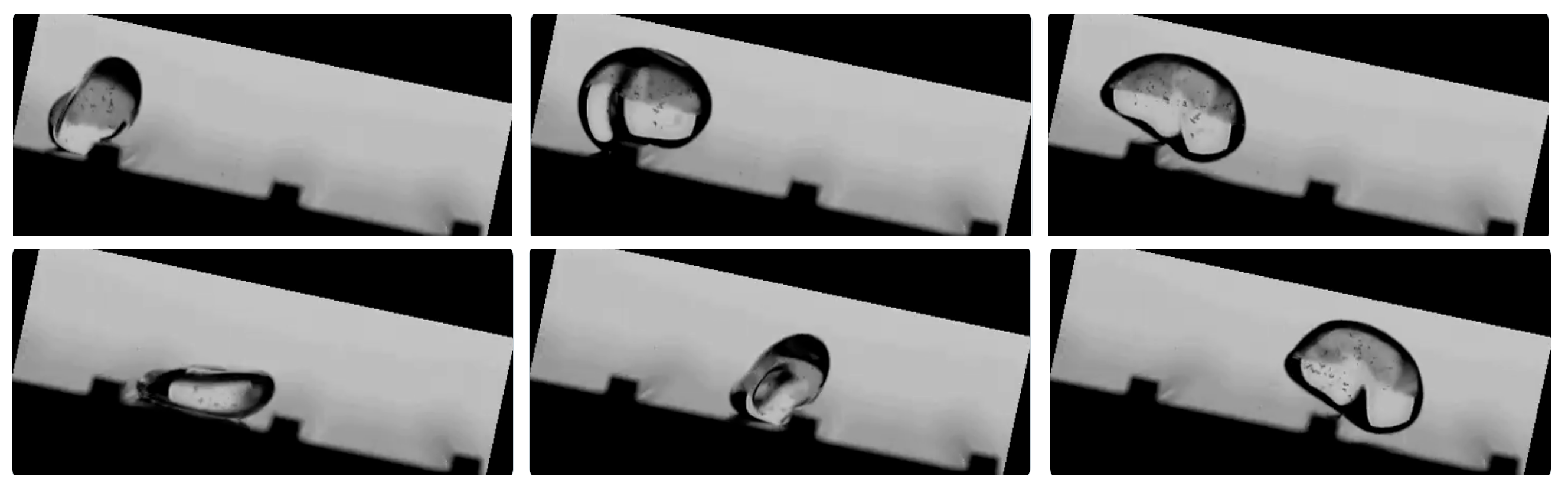}    
    }
    \caption{Droplet hurdles race, courtesy Hélène de Maleprade, Rachid Bendimerad, Christophe Clanet and David Quéré.}
    \label{fig:hurdlesreal}
\end{figure}

\begin{figure}
    \centerline{
    \includegraphics[width=73mm]{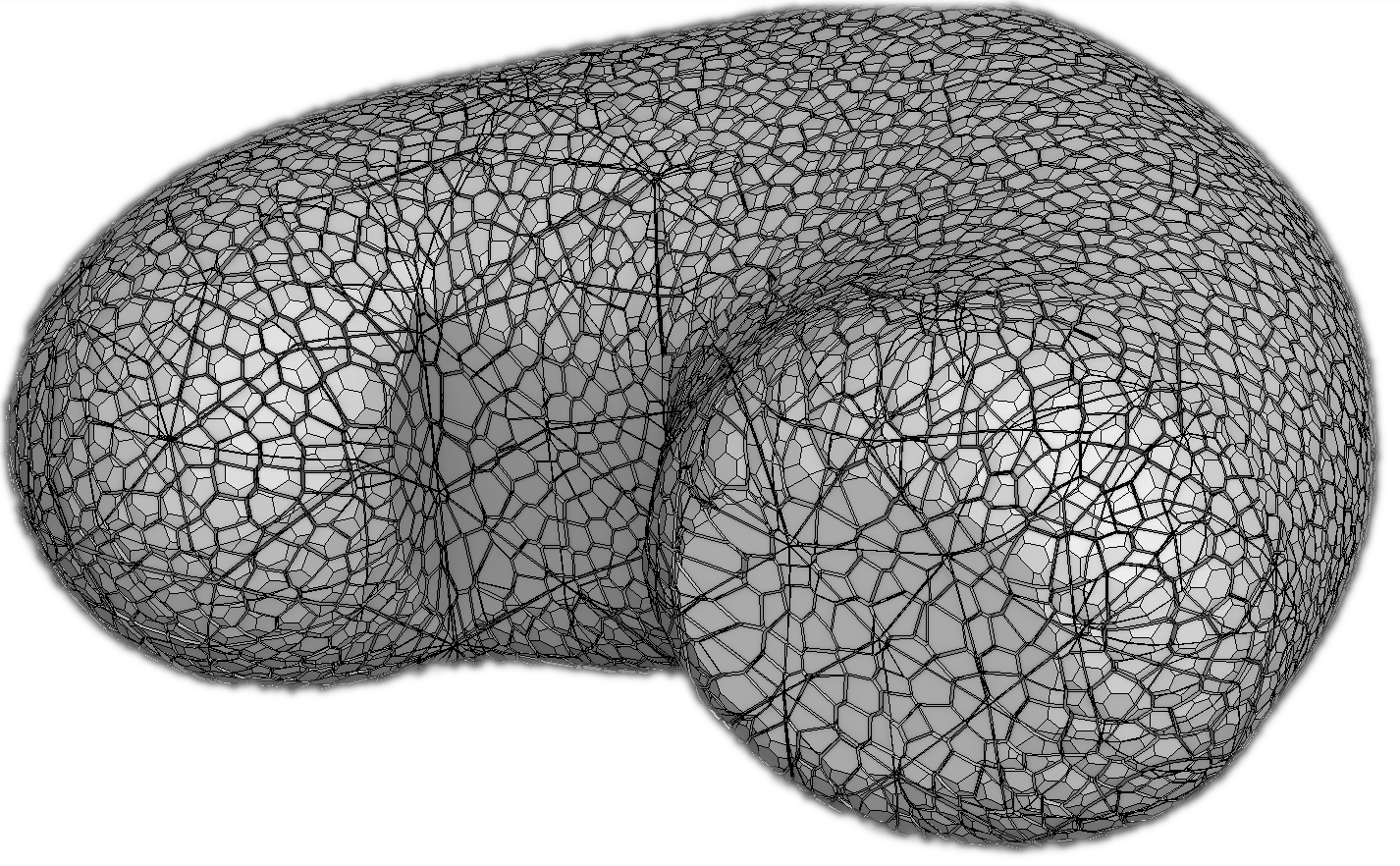}    
    \includegraphics[width=73mm]{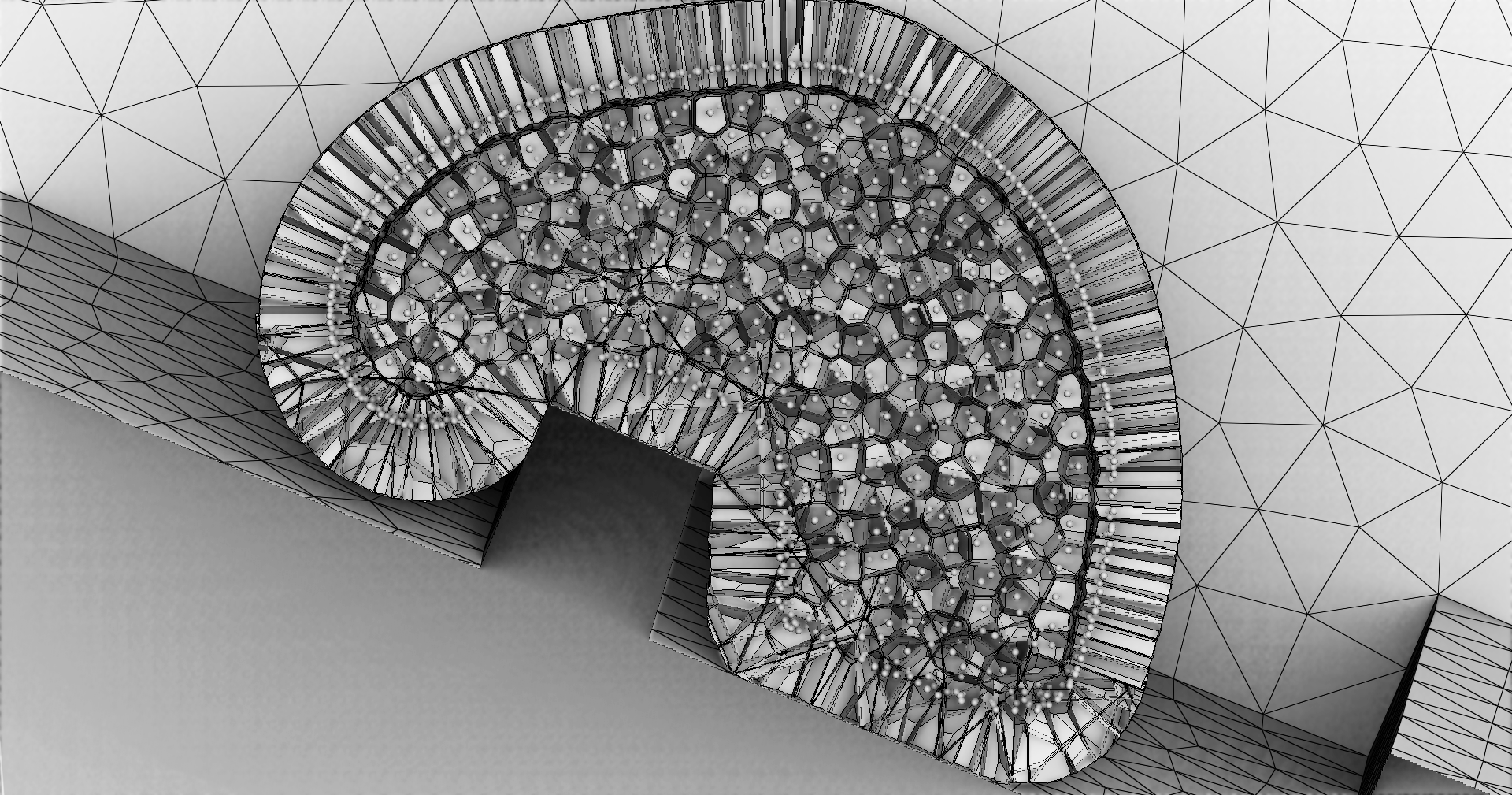}
    }
    \caption{Simulated droplet hurdles race: Mesh details and cross-section.}
    \label{fig:hurdles2}
\end{figure}

\begin{figure}
    \centerline{
    \includegraphics[width=73mm]{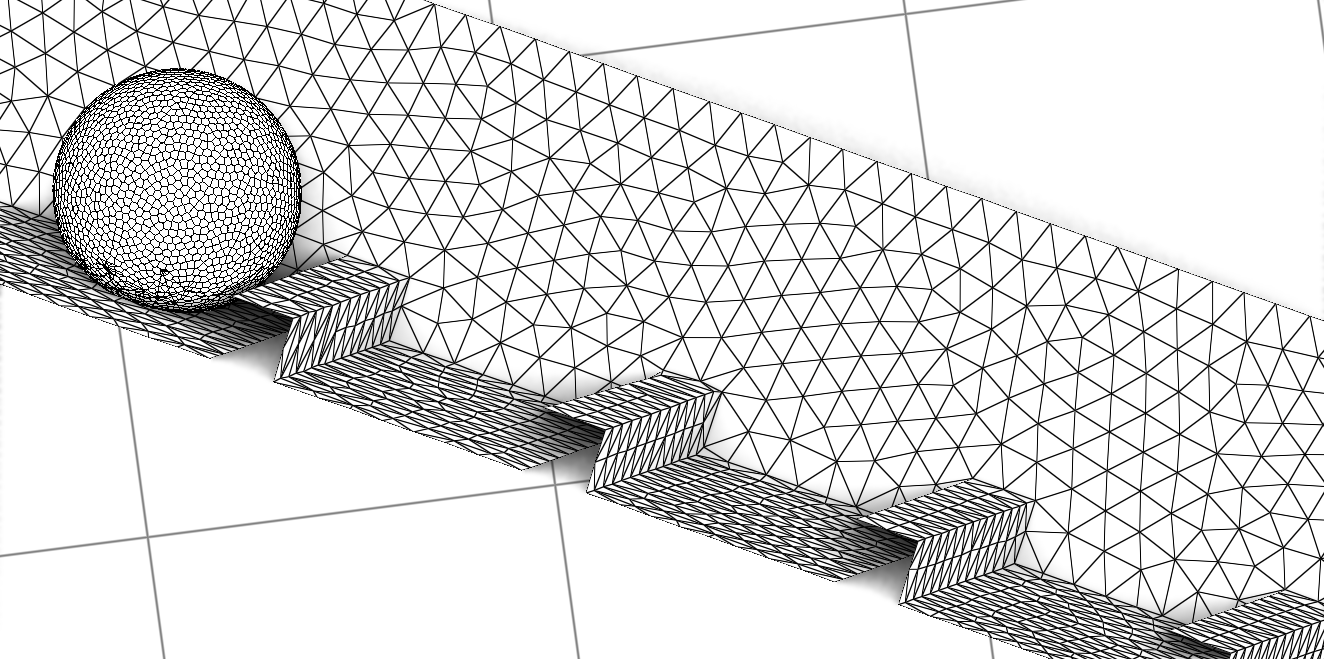}    
    \includegraphics[width=73mm]{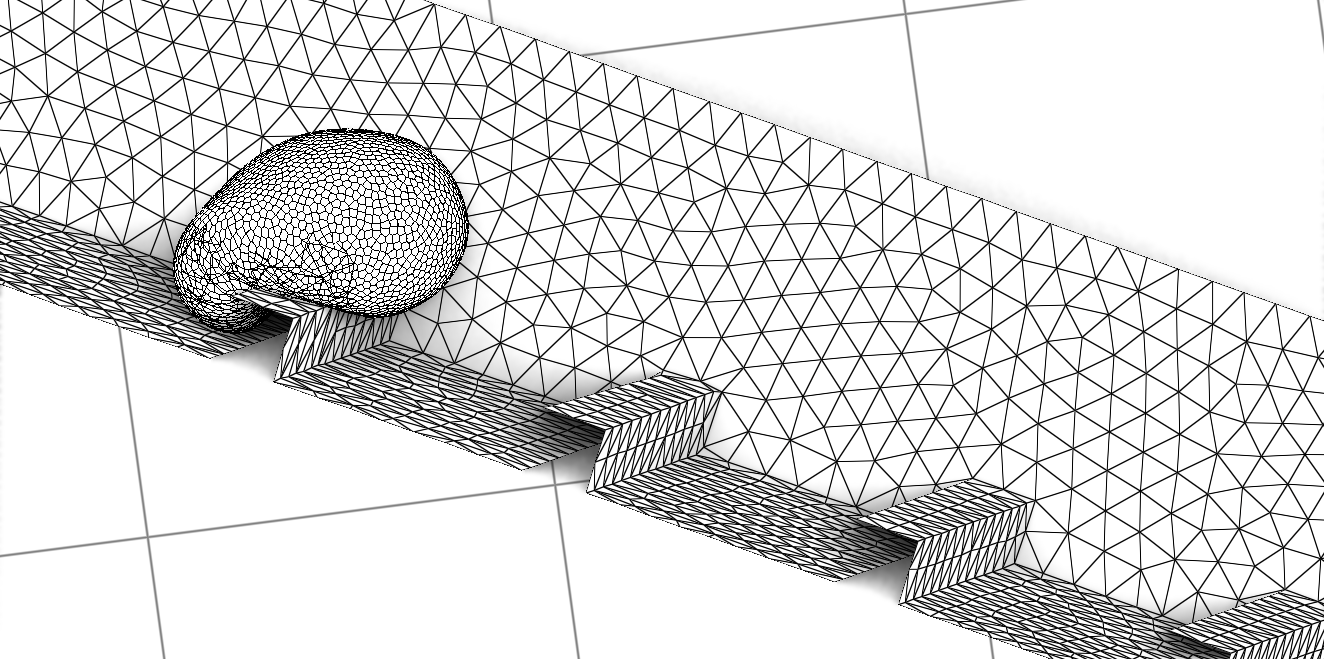}
    }
    \vspace{2mm}
    \centerline{
    \includegraphics[width=73mm]{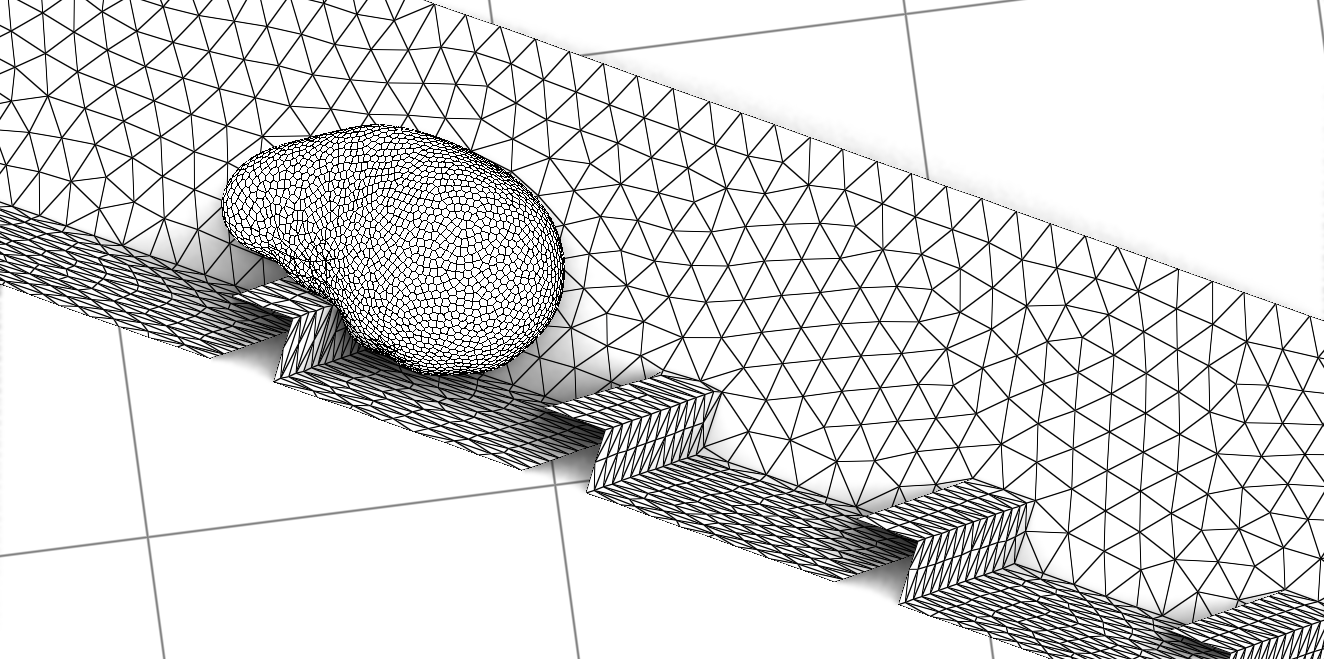}    
    \includegraphics[width=73mm]{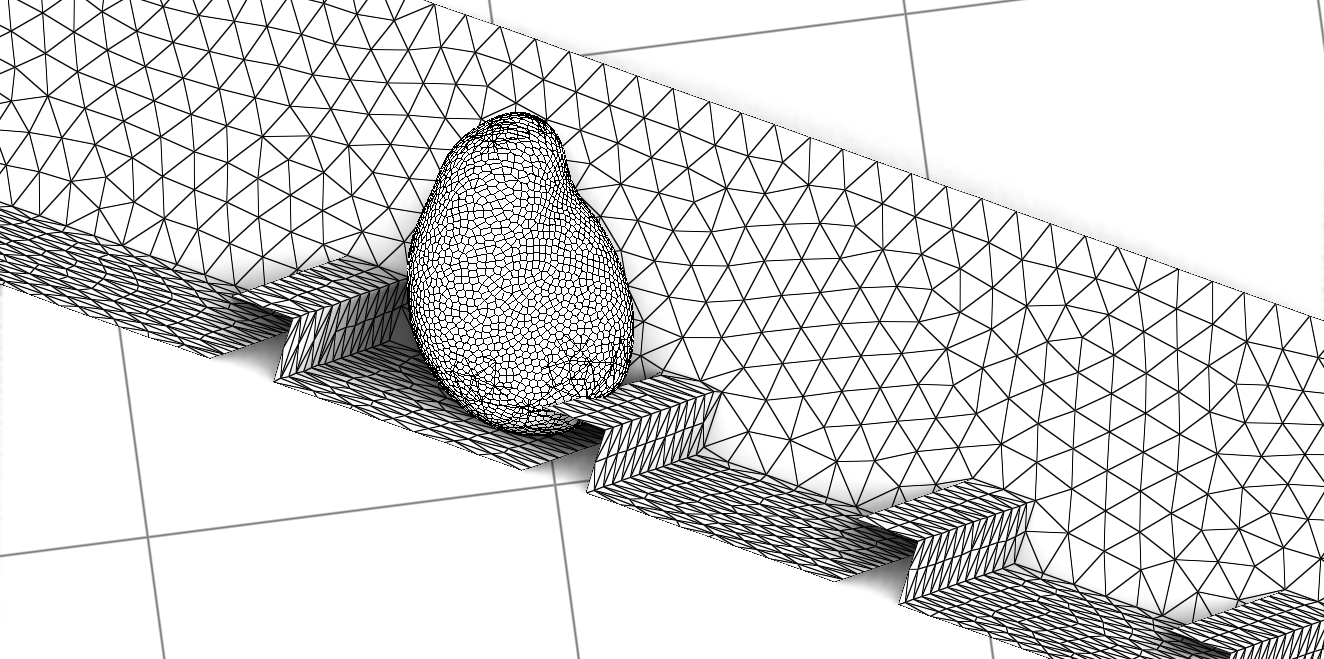}
    }
    \vspace{2mm}
    \centerline{
    \includegraphics[width=73mm]{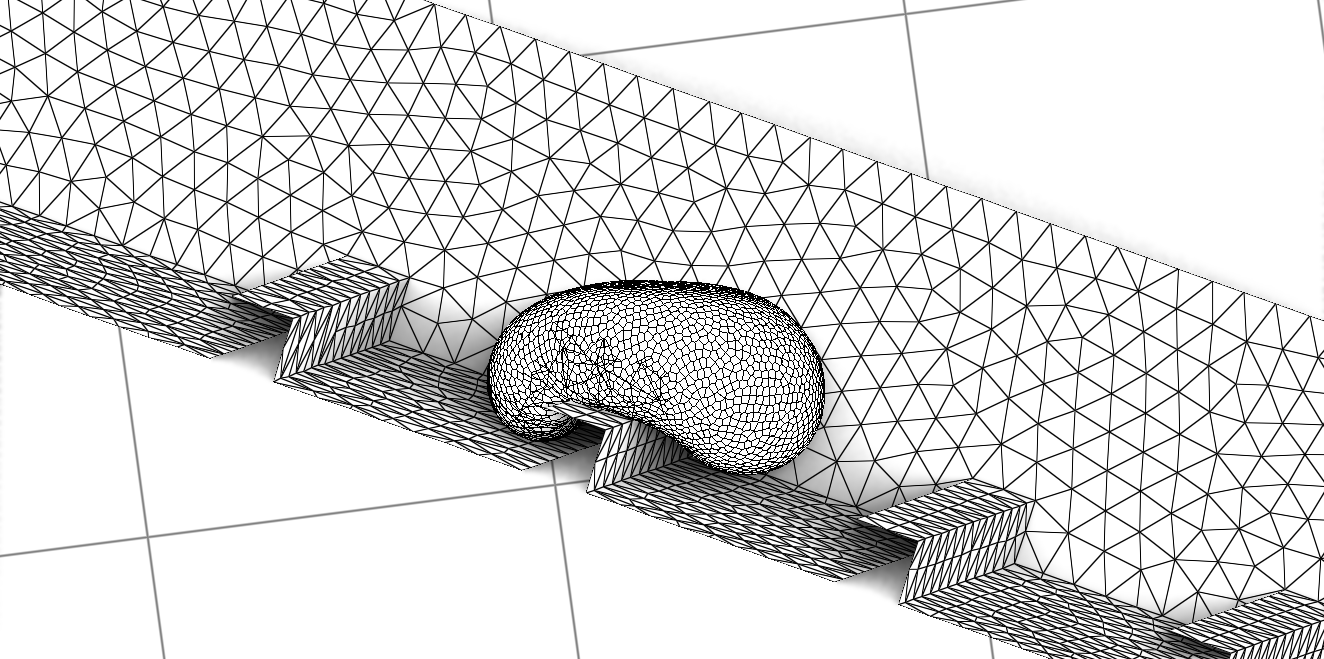}    
    \includegraphics[width=73mm]{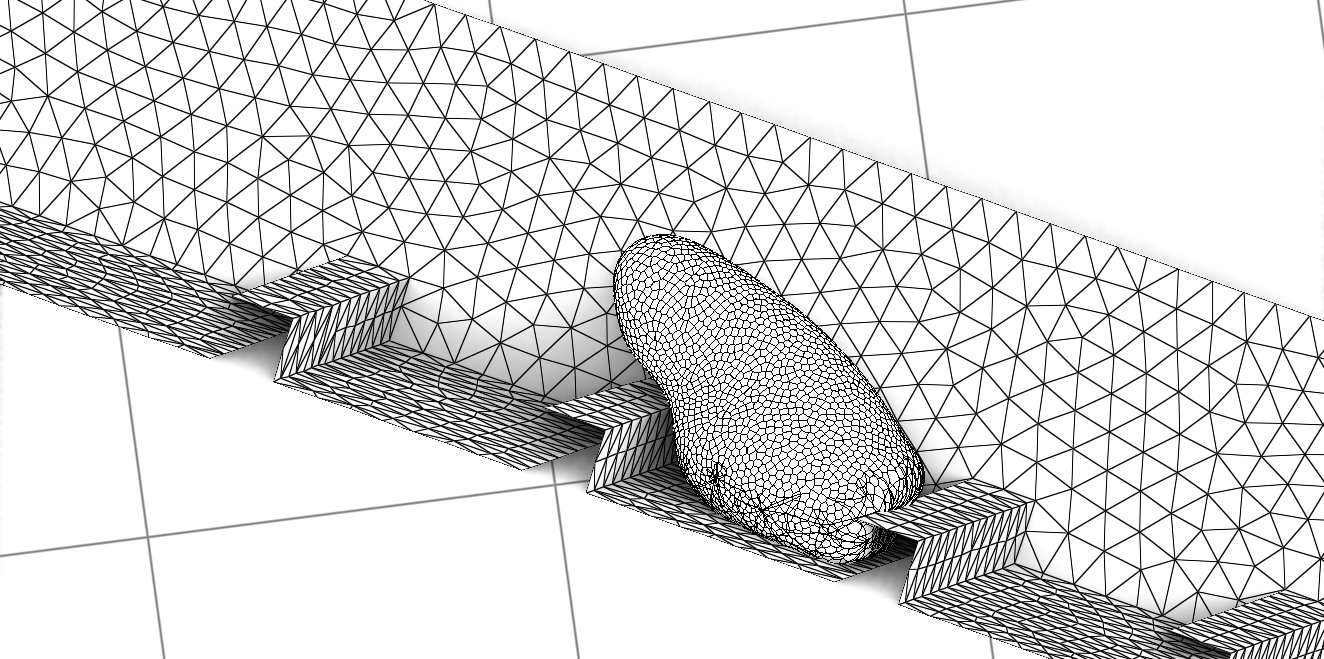}
    }
    \vspace{2mm}
    \centerline{
    \includegraphics[width=73mm]{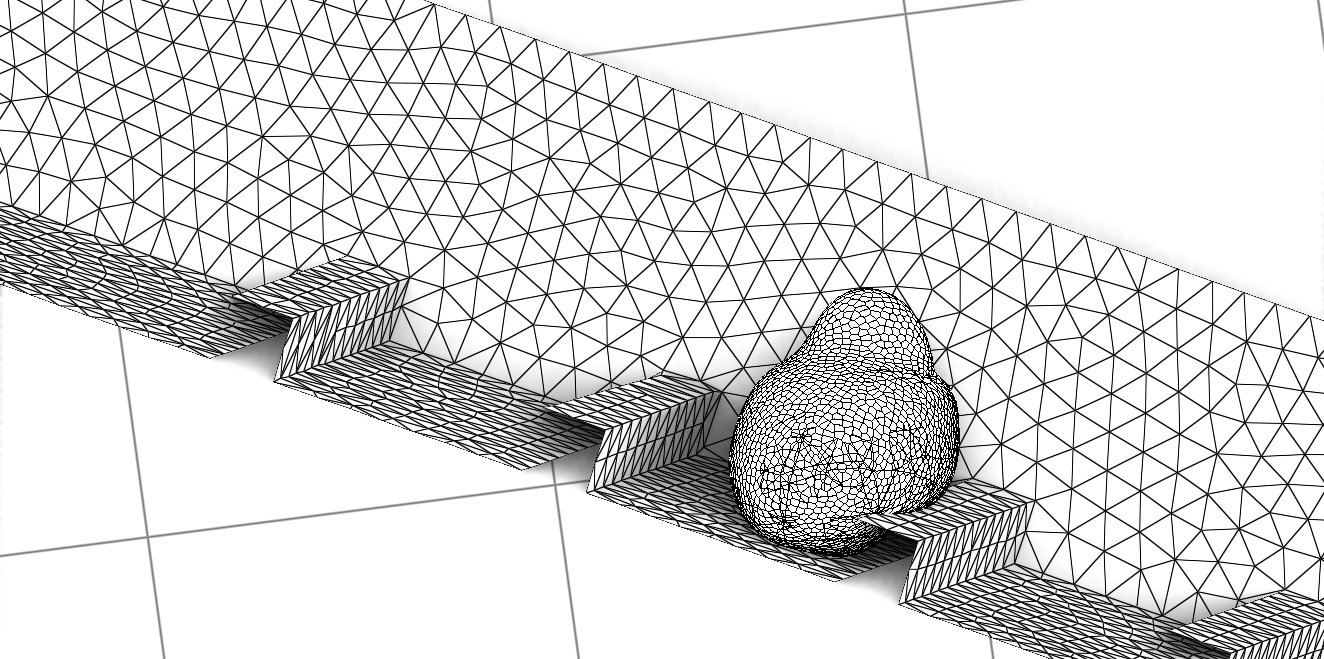}    
    \includegraphics[width=73mm]{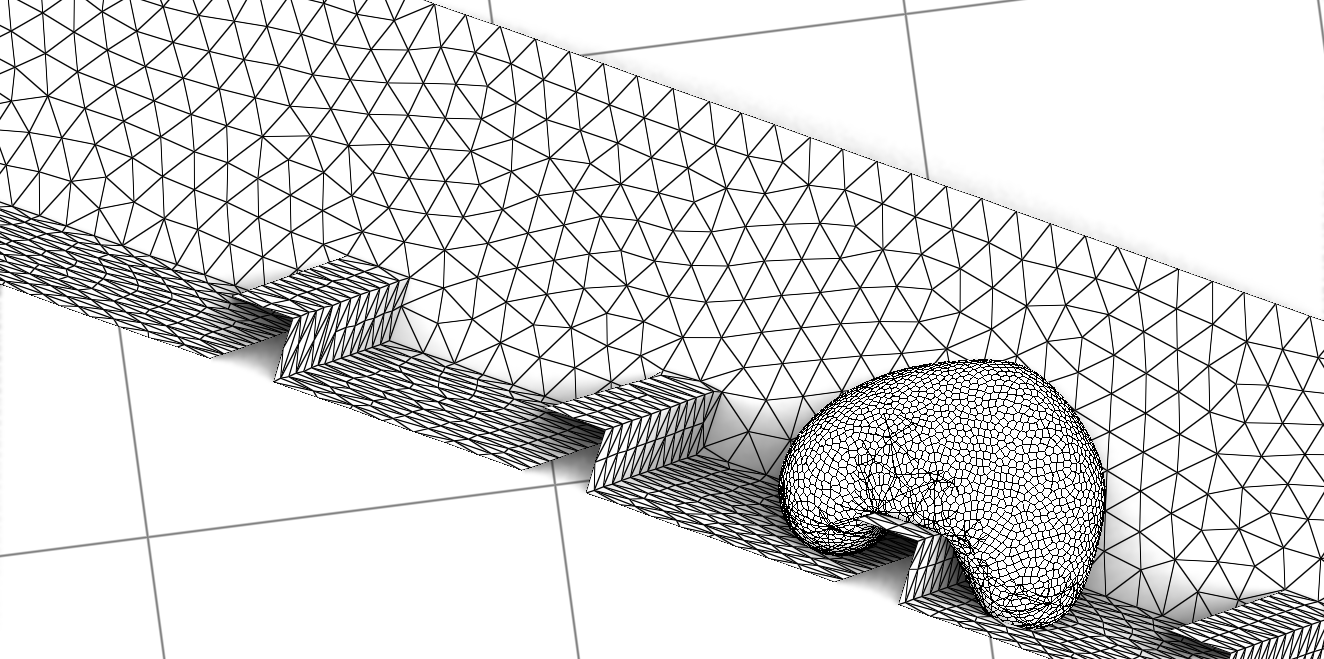}
    }
    \vspace{2mm}
    \centerline{
    \includegraphics[width=73mm]{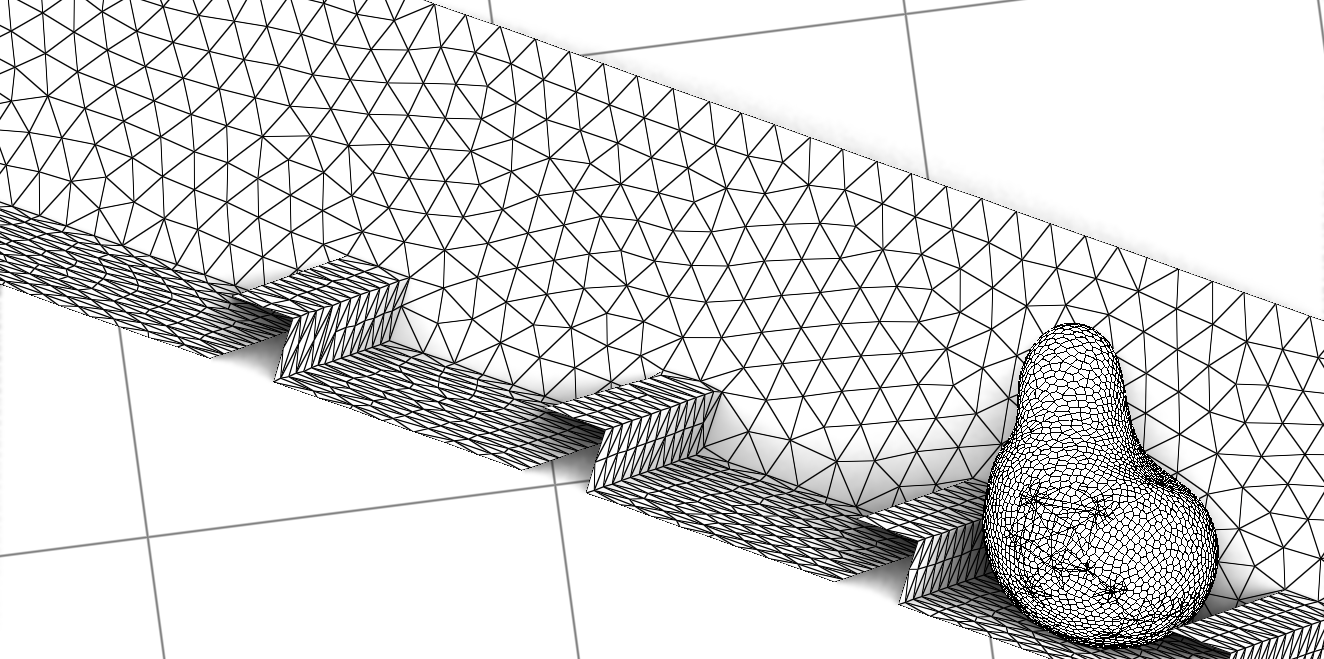}    
    \includegraphics[width=73mm]{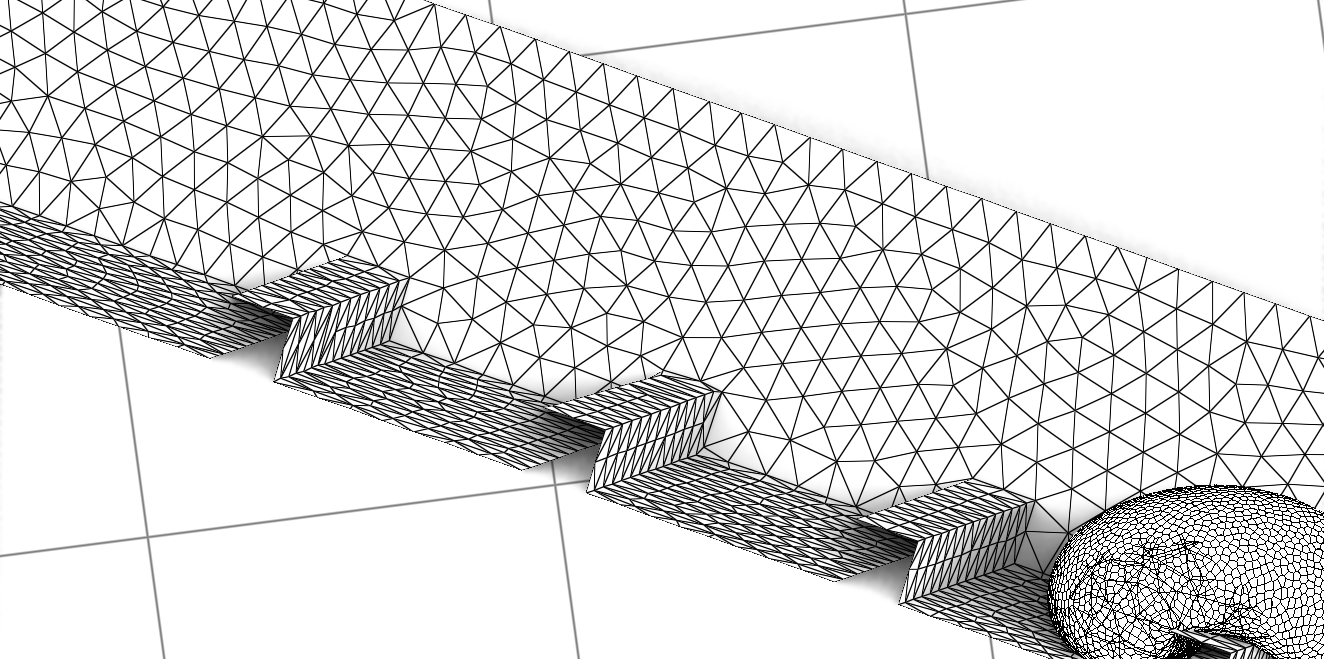}
    }
    \caption{Simulated droplet hurdles race}
    \label{fig:hurdles}
\end{figure}

The method is tested in four different configurations (see videos in  the webpage \url{https://members.loria.fr/blevy/papers/POT}):
\begin{itemize}
    \item a simulated crown splash, shown in Figure \ref{fig:crown}, computed with 500000 cells. The simulated mesh captures all the changes of topology, and preserves the fine details of the motion throughout the simulation;
    \item a fluid in a zigzag domain (Figure \ref{fig:honey}), demonstrating the interactions between 
    the fluid and the boundary of the domain;
    \item the same zigzag domain (Figure \ref{fig:mercury}) with a higher surface tension;
    \item a simulation of the experimental results in \cite{hurdles} shown in Figure \ref{fig:hurdlesreal}, where a droplet slides along a slope and bounces over obstacles. An example of a simulation mesh
    is shown in Figure \ref{fig:hurdles2}, as well as a cross-section, revealing the internal structure of the mesh, and the effect of surface tension that clusters the points $\bx_i$ near the surface (due to the volume constraint, the associated cells are more elongated than the others). Some frames of the full animation are shown in Figure \ref{fig:hurdles}. The fluid-boundary interactions and the typical oscillations of the droplet are reproduced in the simulation.
\end{itemize}
 
The table below indicates the parameters for the four simulations: 
$$
   \begin{array}{l|l|l|l|l|l|c}
     & g & m & \mu & \gamma & N & \mbox{comput. time} \\
   \hline
   \mbox{Splash}   & 10  & 3 & 0.001 & 1.5 & 500000 & 98s \\
   \mbox{Zigzag 1} & 1   & 3 & 0.001 & 0.5 & \ 25000 & 12s \\
   \mbox{Zigzag 2} & 1   & 3 & 0.001 & 2   & \ \ 5000 & 2.8s \\
   \mbox{Hurdles}  & 0.3 & 3 & 0.001 & 3   & \ \ 5000 & 3.5 
   \end{array}
$$
The table reports the parameters for gravity ($g$), viscosity ($\mu$), surface tension ($\gamma$), the number of cells ($N$) and the computation time for one timestep. The domain $\Omega$ is normalized in the $[0,1]^3$ box. For fine-scale tension-dominated simulations (zigzag and hurdles), the simulation space is globally scaled-down by applying a scaling factor to $g$. In the four simulations, the threshold for Newton iterations $\epsilon_K$ is set to $1\%$. During the whole simulation, each cell volume $|V_i|$ remains within a $1\%$ tolerance from the specified volume. Note that unlike some methods that use an Eulerian mesh, volume errors do not accumulate, the $1\%$ bound is met by all cells at all timesteps. The parameter $\epsilon_p$ for the "pressure" is set to $0.004$, and the timestep is set to $0.004$. The spheres are approximated using $N_u = 162$ half-spaces.\\

I shall now report timing breakdown, measured with the crown splash example (Intel Core i9-9880H, 2.3 GHz, multithreaded implementation):
$$
\begin{array}{l|l}
  \mbox{phase} & \mbox{time} \\
  \hline
  \mbox{Laguerre (Bowyer-Watson)}   & 9.7\% \\
  \mbox{Linear solve} & 2.2\% \\
  \mbox{Evaluate gradient} & 32.8\% \\
  \mbox{Evaluate and assemble Hessian}  & 41.5\% \\
  \mbox{Euler update and implicit viscosity lin. solve} & 13.8\%
\end{array}
$$
The phases that take the largest amount of time are evaluating the gradient and Hessian. It is because these phases involve all the geometric computations (query the Axis-Aligned Bounding Box tree, compute intersections between convex polytopes and compute volumes and areas). \\

Concerning the geometric predicate used by the convex polytope clipping algorithm (the sign of a 4x4 determinant), during 100 timesteps of the crown splash example, it was called 15752996161 times (15.7 billions), and the arithmetic filter triggered the arbitrary precision mode 70034 times (that is, 0.0004 \%). Clearly, degenerate configurations represent a tiny proportion, but given the huge number of computed intersections, they appear in each simulation. Our robust intersection algorithm properly handles them, with negligible overhead, thanks to the arithmetic filters (see also \cite{DBLP:journals/cad/Levy16} and references herein).
 
\section{Conclusions and future works} 

The experimental results tend to confirm that semi-discrete partial optimal transport can be used to implement a Lagrangian scheme for free-surface fluid simulation. The precise analysis, comparison to state of the art codes and calibration of the parameters will be the topic of another article. Note also that the convergence to the Merigot-Gallouet scheme to the solution of the incompressible Euler equation is guaranteed \cite{Gallouet2017}, but it is no longer the case for Navier-Stokes, with the added viscosity and surface tension terms. 

I used here a simple semi-implicit time stepping, it would be interesting to implement a simplectic scheme: in the end this will result in a simulator with both precise conservation of volume and precise conservation of the Hamiltonian. However, this will require a careful analysis for sub-stepping. 

Adaptivity is another aspects that would be interesting to study, that is, creating and deleting cells dynamically, during the simulation, using larger cells in zones where the flow is simple, and smaller cells near the boundary of the fluid, to better capture the fluid shape and to more accurately represent surface tension. To even better adapt the geometry of the flow, anisotropic cells could be used, defined by a tensor attached to each point, and cells that are intersections between Laguerre cells and ellipsoids. 

Performance can be improved, using a GPU implementation of the clipped Laguerre diagram \cite{DBLP:journals/cad/YanWLL13,DBLP:journals/tog/RayS0L18,basselin:hal-03169575}. This will require to find a way of making the representation more compact (convex polytopes with 200 faces do not fit well in GPU caches). Direct representation of curved polytopes \cite{DBLP:journals/siamnum/LeclercMSS20} may be an option.

Finally, it is important to stress the fact that the so-defined free surface Lagrangian mesh \emph{smoothly} depends on the parameters $\bx_i$. Namely, the function $K$ is differentiable up to the second order in both $\psi$ and $\bx$, and the second-order derivatives have simple expressions (see \cite{degournay:hal-01721681}). As a consequence, this representation can be used to optimize an objective function that depends on a shape of prescribed volume, where the shape is parameterized by the $\bx_i$'s. Then, the objective function will \emph{smoothly} depend on the $\bx_i$'s, making it possible to optimize it with efficient numerical methods (Newton, LBFGS, \ldots).

In future works, I plan to explore different applications, in astrophysics (extensions of \cite{levy:hal-03081581}), and also in shape and topology optimization that may benefit from the differentiability of the representation. It is also interesting to notice that the current trend of numerical methods called "artificial intelligence" mostly rely on representations that are differentiable with respect to the parameters. I think the representation described here can be used to represent latent spaces of shapes, and efficiently fit them to databases of 3D objects.

\section*{Acknowedgements}
I with to thank Quentin Mérigot, Yann Brenier and Jean-David Benamou for many discussions, Hélène de Maleprade for sharing the photo of her amazing "droplet hurdles race" experiment (with Rachid Bendimerad, Christophe Clanet and David Quéré), and Kevin Mattheus Moerman for indicating me Hélène's work on Twitter (and challenging me to simulate it !).

\bibliographystyle{elsarticle-num}
\bibliography{POT.bib}

\end{document}